\documentclass[aps,prx,reprint,amsmath,amssymb,superscriptaddress,nofootinbib]{revtex4-2}

\usepackage{graphicx}
\graphicspath{{./}}
\usepackage{bm}
\usepackage{tikz}
\usetikzlibrary{decorations.pathmorphing,decorations.markings,arrows.meta,positioning,calc}
\usepackage[hidelinks]{hyperref}

\newcommand{\ket}[1]{\lvert #1 \rangle}
\newcommand{\bra}[1]{\langle #1 \rvert}
\newcommand{\ketbra}[2]{\lvert #1 \rangle\!\langle #2 \rvert}
\newcommand{\avg}[1]{\langle #1 \rangle}
\newcommand{\Tr}{\mathrm{Tr}}
\newcommand{\Lop}{\mathcal{L}}
\newcommand{\Dop}{\mathcal{D}}
\newcommand{\Vint}{V^{\mathrm{int}}}
\newcommand{\ad}{a^{\dagger}}
\newcommand{\bd}{b^{\dagger}}
\newcommand{\sigm}{\sigma^{-}}
\newcommand{\sigp}{\sigma^{+}}
\newcommand{\dg}{^{\dagger}}
\newcommand{\Lint}{\Lop_{\mathrm{int}}}
\newcommand{\bvec}[1]{\mathbf{#1}}
\newcommand{\normord}[1]{\mathord{:}\mspace{2mu}#1\mspace{2mu}\mathord{:}}
\newcommand{\contr}[2]{\langle #1 #2 \rangle_{\mathrm{ss}}}

\usetikzlibrary{decorations.pathmorphing,decorations.markings,arrows.meta,calc}

\def\ldmarkpos{0.5}
\tikzset{
  ld mark at/.code={\def\ldmarkpos{#1}},
  ld tipR/.style={fill,draw=none,line join=round},
  ld arrow right/.style={postaction={decorate,decoration={markings,
      mark=at position \ldmarkpos with {\pgftransformresetnontranslations
        \path[ld tipR] (-1.15mm,-1.05mm) -- (1.15mm,0mm) -- (-1.15mm,1.05mm) -- cycle;}}}},
  ld arrow left/.style={postaction={decorate,decoration={markings,
      mark=at position \ldmarkpos with {\pgftransformresetnontranslations
        \path[ld tipR] (1.15mm,-1.05mm) -- (-1.15mm,0mm) -- (1.15mm,1.05mm) -- cycle;}}}},
  ld arrow right bg/.style={postaction={decorate,decoration={markings,
      mark=at position \ldmarkpos with {\pgftransformresetnontranslations
        \fill[white] (-1.45mm,-1.1mm) rectangle (1.45mm,1.1mm);
        \path[ld tipR] (-1.15mm,-1.05mm) -- (1.15mm,0mm) -- (-1.15mm,1.05mm) -- cycle;}}}},
  ld arrow left bg/.style={postaction={decorate,decoration={markings,
      mark=at position \ldmarkpos with {\pgftransformresetnontranslations
        \fill[white] (-1.45mm,-1.1mm) rectangle (1.45mm,1.1mm);
        \path[ld tipR] (1.15mm,-1.05mm) -- (-1.15mm,0mm) -- (1.15mm,1.05mm) -- cycle;}}}},
  ld wavy/.style={draw=none,preaction={draw,line width=0.8pt,decorate,
    decoration={snake,amplitude=1.5pt,segment length=6pt}}},
  ld coil/.style={draw=none,preaction={draw,line width=0.55pt,decorate,
    decoration={coil,aspect=0.5,amplitude=1.8pt,segment length=3.2pt}}},
  ld a/.style={ld wavy,ld arrow right bg},
  ld ad/.style={ld wavy,ld arrow left bg},
  ld b/.style={ld coil,ld arrow right bg},
  ld bd/.style={ld coil,ld arrow left bg},
  ld sm/.style={line width=0.9pt,ld arrow right},
  ld sp/.style={line width=0.9pt,ld arrow left},
  ld double/.style={double,double distance=1.6pt,line width=0.6pt},
  ld d1/.style={ld double,postaction={decorate,decoration={markings,
      mark=at position \ldmarkpos with {\filldraw[fill=white,line width=0.7pt] (0,0) circle (2.4pt);}}}},
  ld d2/.style={ld double,ld arrow right bg},
  ld d3/.style={ld double,ld arrow left bg},
  ld vtx/.style={circle,fill=black,inner sep=0pt,minimum size=5.6pt},
  ld order/.style={font=\scriptsize,inner sep=1pt,anchor=south},
  ld label/.style={font=\footnotesize,inner sep=1.5pt},
  ld ext/.style={font=\small,inner sep=2pt},
}

\newcommand{\ldglyph}[2][0.9cm]{%
  \tikz[baseline=-0.6ex]\draw[#2] (#1,0) -- (0,0);}

\begin{document}

\title{Operator-language Feynman rules for driven-dissipative quantum systems:\\ from mean field to non-Gaussian photon correlations}

\author{Peter Ehlers}
\author{Phi Hung Nguyen}
\author{Daniel Soh}
\email[Corresponding author: ]{danielsoh@arizona.edu}
\affiliation{James C.\ Wyant College of Optical Sciences, University of Arizona, Tucson, Arizona 85721, USA}

\date{\today}

\begin{abstract}
We construct Feynman rules for Lindblad master equations in the operator
language of quantum optics: a fixed algebra of elementary operator moves and
closed-form propagators for perturbation theory about an exactly solvable, in
most cases Gaussian, generator. The expansion separates into a kinematic part
(the shift algebra on the Wick-ordered eigenoperators
$O_{mn}=\normord{(c\dg)^mc^n}$, from which the vertex table for a specified
interaction is assembled) and a dynamical part that enters only through the
resolvent of the unperturbed generator, diagonal and additive for any stable,
diagonalizable Gaussian $\Lop_0$, including thermal and squeezed reservoirs
and correlated decay, and equally available for a two-level emitter, driven
or undriven. In the driven Kerr cavity the loop expansion is the
semiclassical expansion and the series is asymptotic. In a strongly driven
two-level atom, dressed lines replace a series of finite radius by one accurate
deep into saturation, one line per Mollow component. For Kerr rings and chains
the cost is polynomial in the number of sites at fixed perturbative order. On a three-site ring the
diagrams recover the non-Gaussian part of the three-photon correlation that
the Gaussian cumulant closure lacks; carried to tenth order they give the
smallest $g^{(3)}$ error of the methods compared for $U\leq0.25\kappa$ at a
cost below that of a fourth-order closure, and third- and fourth-order
closures are more accurate at moderate coupling. The same rules,
extended to the counting-field-tilted generator, give the photon-counting
cumulants of a disordered eight-site chain beyond direct diagonalization;
quantum-jump trajectories at three couplings agree with them to within
$1.6$ standard errors, within one at the strongest coupling, where they
place the third cumulant six standard errors above the Gaussian value.
\end{abstract}

\maketitle

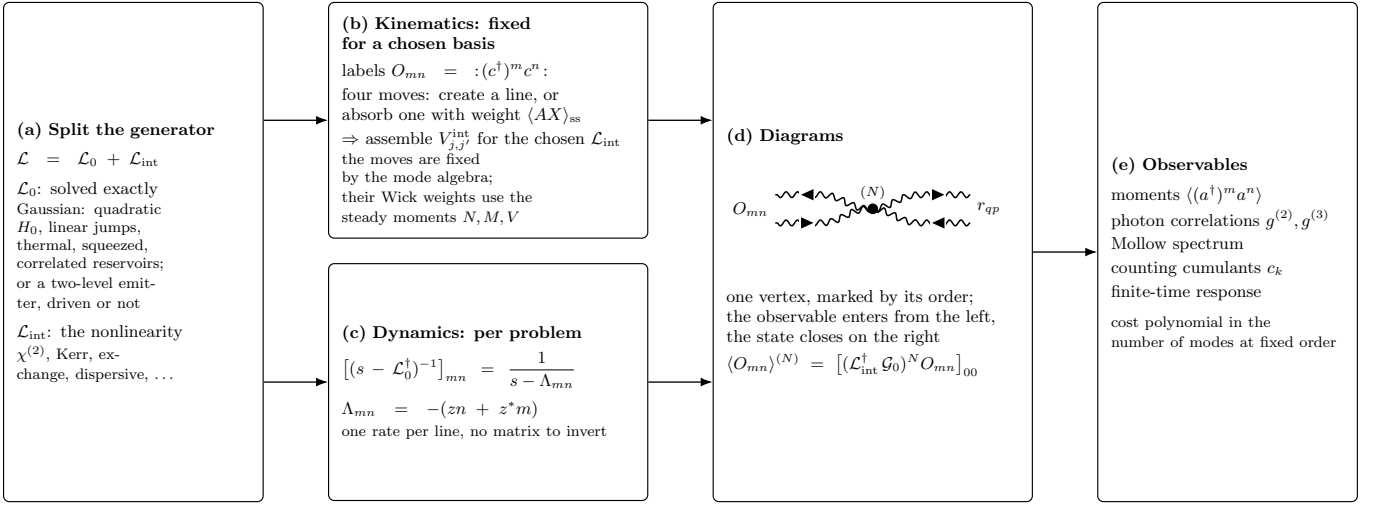
\begin{figure*}[t]
\centering
\resizebox{\textwidth}{!}{%
\begin{tikzpicture}[x=1cm,y=1cm,
  box/.style={draw,rounded corners=3pt,line width=0.6pt,inner sep=7pt,align=left,font=\small,anchor=north west},
  arr/.style={-{Latex[length=2.2mm]},line width=0.7pt}]
\def\xa{0}\def\xb{6.0}\def\xd{13.1}\def\xe{20.2}
\node[box,text width=5.4cm] (kin) at (\xb,3.0) {%
  \textbf{(b) Kinematics: fixed for a chosen basis}\\[4pt]
  labels $O_{mn}=\normord{(c\dg)^mc^n}$\\[2pt]
  four moves: create a line, or\\
  absorb one with weight $\contr{A}{X}$\\[2pt]
  $\Rightarrow$ assemble $\Vint_{j,j'}$ for the chosen $\Lint$\\[1pt]
  {\footnotesize the moves are fixed by the mode algebra;\\
  their Wick weights use the steady moments $N,M,V$}};
\path let \p1=(kin.north), \p2=(kin.south) in
  node[box,text width=5.4cm,minimum height={\y1-\y2}] (dyn)
  at ($(kin.south west)+(0,-0.45)$) {%
  \textbf{(c) Dynamics: per problem}\\[4pt]
  $\big[(s-\Lop_0\dg)^{-1}\big]_{mn}=\dfrac{1}{s-\Lambda_{mn}}$\\[6pt]
  $\Lambda_{mn}=-(zn+z^{*}m)$\\[1pt]
  {\footnotesize one rate per line, no matrix to invert}};
\path let \p1=(kin.north), \p2=(dyn.south) in
  node[box,text width=4.3cm,minimum height={\y1-\y2}] (split) at (\xa,3.0) {%
  \textbf{(a) Split the generator}\\[5pt]
  $\Lop=\Lop_0+\Lint$\\[5pt]
  $\Lop_0$: solved exactly\\[1pt]
  {\footnotesize Gaussian: quadratic $H_0$, linear jumps,\\
  thermal, squeezed, correlated reservoirs;\\
  or a two-level emitter, driven or not}\\[5pt]
  $\Lint$: the nonlinearity\\[1pt]
  {\footnotesize $\chi^{(2)}$, Kerr, exchange, dispersive, \ldots}}
  node[box,text width=5.4cm,minimum height={\y1-\y2}] (dia) at (\xd,3.0) {}
  node[box,text width=4.3cm,minimum height={\y1-\y2}] (out) at (\xe,3.0) {%
  \textbf{(e) Observables}\\[5pt]
  moments $\avg{(\ad)^ma^n}$\\[2pt]
  photon correlations $g^{(2)},g^{(3)}$\\[2pt]
  Mollow spectrum\\[2pt]
  counting cumulants $c_k$\\[2pt]
  finite-time response\\[6pt]
  {\footnotesize cost polynomial in the\\
  number of modes at fixed order}};
\node[box,draw=none,text width=5.4cm,anchor=center] (diatxt) at (dia.center) {%
  \textbf{(d) Diagrams}\\[2.6cm]
  one vertex, marked by its order;\\
  the observable enters from the left,\\
  the state closes on the right\\[4pt]
  $\avg{O_{mn}}^{(N)}=\big[(\Lint\dg\,\mathcal G_0)^N O_{mn}\big]_{00}$};
\coordinate (dc) at ($(diatxt.north)+(0,-1.75)$);
\begin{scope}[shift={(dc)}]
  \draw[ld ad,ld mark at=0.68] (0,0) .. controls (-0.42,0) and (-0.5,0.26) .. (-0.92,0.26) -- (-1.8,0.26);
  \draw[ld a,ld mark at=0.68] (0,0) .. controls (-0.42,0) and (-0.5,-0.26) .. (-0.92,-0.26) -- (-1.8,-0.26);
  \draw[ld a,ld mark at=0.32] (1.8,0.26) -- (0.92,0.26) .. controls (0.5,0.26) and (0.42,0) .. (0,0);
  \draw[ld ad,ld mark at=0.32] (1.8,-0.26) -- (0.92,-0.26) .. controls (0.5,-0.26) and (0.42,0) .. (0,0);
  \node[ld vtx] at (0,0) {};
  \node[ld order,fill=white,fill opacity=0.85,text opacity=1,rounded corners=1pt] at (0,0.13) {$(N)$};
  \node[ld ext,anchor=east] at (-1.85,0) {$O_{mn}$};
  \node[ld ext,anchor=west] at (1.85,0) {$r_{qp}$};
\end{scope}
\draw[arr] (split.east|-kin.west) -- (kin.west);
\draw[arr] (split.east|-dyn.west) -- (dyn.west);
\draw[arr] (kin.east) -- (dia.west|-kin.east);
\draw[arr] (dyn.east) -- (dia.west|-dyn.east);
\draw[arr] (dia.east|-out.west) -- (out.west);
\end{tikzpicture}}
\caption{The construction in one picture. (a)~A Lindbladian is split into a
Gaussian part $\Lop_0$, diagonalized exactly in Sec.~\ref{sec:gaussian}, and a
nonlinear remainder $\Lint$. (b)~Once the eigenoperator basis and its steady
covariance are chosen, every interaction acts through four elementary
moves whose Wick-basis weights are steady-state contractions. The move
algebra is fixed by the mode algebra, while the vertex table for a specified
$\Lint$ is assembled mechanically from those moves
(Sec.~\ref{sec:vertices}). (c)~The dynamics
enters only through the resolvent of $\Lop_0$, a scalar on every label and,
for Gaussian $\Lop_0$, a sum of one rate per line
(Sec.~\ref{sec:propagator}). (d)~Each order of perturbation theory is a
single-vertex diagram built from these two ingredients, evaluated by gluing
first-order tables (Sec.~\ref{sec:ssrules}). (e)~Physical moments, photon
correlations, spectra and counting statistics follow from the same diagram
sets with different weights (Sec.~\ref{sec:recombine}), at a cost polynomial
in the number of modes.}
\label{fig:overview}
\end{figure*}

\section{Introduction}

Diagrammatic perturbation theory is the organizing principle of interacting
quantum field theory. Each term of a Dyson series corresponds to a picture,
the pictures at a given order can be enumerated systematically, and the value
of each follows from a short list of rules for its parts.

Open quantum systems described by a Lindblad master
equation~\cite{lindblad1976generators,gorini1976completely} have no equally
standard diagrammatic formulation in the operator language in which many
quantum-optics problems are posed. Perturbative treatments commonly either
truncate the density matrix in a chosen basis and integrate numerically, or
linearize about a mean-field solution and retain Gaussian fluctuations
\cite{Gardiner2004,Carmichael1999,Clerk2010}. The first becomes exact as the
basis truncation is converged, but its cost grows exponentially with the
number of modes; the second is analytic and inexpensive, but by itself
provides no systematic hierarchy of non-Gaussian corrections. Between them
lies a broad class of polynomial-cost approximations for driven-dissipative
lattices~\cite{weimer2021simulation,fazio2025manybody}: cumulant and cluster
closures~\cite{kira2008cluster,casteels2016truncated,plankensteiner2022quantumcumulants,verstraelen2023quantum},
phase-space sampling with the positive-$P$ and truncated Wigner
representations~\cite{drummond1980generalised,deuar2021fully},
corner-space renormalization~\cite{finazzi2015corner}, linked-cluster
expansions~\cite{biella2018linked}, variational matrix-product and
tensor-network steady
states~\cite{cui2015variational,mascarenhas2015matrix,kshetrimayum2017simple,mckeever2021stable},
neural-network density
matrices~\cite{vicentini2019variational,hartmann2019neural,nagy2019variational},
and open-system dynamical mean-field theory~\cite{scarlatella2021dynamical}.
These approaches fix a level of description in advance---for example a
closure order, bond dimension, or corner size---and are improved by raising
that level.

A diagrammatic expansion is instead organized order by order in the
interaction, with each order built from the one before. Keldysh field theory
provides such an expansion
\cite{keldysh1965diagram,kamenev2011field,sieberer2016keldysh,sieberer2013dynamical,maghrebi2016nonequilibrium}
and has been used to formulate diagram rules for Lindblad problems
directly~\cite{muller2017keldysh,lang2020nonequilibrium,reimer2019density,ferguson2021open},
but in a functional-integral language different from the operator language in
which many quantum-optics problems are naturally expressed. Superoperator
perturbation theory in the operator language also exists: Li, Petruccione and
Koch~\cite{li2014perturbative,li2016resummation} expand about an exactly
solved local generator and resum the series to reach open lattices at strong
on-site nonlinearity. What has been missing, to our knowledge, is a reusable
operator-language vocabulary of lines and vertices together with propagators
available in closed form, so that diagrams can be evaluated by a fixed set of
rules rather than by re-deriving the perturbation series for each problem.

The purpose of this paper is to construct such an expansion directly in the
operator language, and to show that it separates into a kinematic part and a
dynamical part (Fig.~\ref{fig:overview}). The kinematic part is an algebra of
index shifts and contractions that can be generated once for a given operator
algebra. The dynamical part is the resolvent of a chosen solvable
unperturbed generator $\Lop_0$. This separation makes the choice of
$\Lop_0$ a choice of which physics to resum into the propagators and which to
retain as interaction vertices.

Three results from the rules organize the second half of the paper. First,
the expansion has a loop counting. For the driven Kerr cavity the loop-free
diagrams sum to mean-field theory at every order, and each loop carries an
additional inverse power of the coherent photon number, so the semiclassical
expansion can be read directly from the pictures
(Sec.~\ref{sec:loops}). The perturbation series in the Kerr interaction is
asymptotic, and the diagrammatic organization makes the origin of this
behavior transparent. Second, the freedom to choose $\Lop_0$ provides a
resummation mechanism. For a strongly driven two-level atom the bare
expansion contains a geometric ladder with finite radius
$\Omega^2<2D$; incorporating the drive into $\Lop_0$ replaces the bare
atomic propagators by dressed ones and removes this restriction. At fourth
order the dressed expansion gives the cavity photon number within $2\%$ of
the exact result deep into saturation, with one dressed line associated with
each Mollow component (Sec.~\ref{sec:emitter}). Third, at fixed perturbative
order the number of operator labels grows polynomially with the number of
modes, allowing the rules to be applied beyond the regime of practical
direct diagonalization. On a driven-dissipative Kerr ring they improve
systematically on the second-order Gaussian cumulant closure and recover the
non-Gaussian part of the three-photon correlation that a Gaussian closure
sets to zero by construction. Carried to tenth order they give the smallest
$g^{(3)}$ error of the closures and diagram sets compared for
$U\leq0.25\kappa$, at a cost below that of a fourth-order closure, whereas
the fourth-order closure is more accurate on $g^{(2)}$ from
$U\approx0.15\kappa$ and on the photon number from $U\approx0.125\kappa$,
and third- and fourth-order closures are more accurate at moderate coupling
on the three-site ring (Sec.~\ref{sec:cumulant}). Extended to the counting-field-tilted generator,
whose Gaussian part is solved in closed form, the same rules give
photon-counting cumulants for a disordered eight-site chain whose tilted
Liouvillian is beyond practical direct diagonalization. Quantum-jump
trajectories at three couplings provide an independent benchmark
(Sec.~\ref{sec:fcs}). Throughout, the perturbative series is reported at a
stated, fixed order and, wherever a converged exact result exists, compared
with it.

The kinematic structure is simplest in vacuum. Take as labels the
normal-ordered monomials
\[
O_{mn}=(\ad)^m a^n,
\]
which span the operator algebra of a single bosonic mode. Any superoperator
acts as a matrix on this label lattice, and multiplication by the mode
operators is generated by four shift relations,
\begin{equation}
  \ad O_{mn}=O_{(m+1)n},\quad [a,O_{mn}]=m\,O_{(m-1)n},
  \label{eq:introshift1}
\end{equation}
\begin{equation}
  O_{mn}a=O_{m(n+1)},\quad [O_{mn},\ad]=n\,O_{m(n-1)}.
  \label{eq:introshift2}
\end{equation}
These identities are statements about the mode algebra alone. They do not
depend on the Hamiltonian, on the jump operators, or on how the generator has
been split, and they generate the elementary vertex moves used throughout the
paper. For a thermal or squeezed Gaussian steady state the corresponding
eigenoperators are Wick-ordered monomials, and the same construction survives
with additional contraction weights fixed by the steady-state covariance.

The dynamical part changes from problem to problem. It is the resolvent of the
unperturbed generator on the eigenoperator lattice,
$(s-\Lop_0)^{-1}$, whose matrix elements are the propagators of the
diagrams. The method therefore requires a label set closed under the vertex
moves and an unperturbed generator whose relevant resolvent can be written in
closed form. When $\Lop_0$ admits the required eigenoperator basis, the
propagator is a scalar on each eigenoperator label. For open Gaussian systems,
whose Hamiltonian is quadratic and whose jump operators are linear in the mode
operators, the eigenvalues are additionally additive in the line content, so
the propagator denominator can be read directly from the picture line by
line. Additivity is convenient but not essential: a diagonal but
non-additive eigenvalue still gives a scalar propagator attached to the full
label. An observable or state that is not itself an eigenoperator is expanded
in this basis, with one diagram for each term
(Sec.~\ref{sec:propagator}).

Gaussian Lindbladians form a broad exactly solvable class in quantum optics:
their complete superoperator spectrum can be constructed, not merely their
correlation functions. Section~\ref{sec:gaussmode} gives the construction for
a single mode. For jump operators linear in creation and annihilation
operators, the Heisenberg eigenoperators are Wick-ordered monomials in the
damped mode, with contractions fixed by two steady-state moments $N$ and
$M$, while the Schr\"odinger eigenoperators are commutator ladders acting on
the steady state. Vacuum loss ($N=M=0$), a thermal bath ($M=0$), and a
squeezed bath ($M\neq0$) are three limits of the same construction.
Section~\ref{sec:general} extends it to many modes, parametric drives,
correlated decay, and Bogoliubov-rotated problems. Finite-dimensional emitter
sectors can be incorporated similarly once their left and right
eigenoperators are known; the driven two-level emitter is treated explicitly
in Sec.~\ref{sec:emitter}.

Two features distinguish the resulting calculus from the vacuum field theory
it resembles. First, because a Lindbladian acts on operators from both sides,
a factor of $a$ and a factor of $\ad$ propagate with the conjugate rates
$z$ and $z^{*}$. They are therefore represented by lines with opposite
arrows, the operator-language counterpart of the two branches of the Keldysh
contour. Second, because $\Lop_0$ is non-Hermitian as a superoperator, its
left and right eigenoperators differ. The observable is decomposed in the
Heisenberg eigenbasis and the initial state in the Schr\"odinger eigenbasis;
the two are biorthogonal under the trace pairing, and this biorthogonality
closes the calculation. In the compact representation used here, the
$N$ successive interaction insertions contributing at perturbative order
$N$ are collapsed into a single effective vertex marked by that order. The
lines of the observable enter from the left, while the lines leaving the
vertex close on one component of the initial state. The equivalent
one-vertex-per-interaction chain representation is obtained by resolving the
intermediate labels explicitly.

The paper is organized as follows. Section~\ref{sec:pictures} records the
algebraic facts about non-Hermitian superoperators used below.
Section~\ref{sec:gaussian} assembles the solvable starting points: the single
Gaussian mode and its vacuum, thermal, and squeezed limits; the coherently
driven cavity; the general Gaussian Lindbladian including correlated
multimode decay; and the emitter with and without drive.
Section~\ref{sec:diagrams} derives the expansion, states the vertex rules for
one mode and for many, classifies the propagators, and gives the Feynman rules
together with the map from eigenoperator moments to physical observables.
Section~\ref{sec:examples} works the driven Kerr cavity and the driven atom,
and Sec.~\ref{sec:ring} treats the Kerr ring and chain, where the method is
compared with mean-field and cumulant closures.
Section~\ref{sec:discussion} states the domain and limitations of the method.
Further examples, finite-time rules, derivations, convergence tests, and
numerical details are collected in the Supplemental Material~\cite{SM}.

\subsection{Relation to existing approaches}
\label{sec:related}

\emph{Damping bases and third quantization.} The eigenoperators of
Sec.~\ref{sec:gaussian} are the damping basis of Briegel and
Englert~\cite{briegel1993quantum}, who introduced the left and right
eigenoperators of a dissipative Liouvillian and their biorthogonality;
Barnett and Stenholm~\cite{barnett2000spectral} completed the spectral
decomposition, and Englert and Morigi~\cite{englert2002five} extended the
basis to thermal and squeezed reservoirs. Section~\ref{sec:gaussian} collects
the single-mode Gaussian family into one construction labelled by the two
steady-state moments $N$ and $M$, Eq.~\eqref{eq:Omn}, and supplies the
inverse map of Sec.~\ref{sec:recombine}. For quadratic Lindbladians the same
spectra follow from third
quantization~\cite{prosen2008third,prosen2010quantization,prosen2010exact,barthel2022solving,mcdonald2023third};
the spectral structure of Liouvillians is reviewed in
Ref.~\cite{minganti2018spectral}. The Wick construction used here is the same
statement in the language of normal-ordered monomials, which is the form the
vertex algebra of Sec.~\ref{sec:vertices} acts on. What is new is not the
spectrum but its use as the sole dynamical input of an expansion whose
kinematics is fixed independently.

\emph{Superoperator perturbation theory.} Li, Petruccione and
Koch~\cite{li2014perturbative} developed perturbation theory for Markovian
open systems at the level of superoperators, with a diagrammatic
representation of the terms, and resummed it in Ref.~\cite{li2016resummation}
for driven-dissipative lattices. The two expansions are complementary. They
take the on-site problem, nonlinearity included, as the solved part and expand
in the intersite coupling, so their unperturbed generator is diagonalized
numerically and its vertices change from problem to problem. We take the
Gaussian part as the solved part and expand in the nonlinearity, so the
propagators are in closed form and additive in the line content. The
elementary move algebra is fixed by
Eqs.~\eqref{eq:introshift1}--\eqref{eq:introshift2}, and the vertex table for
a chosen interaction is assembled from those moves. Theirs is the natural
tool at strong on-site interaction and weak hopping; ours at weak
nonlinearity with arbitrary hopping, correlated decay, or reservoir
squeezing. The single-vertex form of every diagram is specific to the present
construction. Resolvent and Drazin-inverse constructions of the perturbed
steady state~\cite{lenarcic2018perturbative,nagib2025exact} address the same
object non-diagrammatically, and a continued-fraction treatment in the
damping basis~\cite{hartmann2026exact} shares our starting point for
few-emitter problems.

\emph{Double-sided diagrams.} Double-sided Feynman diagrams enumerate
Liouville pathways in nonlinear spectroscopy~\cite{Mukamel1995}. They track
the state of the density matrix between light--matter interactions, with the
bath entering through phenomenological dephasing rates. The diagrams here
track operator content: a line is a factor of $a$ or $\ad$, the elementary
moves follow from the mode algebra and assemble the interaction-specific
vertex table, and the dissipation is inside the propagator exactly. For a few-level emitter driven by external pulses the
Liouville-pathway picture remains the more natural one.

\emph{Keldysh and real-time techniques.} Keldysh field
theory~\cite{kamenev2011field,sieberer2016keldysh} is the other systematic
diagrammatic route, and the correspondence is close: the two arrow
orientations are the operator-language counterpart of the two contour
branches, and the loop counting of Sec.~\ref{sec:loops} is the usual
semiclassical expansion. Diagram rules for Lindblad dynamics have been
formulated on the contour for correlated gain and
loss~\cite{muller2017keldysh}, for strongly interacting
photons~\cite{lang2020nonequilibrium}, for the density-operator
propagator~\cite{reimer2019density,schoeller1994mesoscopic,koenig1996zero},
and for the steady state beyond the golden rule~\cite{ferguson2021open}.
Most recently, Blommel \emph{et al.}~\cite{blommel2025manybody} formulated a
Keldysh--Lindblad many-body perturbation theory in which dissipative
interactions are represented by contour interaction lines and enter the
Kadanoff--Baym self-energy structure. The present construction uses a
different representation: observables are expanded directly in a Liouvillian
eigenoperator basis, the free Liouvillian resolvent supplies the propagator,
and algebraic operator shifts assemble the interaction-specific vertices.
Diagrammatic Monte Carlo has been applied to dissipative impurity
models~\cite{vanhoecke2024diagrammatic}, and tensor-train compression to
high-order diagrammatic sums~\cite{nunez2022learning}. In the operator
formulation the objects in the diagrams are the operators one measures, so no
functional integral, response field, or Wigner transform is needed, and the
reservoir structure that a contour treatment carries in its self-energies is
absorbed into the contractions that enter the Wick-basis vertex assembly.

\emph{Resummations of the drive and counting statistics.} Resumming a strong
coherent drive into dressed states is the dressed-atom description of
resonance fluorescence~\cite{cohentannoudji1977dressed,mollow1969power};
Sec.~\ref{sec:emitter} shows that the dressed propagator enters the diagram
algebra as a line. The counting-field formalism of Sec.~\ref{sec:fcs} is
standard~\cite{esposito2009nonequilibrium,bagrets2003full,flindt2008counting,garrahan2010thermodynamics,landi2024current},
and its application to single Kerr resonators is
known~\cite{kewming2022diverging}; the explicit tilted adjoint eigenoperator and its integration into the operator-space diagram rules are new to our knowledge.

\section{Lindbladians as non-Hermitian superoperators}
\label{sec:pictures}

\subsection{Two pictures}

We take as given a Lindblad master equation
\begin{align}
  \Lop\rho &= -i[H,\rho] + \sum_j \gamma_j \Dop[L_j]\rho, \nonumber\\
  \Dop[L]\rho &= L\rho L\dg - \tfrac{1}{2}\{L\dg L,\rho\},
  \label{eq:lindblad}
\end{align}
with $H$ Hermitian and $L_j$ arbitrary. Every measured quantity has the form
$\Tr[O f(\Lop)\rho]$ for some observable $O$, initial state $\rho$, and
function $f$. The identity
\begin{equation}
  \Tr\!\left[O\,\Lop\rho\right] = \Tr\!\left[(\Lop\dg O)\,\rho\right]
  \label{eq:duality}
\end{equation}
defines the adjoint (Heisenberg-picture) generator
\begin{equation}
  \Lop\dg O = i[H,O] + \sum_j \gamma_j\left(L_j\dg O L_j
    - \tfrac{1}{2}\{L_j\dg L_j, O\}\right).
  \label{eq:adjoint}
\end{equation}
Letting $f(\Lop)$ act to the right on $\rho$ is the Schr\"odinger picture;
letting it act to the left on $O$ is the Heisenberg picture. Both give the same
expectation values, and we move freely between them.

For a single decay channel with $H_0=\omega L\dg L$, Eq.~\eqref{eq:adjoint}
reduces to
\begin{equation}
  \Lop_0\dg O = -[O,L\dg]\,z\,L - L\dg z^{*}[L,O],
  \qquad z = \tfrac{\gamma}{2}+i\omega,
  \label{eq:compact}
\end{equation}
which is the template for Sec.~\ref{sec:gaussian}. The complex rate $z$
combines dissipation and detuning, and the appearance of $z$ on one side and
$z^{*}$ on the other is the origin of the two arrow orientations of the
diagram lines.

\subsection{Failure of the Leibniz rule}

A Hamiltonian generator is a derivation: $i[H,AB]=(i[H,A])B+A(i[H,B])$. A
Lindbladian is not, and the obstruction is the reason interaction vertices in
this formalism are not products of free-field vertices. A short calculation
gives
\begin{equation}
  \Lop\dg(AB) = (\Lop\dg A)B + A(\Lop\dg B)
    + \sum_j \gamma_j\,[L_j\dg,A]\,[B,L_j].
  \label{eq:leibniz}
\end{equation}
The Hamiltonian and anticommutator pieces cancel between the two sides,
leaving the recycling terms, whose difference is
\begin{align}
  &L\dg ABL - L\dg ALB - AL\dg BL + AL\dg LB \nonumber \\
  &\quad = (L\dg A - AL\dg)(BL - LB) = [L\dg,A][B,L].
\end{align}
The correction is bilinear in the commutators of the jump operators with the
two factors. It vanishes whenever one of $A$, $B$ commutes with all $L_j$, in
particular for the identity, so $\Lop\dg$ annihilates constants. When it does
not vanish it produces the fusion vertices of Sec.~\ref{sec:vertices}, in
which two existing lines are joined rather than a new line created.

\subsection{Left and right eigenoperators}

$\Lop$ preserves Hermiticity, positivity, and trace, but it is not a Hermitian
superoperator, so its left and right eigenoperators differ. Writing
\begin{equation}
  \Lop\dg O_{\mu} = \Lambda_{\mu}O_{\mu},
  \qquad
  \Lop\, r_{\mu} = \Lambda_{\mu} r_{\mu},
\end{equation}
the two families are biorthogonal under the trace pairing,
$\Tr[O_{\mu}r_{\nu}] \propto \delta_{\mu\nu}$, and this pairing, not a
Hilbert--Schmidt inner product, is used to expand states and observables. A
calculation of $\Tr[Oe^{\Lop\Delta t}\rho_0]$ requires decomposing $O$ in the
Heisenberg (left) eigenbasis and $\rho_0$ in the Schr\"odinger (right)
eigenbasis. The method is efficient when both decompositions are short.

\section{Solvable Gaussian starting points}
\label{sec:gaussian}

This section collects the exactly diagonalized Lindbladians that can be taken
as $\Lop_0$: the single Gaussian mode and its vacuum, thermal, and squeezed
limits, the coherently driven cavity, the general Gaussian Lindbladian with
correlated multimode decay, and the two-level emitter with and without drive.
The continuum limit and pure dephasing are treated in the Supplemental
Material~\cite{SM}. A reader interested only in the diagrammatic rules may
proceed to Sec.~\ref{sec:diagrams} and refer back as needed.

\subsection{A single Gaussian mode}
\label{sec:gaussmode}

A bosonic Lindbladian is Gaussian when its Hamiltonian is at most quadratic and
every jump operator is at most linear in $a$ and $\ad$. After the steady
displacement $\avg{a}_{\mathrm{ss}}$ is subtracted (Sec.~\ref{sec:drive}),
$\Lop_0\dg$ maps linear operators to linear operators
(Sec.~\ref{sec:general}). Suppose it has a damped mode $c$ with $[c,c\dg]=I$,
\begin{equation}
  \Lop_0\dg c = -z\,c,
  \qquad
  \Lop_0\dg c\dg = -z^{*}c\dg,
  \qquad
  \mathrm{Re}\,z>0 .
  \label{eq:dampedmode}
\end{equation}
The steady Gaussian fluctuations are then fixed by two ordered second moments,
\begin{equation}
  N = \avg{c\dg c}_{\mathrm{ss}},
  \qquad
  M = \avg{cc}_{\mathrm{ss}},
  \label{eq:NM}
\end{equation}
together with $\avg{c\dg c\dg}_{\mathrm{ss}}=M^{*}$ and, from $[c,c\dg]=I$,
$\avg{cc\dg}_{\mathrm{ss}}=N+1$. Positivity of $\rho_{\mathrm{ss}}$ requires
$|M|^2\leq N(N+1)$. (The symbol $N$ denotes this moment; the order of
perturbation theory is written $N$ only with the word ``order'' or as a
superscript $(N)$.) Table~\ref{tab:NM} lists the values for the steady states
met below.

\begin{table}[b]
\caption{Steady-state moments $N=\avg{c\dg c}$ and $M=\avg{cc}$ of the damped
mode for the steady states considered in Sec.~\ref{sec:singlemode};
$\bar n$ is a thermal occupation and $\xi=re^{i\phi}$ a squeezing parameter.}
\label{tab:NM}
\begin{ruledtabular}
\begin{tabular}{lcc}
Steady state & $N$ & $M$ \\ \hline
Vacuum & $0$ & $0$ \\
Thermal & $\bar n$ & $0$ \\
Squeezed vacuum & $\sinh^2r$ & $-\tfrac12e^{i\phi}\sinh 2r$ \\
Squeezed thermal & $(\bar n+\tfrac12)\cosh 2r-\tfrac12$ & $-(\bar n+\tfrac12)e^{i\phi}\sinh 2r$ \\
\end{tabular}
\end{ruledtabular}
\end{table}

\emph{Left eigenoperators.} Define the Gaussian Wick generating function
\begin{align}
  \mathcal{W}(s,t) &= e^{-\frac12M^{*}s^2-Nst-\frac12Mt^2}\,e^{sc\dg}e^{tc}
   \nonumber\\
   &= \sum_{m,n\geq0}\frac{s^mt^n}{m!\,n!}\,O^{(N,M)}_{mn} .
  \label{eq:wickgen}
\end{align}
Its coefficients are the Wick-ordered monomials
$O^{(N,M)}_{mn}=\normord{(c\dg)^mc^n}$,
\begin{widetext}
\begin{equation}
  O^{(N,M)}_{mn} = \sum_{\substack{p,q,k\geq0\\ 2p+k\leq m,\ 2q+k\leq n}}
    \frac{m!\,n!\,(-M^{*})^p(-M)^q(-N)^k}{2^{p+q}\,p!\,q!\,k!\,(m-2p-k)!\,(n-2q-k)!}
    \,(c\dg)^{m-2p-k}\,c^{\,n-2q-k},
  \label{eq:Omn}
\end{equation}
\end{widetext}
with
\begin{equation}
  \Lop_0\dg O^{(N,M)}_{mn} = \Lambda_{mn}O^{(N,M)}_{mn},
  \qquad
  \Lambda_{mn} = -(zn+z^{*}m),
  \label{eq:Lambdamn}
\end{equation}
where $p$ counts $c\dg c\dg$ contractions, $q$ counts $cc$ contractions, and
$k$ counts $c\dg c$ contractions. The first few are $O_{00}=I$,
$O_{10}=c\dg$, $O_{01}=c$, and
\begin{align}
  O_{20} &= (c\dg)^2-M^{*}, \qquad O_{02} = c^2-M, \nonumber\\
  O_{11} &= c\dg c-N, \nonumber\\
  O_{21} &= (c\dg)^2c-2Nc\dg-M^{*}c, \nonumber\\
  O_{12} &= c\dg c^2-2Nc-Mc\dg .
\end{align}
Every nonconstant eigenoperator has zero steady-state mean,
$\avg{O^{(N,M)}_{mn}}_{\mathrm{ss}}=0$ for $m+n>0$. Because the jump
operators are linear, $\Lop_0\dg$ acting on a product of $c$'s and $c\dg$'s
returns the product with the additive rate plus c-number corrections that
lower the degree by two [Eq.~\eqref{eq:triangular}]; the steady state fixes
those corrections to be the rates times $N$ and $M$
[Eq.~\eqref{eq:contraction}], and subtracting all pair contractions cancels
them at every degree. The spectrum is therefore that of vacuum loss: $N$ and
$M$ deform the eigenoperators but not the eigenvalues.

\emph{Right eigenoperators.} Write $\mathrm{ad}_X(\rho)=[X,\rho]$. The
Schr\"odinger-picture eigenoperators are commutator ladders on the steady
state,
\begin{align}
  R^{(N,M)}_{mn} &= \frac{1}{m!\,n!}\,
    \big(\mathrm{ad}_{c\dg}\big)^m\big(-\mathrm{ad}_c\big)^n\rho_{\mathrm{ss}},
  \nonumber\\
  \Lop_0R^{(N,M)}_{mn} &= -(zm+z^{*}n)\,R^{(N,M)}_{mn},
  \label{eq:Rmn}
\end{align}
where the two commutator superoperators commute because $[c,c\dg]$ is a
c-number. They are biorthogonal to the left eigenoperators with transposed
indices,
\begin{equation}
  \Tr\!\left[O^{(N,M)}_{nm}\,R^{(N,M)}_{m'n'}\right] = \delta_{mm'}\delta_{nn'} .
  \label{eq:dual}
\end{equation}
To see this, move each commutator onto the observable with
$\Tr(O[X,\rho])=\Tr([O,X]\rho)$ and use $[O_{pq},c\dg]=qO_{p(q-1)}$ and
$[O_{pq},-c]=pO_{(p-1)q}$, which hold for Wick monomials because $[c,c\dg]$ is
a c-number; the chain ends on $\avg{O_{pq}}_{\mathrm{ss}}$, which vanishes
unless $p=q=0$. The eigenvalue equation in Eq.~\eqref{eq:Rmn} then follows from
biorthogonality with a complete left eigenbasis. Note the index transposition
in Eq.~\eqref{eq:dual}: the eigenvalue attached to $R_{mn}$ is $-(zm+z^{*}n)$,
not $\Lambda_{mn}$; it reappears in the Feynman rules of
Sec.~\ref{sec:ssrules}. Only $R_{00}=\rho_{\mathrm{ss}}$ is a density
operator; every other $R_{mn}$ is traceless and need be neither Hermitian nor
positive. $N$ and $M$ enter the right eigenoperators only through
$\rho_{\mathrm{ss}}$.

\emph{Reading off the bath.} The pair $(N,M)$ determines the thermal
occupation and squeezing of the steady state. With
$\nu=[(N+\tfrac12)^2-|M|^2]^{1/2}$,
\begin{equation}
  \bar n = \nu-\tfrac12,
  \qquad
  \tanh 2r = \frac{|M|}{N+\tfrac12},
  \qquad
  e^{i\phi} = -\frac{M}{|M|},
  \label{eq:symplectic}
\end{equation}
so $M=0$ is the thermal limit, $\bar n=N$, and $\nu=\tfrac12$ is a pure squeezed
vacuum. The construction assumes a canonical damped mode $c$; when the drift
has two real rates with Hermitian eigenmodes, as for a parametric oscillator
pumped more strongly than it is detuned, the multimode form of
Sec.~\ref{sec:general} applies with $K=1$.

\subsection{Vacuum, thermal, and squeezed limits}
\label{sec:singlemode}

The three reservoirs met in practice are three points of the $(N,M)$ family;
the derivations are given in the Supplemental Material~\cite{SM}. For vacuum
loss, $\Lop_0\rho=-i\omega[\ad a,\rho]+\gamma\Dop[a]\rho$, the damped mode is
$c=a$ with $z=\tfrac{\gamma}{2}+i\omega$, the steady state is the vacuum,
$N=M=0$, and Eq.~\eqref{eq:Omn} reduces to the normal-ordered monomials
\begin{equation}
  O_{mn} = (\ad)^m a^n,
  \qquad
  \Lambda_{mn} = -(zn + z^{*}m).
  \label{eq:Omnvac}
\end{equation}
Additivity of $\Lambda_{mn}$ in $m$ and $n$ is the structural feature used
throughout: each factor of $a$ carries its own rate $z$ and each factor of
$\ad$ its own $z^{*}$, with no cross terms, and the same additivity holds for
every $(N,M)$. The right eigenoperators of Eq.~\eqref{eq:Rmn} become finite
sums,
\begin{equation}
  r_{mn} = \sum_{k=0}^{\min(m,n)}
     \frac{(-1)^k}{k!\,(m-k)!\,(n-k)!}\,(\ad)^{m-k}\ketbra{0}{0}a^{\,n-k},
  \label{eq:rmn}
\end{equation}
for example $r_{11}=\ketbra{1}{1}-\ketbra{0}{0}$. For a bath of mean
occupation $\bar n$, with dissipators $\gamma(\bar n+1)\Dop[a]$ and
$\gamma\bar n\Dop[\ad]$, the damped mode and $z$ are unchanged and $N=\bar n$,
$M=0$. The eigenoperators $O^{\mathrm{th}}_{mn}$ are Eq.~\eqref{eq:Omn} with
$M=0$, a finite sum, and the eigenvalues are unchanged, so the propagators of
the expansion are temperature independent and only the vertices and the
external operator decomposition change; on the Schr\"odinger side the thermal
right eigenoperators are infinite sums in the vacuum basis~\cite{SM}. A
squeezed reservoir couples through the Bogoliubov mode
$b=\cosh r\,a+e^{i\phi}\sinh r\,\ad$ and has thermal occupation
$\bar n_{\rm B}$; the damped mode is again $c=a$ with
$z=\tfrac{\gamma}{2}+i\omega$, and squeezing and temperature enter only
through the steady moments,
\begin{align}
  N&=\bar n_{\rm B}+(2\bar n_{\rm B}+1)\sinh^2 r, \nonumber\\
  M&=-\frac{\gamma}{\gamma+2i\omega}\,(2\bar n_{\rm B}+1)\,e^{i\phi}\sinh r\cosh r .
  \label{eq:NMsqueezed}
\end{align}
The eigenoperators are Eq.~\eqref{eq:Omn} with these values, now with all
three contractions present, for example $O_{20}=(\ad)^2-M^{*}$ and
$O_{21}=(\ad)^2a-2N\ad-M^{*}a$. The thermal and vacuum cases are the limits
$r=0$ and $r=\bar n_{\rm B}=0$ of Eq.~\eqref{eq:NMsqueezed}. In all three
cases $N$ and $M$ deform the eigenoperators but not the spectrum.

\subsection{Coherent drive and the displaced frame}
\label{sec:drive}

A coherently driven cavity in the rotating frame is governed by
\begin{equation}
  H = \Delta_c \ad a + i(\eta \ad - \eta^{*}a),
  \qquad \Lop\rho = -i[H,\rho]+\kappa \Dop[a]\rho .
  \label{eq:driveH}
\end{equation}
The three contributions to the Heisenberg equation for $a$ are
$i[\Delta_c\ad a, a] = -i\Delta_c a$, $i[H_d,a]=\eta$, and
$\kappa\Dop\dg[a]a = -\tfrac{\kappa}{2}a$, so that
\begin{equation}
  \dot a = -\left(\tfrac{\kappa}{2}+i\Delta_c\right)a + \eta
        \equiv -z_c a + \eta .
\end{equation}
The steady-state amplitude is $\alpha = \eta/z_c$, and the fluctuation operator
$b = a-\alpha$ obeys $\dot b = -z_c b$. The drive is therefore removed exactly
by a displacement, and Sec.~\ref{sec:singlemode} applies with $a\to b$. This
is why the driven cavity is a Gaussian starting point, and it is the frame in
which the Kerr example of Sec.~\ref{sec:kerr} is worked.

The same displacement fails for a two-level emitter. The argument relies on
$[a,\ad]=I$ being a c-number, whereas $[\sigm,\sigp]=-\sigma_z$ is an
operator, so the driven atomic problem does not reduce to the undriven one
except under the weak-excitation approximation $\sigma_z\approx-I$, which
bosonizes the emitter; the driven atom is treated in
Sec.~\ref{sec:drivenatom}. Squeezing terms $\ad_i\ad_j+\mathrm{h.c.}$ in
$H_0$ are also not removed by a canonical transformation: the Bogoliubov
rotation that diagonalizes $H_0$ turns the jump operator $a$ into a
combination $u\,\tilde a + v\,\tilde a\dg$, so the dissipator no longer
conserves quanta in the new frame. Anomalous terms are handled directly in
Sec.~\ref{sec:general}.

\subsection{The general Gaussian Lindbladian}
\label{sec:general}

Two facts underlie the single-mode construction of Sec.~\ref{sec:gaussmode}
and hold for any stable, diagonalizable Gaussian Lindbladian, including parametric drives,
correlated squeezed reservoirs, and the Bogoliubov frames of
Sec.~\ref{sec:drive}. We state them in general and then specialize to
particle-conserving jump operators, the case used in Sec.~\ref{sec:ring}.

Collect all mode operators, creation and annihilation alike, into the
$2K$-component vector
\begin{equation}
  \bvec{A} = \left(a_1,\ldots,a_K,\ a_1\dg,\ldots,a_K\dg\right)^{T},
\end{equation}
measured from their steady means, take $H_0 = \tfrac{1}{2}\bvec{A}\dg\mathcal{H}\bvec{A}$ with
$\mathcal{H}=\mathcal{H}\dg$, and let the jump operators be arbitrary linear
forms $L_\mu = \bvec{c}_\mu\dg\bvec{A}$, mixing $a$ and $\ad$ freely.

\emph{(i) The linear sector closes.} Writing the dissipator in the equivalent
form
\begin{equation}
  \Dop\dg[L]O = \tfrac{1}{2}L\dg[O,L] + \tfrac{1}{2}[L\dg,O]L,
  \label{eq:Ddagcompact}
\end{equation}
one sees that if $O$ and $L$ are both linear then $[O,L]$ is a c-number and
$\Dop\dg[L]O$ is again linear. The same holds for $i[H_0,O]$ with quadratic
$H_0$. Hence there is a $2K\times2K$ drift matrix $\mathcal{G}$ with
\begin{equation}
  \Lop_0\dg\bvec{A} = -\mathcal{G}\,\bvec{A}.
  \label{eq:Gdef}
\end{equation}
We assume $\mathcal G$ diagonalizable. This excludes exceptional points of the
dissipative dynamics, at which $\Lop_0$ has no eigenbasis and the construction
does not apply; the assumption is not repeated below. Let $\beta_\mu$,
$\mu=1,\ldots,2K$, be the linear left eigenmodes,
\begin{equation}
  \Lop_0\dg\beta_\mu = -\zeta_\mu\beta_\mu,
  \qquad \mathrm{Re}\,\zeta_\mu>0 .
  \label{eq:betamodes}
\end{equation}
Since $\Lop_0\dg$ preserves Hermiticity, the modes come in conjugate pairs:
if $\beta$ has rate $\zeta$, then $\beta\dg$ has rate $\zeta^{*}$. A steady
state exists only if every $\mathrm{Re}\,\zeta_\mu>0$; otherwise the system
is parametrically unstable.

\emph{(ii) Products close up to c-numbers.} For linear $B_1$, $B_2$ and
linear jump operators, the correction term in Eq.~\eqref{eq:leibniz} is a
c-number, so for a monomial of degree $d$,
\begin{widetext}
\begin{equation}
  \Lop_0\dg\left(\beta_{\mu_1}\cdots\beta_{\mu_d}\right)
   = -\Big(\sum_{p}\zeta_{\mu_p}\Big)\beta_{\mu_1}\cdots\beta_{\mu_d}
     + \sum_{p<q} C_{\mu_p\mu_q}\;
       \beta_{\mu_1}\cdots\widehat{\beta_{\mu_p}}\cdots\widehat{\beta_{\mu_q}}\cdots\beta_{\mu_d},
  \qquad
  C_{\mu\nu} = \sum_\lambda \gamma_\lambda\,[L_\lambda\dg,\beta_\mu]\,[\beta_\nu,L_\lambda],
  \label{eq:triangular}
\end{equation}
\end{widetext}
where a hat denotes omission. The generator is therefore triangular, lowering
the degree by two, and is diagonalized by subtracting the lower terms. The
subtraction coefficients follow from the steady state: applying
$\Tr[\,\cdot\,\rho_{\mathrm{ss}}]$ to Eq.~\eqref{eq:triangular} at $d=2$ and
using $\Tr[(\Lop_0\dg O)\rho_{\mathrm{ss}}]=0$ gives
\begin{equation}
  \frac{C_{\mu\nu}}{\zeta_\mu+\zeta_\nu} = \avg{\beta_\mu\beta_\nu}_{\mathrm{ss}}
  \;\equiv\; V_{\mu\nu} \;\equiv\; \contr{\beta_\mu}{\beta_\nu},
  \label{eq:contraction}
\end{equation}
the ordered steady-state covariance, and consequently
$\Lop_0\dg(\beta_\mu\beta_\nu-V_{\mu\nu})=-(\zeta_\mu+\zeta_\nu)(\beta_\mu\beta_\nu-V_{\mu\nu})$.
In a quadrature basis $V$ follows equally from the Lyapunov equation
$\mathsf{A}V+V\mathsf{A}^{T}+\mathsf{D}=0$ with drift $\mathsf{A}$ and
diffusion $\mathsf{D}$. For quadratic Lindbladians the same spectral
information is supplied by third
quantization~\cite{prosen2008third,prosen2010quantization,mcdonald2023third}.

\emph{Left eigenoperators.} Solving the recursion pairing by pairing gives, for
a multi-index $\bvec n=(n_1,\ldots,n_{2K})$,
\begin{equation}
  O_{\bvec n} = \normord{\prod_{\mu=1}^{2K}\beta_\mu^{\,n_\mu}}_V,
  \qquad
  \Lambda_{\bvec n} = -\sum_{\mu=1}^{2K} n_\mu\zeta_\mu ,
  \label{eq:generaleigen}
\end{equation}
where $\normord{\ \cdot\ }_V$ keeps the operator order and subtracts every pair
contraction $V_{\mu\nu}$, i.e.\ the multivariate Hermite polynomial built on the
steady-state covariance. A pairing shifts the diagonal eigenvalue by
$\sum_{(\mu,\nu)\in\text{pairing}}(\zeta_\mu+\zeta_\nu)$, which by
Eq.~\eqref{eq:contraction} cancels the off-diagonal term generated at the
next level down.

\emph{Right eigenoperators.} Choose dual linear operators $\tilde\beta_\mu$
with $[\beta_\mu,\tilde\beta_\nu]=\delta_{\mu\nu}I$, which exist because the
commutator matrix of the $\beta_\mu$ is invertible. Then
\begin{align}
  R_{\bvec n} &= \Big[\prod_{\mu=1}^{2K}\frac{\mathrm{ad}_{\tilde\beta_\mu}^{\,n_\mu}}{n_\mu!}\Big]\rho_{\mathrm{ss}},
  \nonumber\\
  \Lop_0R_{\bvec n} &= \Lambda_{\bvec n}R_{\bvec n},
  \qquad
  \Tr\!\left(O_{\bvec m}R_{\bvec n}\right)=\delta_{\bvec m\bvec n},
  \label{eq:Rgeneral}
\end{align}
by the argument given below Eq.~\eqref{eq:dual}; the commutator superoperators
commute because commutators of linear operators are c-numbers.

\emph{Reductions.} For $K=1$ with a canonical damped mode, $\beta_1=c\dg$ and
$\beta_2=c$ with duals $\tilde\beta_1=-c$ and $\tilde\beta_2=c\dg$, and the
covariance has entries $V_{c\dg c}=N$, $V_{cc}=M$, $V_{c\dg c\dg}=M^{*}$,
$V_{cc\dg}=N+1$. Equation~\eqref{eq:generaleigen} is then Eq.~\eqref{eq:Omn},
and Eq.~\eqref{eq:Rgeneral} is Eq.~\eqref{eq:Rmn} in the same-index convention,
$R_{\bvec n}=R_{n_2n_1}$. In general the covariance contains a normal block $N_{ij}=\avg{c_i\dg c_j}_{\mathrm{ss}}$ and an anomalous block
$M_{ij}=\avg{c_ic_j}_{\mathrm{ss}}$: thermal modes have $M_{ij}=0$ and diagonal
$N_{ij}$, correlated thermal noise has off-diagonal $N_{ij}$, and single- or
two-mode squeezing has $M_{ij}\neq0$.

Equation~\eqref{eq:generaleigen} keeps $\Lambda_{\bvec n}$ additive. Every
line carries its own rate, the propagator factorizes, and the diagrams of
Sec.~\ref{sec:diagrams} are unchanged in form. What changes is the vertex
table: with a nonvanishing covariance, contraction channels open that are
forbidden in vacuum (Sec.~\ref{sec:vertices}).

\emph{Particle-conserving reduction.}
\label{sec:multimode}
When the jump operators conserve particle number, $\mathcal{G}$ is block
diagonal and the construction reduces to one in the annihilation operators
alone. Consider $K$ modes $\bvec{a}=(a_1,\dots,a_K)^{T}$ with
$[a_k,a_{k'}\dg]=\delta_{kk'}$, a quadratic Hamiltonian
$H_0=\bvec{a}\dg\Omega\bvec{a}$ with $\Omega=\Omega\dg$, and decay channels
$L_j=\sqrt{\gamma_j}\,\ell_j$ built from arbitrary normalized mode
combinations $\ell_j = \bvec{v}_j\dg\bvec{a}$. No orthogonality between
different $\bvec{v}_j$ is assumed; overlapping channels describe collective
or correlated loss. The channel data enter through the positive semidefinite
decay matrix $X = \sum_j \tfrac{\gamma_j}{2}\,\bvec{v}_j\bvec{v}_j\dg$, whose
off-diagonal entries encode correlated decay. With $\Gamma = X + i\Omega$ and
$\Tr_k$ a trace over the mode index, the Hamiltonian and dissipative parts
merge into
\begin{align}
  \Lop_0\rho &= \Tr_k\!\left[(\Gamma+\Gamma\dg)\cdot\bvec{a}\rho\bvec{a}\dg\right]
     - (\bvec{a}\dg\Gamma\bvec{a})\rho - \rho(\bvec{a}\dg\Gamma\dg\bvec{a}),
     \label{eq:Lmulti}
\end{align}
and in the Heisenberg picture
\begin{equation}
  \Lop_0\dg O = -[O,\bvec{a}\dg]\,\Gamma\,\bvec{a}
                - \bvec{a}\dg\,\Gamma\dg\,[\bvec{a},O],
  \label{eq:Lmultiadj}
\end{equation}
the multimode form of Eq.~\eqref{eq:compact}. In particular
$\Lop_0\dg\bvec{a} = -\Gamma\bvec{a}$, so $\Gamma$ plays the role of the
single complex rate $z$, and $\mathcal{G}=\mathrm{diag}(\Gamma,\Gamma^{*})$.

$\Gamma$ is not Hermitian, so its left and right eigenvectors differ. With
$\Gamma=\sum_k\zeta_k\bvec{v}_{k,R}\bvec{v}_{k,L}\dg$,
$\Gamma\bvec{v}_{k,R}=\zeta_k\bvec{v}_{k,R}$,
$\bvec{v}_{k,L}\dg\Gamma=\zeta_k\bvec{v}_{k,L}\dg$, and
$\bvec{v}_{k,L}\dg\bvec{v}_{k',R}=\delta_{kk'}$, the left modes
$a_{k,L}=\bvec{v}_{k,L}\dg\bvec{a}$ decay independently,
$\Lop_0\dg a_{k,L} = -\zeta_k a_{k,L}$, and the Heisenberg eigenoperators are
the monomials in the left modes alone,
\begin{align}
  O_{\bvec{m},\bvec{n}} &= \prod_k (a_{k,L}\dg)^{m_k}\prod_k a_{k,L}^{n_k},
    \nonumber\\
  \Lambda_{\bvec{m},\bvec{n}} &= -\sum_k(\zeta_k n_k + \zeta_k^{*}m_k).
  \label{eq:Omultimode}
\end{align}
The proof uses $[a_{k,L},a_{k',R}\dg]=\delta_{kk'}$ in
Eq.~\eqref{eq:Lmultiadj}: the commutator removes one factor of $a_{k,L}$
with weight $n_k$, and the trailing $a_{k,L}$ puts it back. The left modes are not canonically normalized among themselves,
$[a_{k,L},a_{k',L}\dg]=\bvec{v}_{k,L}\dg\bvec{v}_{k',L}\neq\delta_{kk'}$ in
general. This does not obstruct Eq.~\eqref{eq:Omultimode}, but the diagram
lines of Sec.~\ref{sec:diagrams} then carry a mode label whose contractions
are governed by the Gram matrix of the left modes, Eq.~\eqref{eq:mmove1},
rather than by orthonormality. The Schr\"odinger-picture eigenoperators
follow from the right modes $a_{k,R}=\bvec{v}_{k,R}\dg\bvec{a}$, which are
dual to the left modes, $[a_{k,L},a_{k',R}\dg]=\delta_{kk'}$ and
$[a_{k,L}\dg,-a_{k',R}]=\delta_{kk'}$. They are the multimode commutator
ladders on the vacuum,
\begin{align}
  R_{\bvec m,\bvec n} &= \prod_k\frac{\big(\mathrm{ad}_{a_{k,R}\dg}\big)^{m_k}
     \big(-\mathrm{ad}_{a_{k,R}}\big)^{n_k}}{m_k!\,n_k!}\;\ketbra{0}{0},
  \nonumber\\
  \Lop_0R_{\bvec m,\bvec n} &= -\sum_k\left(\zeta_km_k+\zeta_k^{*}n_k\right)R_{\bvec m,\bvec n},
  \label{eq:Rmultimode}
\end{align}
biorthogonal to the left family with transposed indices,
$\Tr[O_{\bvec n,\bvec m}R_{\bvec m',\bvec n'}]=\delta_{\bvec m\bvec m'}\delta_{\bvec n\bvec n'}$,
by the argument of Eq.~\eqref{eq:dual} applied mode by mode. For $K=1$ they
are the $r_{mn}$ of Eq.~\eqref{eq:rmn}; for a normal $\Gamma$ the right and
left modes coincide; and in vacuum Eq.~\eqref{eq:generaleigen} reduces to
Eq.~\eqref{eq:Omultimode}.

\subsection{Two-level systems}
\label{sec:tls}

For a two-level emitter the starting point is
\begin{equation}
  \Lop_0\rho = -i\epsilon[\sigp\sigm,\rho] + \gamma\Dop[\sigm]\rho .
\end{equation}
At zero temperature the Hamiltonian term rotates coherences, the jump term
transfers population $\ketbra{e}{e}\to\ketbra{g}{g}$, and the anticommutator
supplies the no-jump damping required for trace preservation. Writing
$n_e=\sigp\sigm$, the matrix elements evolve as
\begin{equation}
  \dot\rho_{ee}=-\gamma\rho_{ee},\quad
  \dot\rho_{gg}=+\gamma\rho_{ee},\quad
  \dot\rho_{eg}=-\left(\tfrac{\gamma}{2}+i\epsilon\right)\rho_{eg},
\end{equation}
so coherences decay at half the population rate in the absence of pure
dephasing. The spectrum is
\begin{equation}
  \left\{0,\ -\tfrac{\gamma}{2}-i\epsilon,\ -\tfrac{\gamma}{2}+i\epsilon,\ -\gamma\right\}
\end{equation}
with Heisenberg eigenoperators $\{I,\sigm,\sigp,\sigp\sigm\}$ and
Schr\"odinger eigenoperators
$\{\ketbra{g}{g},\ \ketbra{e}{g},\ \ketbra{g}{e},\ \ketbra{e}{e}-\ketbra{g}{g}\}$.
At finite temperature $N$ the rates become $\Gamma_1=\gamma(2N+1)$ and
$\Gamma_2=\Gamma_1/2$, the population eigenoperator becomes $\sigp\sigm-p_e$
with $p_e=N/(2N+1)$, and, unlike the bosonic case, the bath shifts the
eigenvalues; the finite-temperature eigenoperators are given in the
Supplemental Material~\cite{SM}.

\subsection{The driven emitter}
\label{sec:drivenatom}

With a drive the atomic problem is no longer diagonal in the bare operators.
Taking
\begin{equation}
  H_a = \frac{\Delta_a}{2}\sigma_z + \frac{\Omega}{2}\sigma_x
\end{equation}
and $\gamma\Dop[\sigm]$, we choose the phase of the drive so that $\Omega$ is
real and nonnegative; the complex conjugates retained below make the
phase-covariant form explicit. The optical Bloch equations for fluctuations
about the steady state read
\begin{align}
  \Lop_a\dg\,\delta\sigm &= -z_a\,\delta\sigm + \frac{i\Omega}{2}\delta\sigma_z, \nonumber\\
  \Lop_a\dg\,\delta\sigp &= -z_a^{*}\,\delta\sigp - \frac{i\Omega^{*}}{2}\delta\sigma_z,
     \label{eq:blochfluc} \\
  \Lop_a\dg\,\delta\sigma_z &= i\Omega^{*}\delta\sigm - i\Omega\,\delta\sigp
     - \gamma\,\delta\sigma_z, \nonumber
\end{align}
with $z_a = \tfrac{\gamma}{2}+i\Delta_a$. The steady state is
\begin{equation}
  \rho_{\mathrm{ss}} =
  \begin{pmatrix} p_e & s_- \\ s_+ & 1-p_e \end{pmatrix},
  \qquad
  D = \Delta_a^2 + \frac{\gamma^2}{4},
\end{equation}
\begin{equation}
  s_{\mp} = -\frac{\Omega(\Delta_a \pm i\gamma/2)}{2\left(D+\Omega^2/2\right)},
  \qquad
  p_e = \frac{\Omega^2/4}{D+\Omega^2/2}.
  \label{eq:atomss}
\end{equation}
Equations~\eqref{eq:blochfluc} say that the three traceless fluctuations close
among themselves. Writing $X=c_-\delta\sigm+c_+\delta\sigp+c_z\delta\sigma_z$
and reading off the coefficients, $\Lop_a\dg$ acts on the column
$\bvec c=(c_-,c_+,c_z)^T$ as the matrix
\begin{equation}
  \mathcal{A} =
  \begin{pmatrix}
    -z_a & 0 & i\Omega \\
    0 & -z_a^{*} & -i\Omega \\
    i\Omega/2 & -i\Omega/2 & -\gamma
  \end{pmatrix},
  \label{eq:blochmatrix}
\end{equation}
so the dressed eigenoperators are its eigenvectors: if
$\mathcal{A}\bvec c^{(k)}=\lambda_k\bvec c^{(k)}$ then
\begin{equation}
  X_k = c^{(k)}_-\delta\sigm + c^{(k)}_+\delta\sigp + c^{(k)}_z\delta\sigma_z,
  \qquad \Lop_a\dg X_k = \lambda_k X_k ,
  \label{eq:Xkdef}
\end{equation}
and $\Tr[X_k\rho_{\mathrm{ss}}]=0$ by construction. Off resonance
Eq.~\eqref{eq:blochmatrix} is diagonalized numerically, a one-time
$3\times3$ cost.

On resonance ($\Delta_a=0$, real $\Omega$, $z_a=\gamma/2$) the eigenvectors are
elementary. The combination $\bvec c=(1,1,0)$ is annihilated by the
$\delta\sigma_z$ column and gives
\begin{equation}
  X_1 = \delta\sigma_x = \delta\sigm + \delta\sigp,
  \qquad \lambda_1 = -\frac{\gamma}{2},
  \label{eq:X1}
\end{equation}
while the orthogonal combination $\delta\sigm-\delta\sigp$ mixes only with
$\delta\sigma_z$, through the $2\times2$ block
$\left(\begin{smallmatrix}-\gamma/2 & i\Omega\\ i\Omega & -\gamma\end{smallmatrix}\right)$,
whose eigenvalues are
\begin{equation}
  \lambda_{\pm} = -\frac{3\gamma}{4} \pm \sqrt{\frac{\gamma^2}{16}-\Omega^2}
  \label{eq:lampm}
\end{equation}
with eigenvectors
\begin{equation}
  X_{\pm} = \delta\sigm - \delta\sigp + d_{\pm}\,\delta\sigma_z,
  \qquad
  d_{\pm} = -\frac{i}{\Omega}\left(\frac{\gamma}{2}+\lambda_{\pm}\right).
  \label{eq:Xpm}
\end{equation}
The full atomic spectrum is $\{0,\lambda_1,\lambda_+,\lambda_-\}$. For
$\Omega\gg\gamma$, $\lambda_{\pm}\simeq-\tfrac{3\gamma}{4}\pm i\Omega$ and
$d_{\pm}\to\pm1$: a central line of width $\gamma/2$ and two sidebands at
$\pm\Omega$ of width $3\gamma/4$, which is the Mollow triplet, and
Eqs.~\eqref{eq:X1} and \eqref{eq:Xpm} are its dressed-state operators. Below
$\Omega=\gamma/4$ the square root is real, $X_\pm$ are two overdamped
combinations, and nothing in the construction changes.

Inverting Eqs.~\eqref{eq:X1} and \eqref{eq:Xpm} gives the expansion of the
measured operators in the dressed basis, which the diagrams need:
\begin{align}
  \sigma^{\mp} &= \avg{\sigma^{\mp}}_{\mathrm{ss}}
   \pm \frac{X_1}{2}
   + \frac{1}{2}\,\frac{d_{\pm}X_{\mp}-d_{\mp}X_{\pm}}{d_{+}-d_{-}}\, ,
   \nonumber\\
  \delta\sigma_z &= \frac{X_+-X_-}{d_+-d_-},
  \label{eq:dressedexpand}
\end{align}
so that $c^{\mp}_1=\pm\tfrac12$ and
$c^{\mp}_{\pm}=\mp d_{\mp}/2(d_+-d_-)$ in the notation of
Eq.~\eqref{eq:dressedmoves}.

Two structural facts matter for the diagrammatics. First, because the
single-emitter operator algebra is four dimensional, products of atomic
eigenoperators collapse back into the same four-element basis. An emitter
therefore contributes at most one line to any diagram, and its vertices are
the structure constants of the dressed algebra, a finite table computed once.
Second, additivity across sectors requires that no jump operator act on the
emitter and the field simultaneously. If $\Lop_0$ factorizes as
$\Lop_0^{\mathrm{cav}}\otimes\mathrm{id}+\mathrm{id}\otimes\Lop_0^{\mathrm{atom}}$,
the correction term in Eq.~\eqref{eq:leibniz} vanishes between sectors and
the eigenvalues add. A shared decay channel, as in waveguide QED with
$L=\sqrt{\kappa}\,a+\sqrt{\gamma}\,\sigm$, makes the correction
$[L\dg,A][B,L]$ operator valued on the atomic side, and the Wick
construction of Sec.~\ref{sec:general} does not apply. The emitter must then
be bosonized under the weak-excitation approximation, or the collective part
of the coupling moved into $\Lint$.

\begin{figure}[t]
\centering
\resizebox{0.95\columnwidth}{!}{%
\begin{tikzpicture}[x=1cm,y=1cm]
\begin{scope}[shift={(0,0)}]
  \draw[ld ad,ld mark at=0.68] (0,0) .. controls (-0.42,0) and (-0.5,0.26) .. (-0.92,0.26) -- (-2,0.26);
  \node[ld label,anchor=east,inner sep=0.5pt] at (-2.05,0.26) {\scriptsize $\times m$};
  \draw[ld a,ld mark at=0.68] (0,0) .. controls (-0.42,0) and (-0.5,-0.26) .. (-0.92,-0.26) -- (-2,-0.26);
  \node[ld label,anchor=east,inner sep=0.5pt] at (-2.05,-0.26) {\scriptsize $\times n$};
  \draw[ld ad,ld mark at=0.32] (2,0.26) -- (0.92,0.26) .. controls (0.5,0.26) and (0.42,0) .. (0,0);
  \node[ld label,anchor=west,inner sep=0.5pt] at (2.05,0.26) {\scriptsize $\times p$};
  \draw[ld a,ld mark at=0.32] (2,-0.26) -- (0.92,-0.26) .. controls (0.5,-0.26) and (0.42,0) .. (0,0);
  \node[ld label,anchor=west,inner sep=0.5pt] at (2.05,-0.26) {\scriptsize $\times q$};
  \node[ld vtx] at (0,0) {};
  \node[ld order,fill=white,fill opacity=0.85,text opacity=1,rounded corners=1pt] at (0,0.13) {$(N)$};
  \node[ld ext,anchor=east] at (-2.5,0) {$O_{mn}$};
  \node[ld ext,anchor=west] at (2.5,0) {$r_{qp}$};
  \node[ld label,font=\scriptsize,anchor=north,align=center] at (-1.75,-0.62) {incoming propagator\\ $(zn+z^{*}m)^{-1}$};
  \node[ld label,anchor=north,align=center] at (0,-1.4) {order-$N$ amplitude $D^{(N)}_{mn\to pq}$};
  \node[ld label,font=\scriptsize,anchor=north,align=center] at (1.75,-0.62) {initial-state weight\\ $\Tr[O_{pq}\rho_0]$};
  \draw[-Latex,gray] (2.0,-2.0) -- (-2.0,-2.0);
  \node[gray,font=\footnotesize,anchor=north] at (0,-2.05) {time};
\end{scope}
\end{tikzpicture}}
\caption{Diagram conventions. Every diagram has a single vertex, marked with the
order $N$ of perturbation theory. The measured eigenoperator $O_{mn}$ [$(\ad)^ma^n$ in vacuum,
Eq.~\eqref{eq:Omn} in general] enters from the left as $m$ lines with a left arrow (factors of $\ad$, rate
$z^{*}$ each) and $n$ lines with a right arrow (factors of $a$, rate $z$
each), and carries the incoming propagator $(zn+z^{*}m)^{-1}$. The lines
leaving to the right are the content $O_{pq}$ reached after $N$ applications
of $\Lint\dg$; they close on the component $r_{qp}$ of the initial state, with
weight $\Tr[O_{pq}\rho_0]$ (note the index transposition of
Eq.~\eqref{eq:dual}). Time runs from the initial state on the right to the
observable on the left. There is thus one set of diagrams for each order and
each initial-state term; in the steady state only the diagrams with no
outgoing lines, closing on $r_{00}$, survive. All line types are collected in
Table~\ref{tab:elements}.}
\label{fig:conventions}
\end{figure}
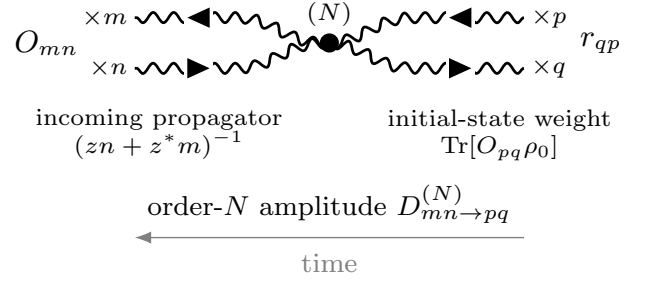

\section{The diagrammatic expansion}
\label{sec:diagrams}

\subsection{Superoperator interaction picture}

Split the generator as
\begin{equation}
  \Lop = \Lop_0 + \Lint,
\end{equation}
where $\Lop_0$ is one of the exactly solvable Gaussian pieces of
Sec.~\ref{sec:gaussian} and $\Lint$ is a weaker interaction, typically a
$\chi^{(2)}$ or $\chi^{(3)}$ Hamiltonian contribution with no additional
dissipative dynamics. No explicit expansion parameter is carried; it resides
in $\Lint$. As in unitary perturbation theory, convergence is not guaranteed,
and the behaviour of the series is examined case by case in
Secs.~\ref{sec:examples} and \ref{sec:ring}.

Differentiating $e^{-\Lop_0\Delta t}e^{(\Lop_0+\Lint)\Delta t}$ produces the
superoperator interaction picture and hence the Dyson series
\begin{widetext}
\begin{equation}
  e^{(\Lop_0+\Lint)\Delta t} = \sum_{N=0}^{\infty}
    \int_0^{\Delta t}\!\!\!d\tau_1\int_0^{\tau_1}\!\!\!d\tau_2\cdots\int_0^{\tau_{N-1}}\!\!\!d\tau_N\;
    e^{\Lop_0(\Delta t-\tau_1)}\Lint e^{\Lop_0(\tau_1-\tau_2)}\Lint\cdots
    e^{\Lop_0(\tau_{N-1}-\tau_N)}\Lint e^{\Lop_0\tau_N}.
  \label{eq:dyson}
\end{equation}
\end{widetext}
Because the spectrum of $\Lop_0$ is unbounded, $e^{-\Lop_0\Delta t}$ is
formal; Eq.~\eqref{eq:dyson} nevertheless holds and is rederived without it in
the Supplemental Material~\cite{SM}.

The object of interest is $\Tr[O_{mn}e^{\Lop(t-t_0)}r_{m'n'}]$, whose $N$th
order term is
\begin{widetext}
\begin{equation}
  \left(\Tr\!\left[O_{mn}e^{\Lop(t-t_0)}r_{m'n'}\right]\right)^{(N)}
   = \int_0^{\Delta t}\!\!\!\cdots\!\int_0^{\tau_{N-1}}
     \Tr\!\left[O_{mn}e^{\Lop_0(\Delta t-\tau_1)}\Lint e^{\Lop_0(\tau_1-\tau_2)}
       \cdots \Lint e^{\Lop_0\tau_N}r_{m'n'}\right]d\tau^N .
  \label{eq:Nth}
\end{equation}
\end{widetext}
To use this one must decompose the observable $O$ in the Heisenberg eigenbasis
of $\Lop_0$ and the initial state $\rho_0$ in the Schr\"odinger eigenbasis, then
sum the contributing overlaps. The method is efficient only when both
decompositions are short. The quantities usually wanted, quadratures and
photon numbers, require only $O_{01}$, $O_{10}$, and $O_{11}$, so the
diagrams start with one or two external lines. On the state side,
calculations are simplest near a steady state: $r_{00}$ alone if the exact
steady state is sought, with $r_{10}$, $r_{01}$ and beyond added for
finite-time corrections.

\subsection{Vertices from the shift relations}
\label{sec:vertices}

This subsection gives the elementary move algebra that is reused across
interactions. Once the eigenoperator basis of $\Lop_0$ is chosen, these moves
depend only on the operator algebra and on the steady-state contractions of
$\Lop_0$. The numerical vertex factors themselves also depend on the chosen
$\Lint$: they are assembled by applying the same moves to the operator products
that occur in its Hamiltonian and dissipative terms. Changing the split between
$\Lop_0$ and $\Lint$ therefore changes the vertex table but not the underlying
move rules.

The interaction is rewritten as a map on the eigenoperator basis,
\begin{equation}
  \Lint\dg O_{mn} = \sum_{j,j'} \Vint_{j,j'}\,O_{(m+j)(n+j')} .
  \label{eq:vertexdef}
\end{equation}
The coefficients $\Vint_{j,j'}$ are the vertex factors. Each output label of
Eq.~\eqref{eq:vertexdef} is one first-order diagram: the lines of $O_{mn}$
enter the vertex, those of $O_{(m+j)(n+j')}$ leave it, and the vertex is marked
$(1)$ (Fig.~\ref{fig:conventions}). The factors follow from writing $\Lint\dg$
in terms of mode operators and multiplying them into the eigenoperator one at a
time. The rule is stated for the general single-mode eigenoperators
$O^{(N,M)}_{mn}$ of Sec.~\ref{sec:gaussmode}, specialized to its vacuum,
thermal, and squeezed limits, and then extended to many modes and to
emitters.

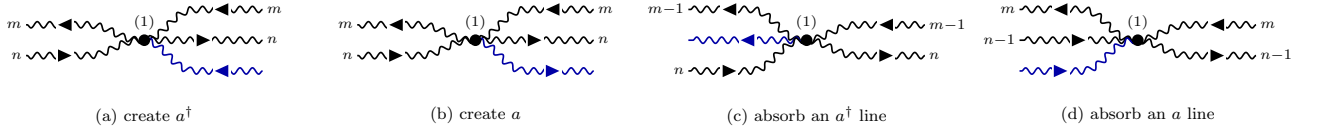
\begin{figure*}[t]
\centering
\resizebox{0.95\textwidth}{!}{%
\begin{tikzpicture}[x=1cm,y=1cm]
\begin{scope}[shift={(0,0)}]
  \draw[ld ad,ld mark at=0.68] (0,0) .. controls (-0.42,0) and (-0.5,0.26) .. (-0.92,0.26) -- (-2,0.26);
  \node[ld label,anchor=east,inner sep=0.5pt] at (-2.05,0.26) {\scriptsize $m$};
  \draw[ld a,ld mark at=0.68] (0,0) .. controls (-0.42,0) and (-0.5,-0.26) .. (-0.92,-0.26) -- (-2,-0.26);
  \node[ld label,anchor=east,inner sep=0.5pt] at (-2.05,-0.26) {\scriptsize $n$};
  \draw[ld ad,ld mark at=0.32] (2,0.52) -- (0.92,0.52) .. controls (0.5,0.52) and (0.42,0) .. (0,0);
  \node[ld label,anchor=west,inner sep=0.5pt] at (2.05,0.52) {\scriptsize $m$};
  \draw[ld a,ld mark at=0.5] (2,0) -- (0,0);
  \node[ld label,anchor=west,inner sep=0.5pt] at (2.05,0) {\scriptsize $n$};
  \begin{scope}[color=blue!65!black]\draw[ld ad,ld mark at=0.32] (2,-0.52) -- (0.92,-0.52) .. controls (0.5,-0.52) and (0.42,0) .. (0,0);\end{scope}
  \node[ld vtx] at (0,0) {};
  \node[ld order,fill=white,fill opacity=0.85,text opacity=1,rounded corners=1pt] at (0,0.13) {$(1)$};
  \node[font=\footnotesize,anchor=north] at (0,-1) {(a)\ create $\ad$};
\end{scope}
\begin{scope}[shift={(5.6,0)}]
  \draw[ld ad,ld mark at=0.68] (0,0) .. controls (-0.42,0) and (-0.5,0.26) .. (-0.92,0.26) -- (-2,0.26);
  \node[ld label,anchor=east,inner sep=0.5pt] at (-2.05,0.26) {\scriptsize $m$};
  \draw[ld a,ld mark at=0.68] (0,0) .. controls (-0.42,0) and (-0.5,-0.26) .. (-0.92,-0.26) -- (-2,-0.26);
  \node[ld label,anchor=east,inner sep=0.5pt] at (-2.05,-0.26) {\scriptsize $n$};
  \draw[ld ad,ld mark at=0.32] (2,0.52) -- (0.92,0.52) .. controls (0.5,0.52) and (0.42,0) .. (0,0);
  \node[ld label,anchor=west,inner sep=0.5pt] at (2.05,0.52) {\scriptsize $m$};
  \draw[ld a,ld mark at=0.5] (2,0) -- (0,0);
  \node[ld label,anchor=west,inner sep=0.5pt] at (2.05,0) {\scriptsize $n$};
  \begin{scope}[color=blue!65!black]\draw[ld a,ld mark at=0.32] (2,-0.52) -- (0.92,-0.52) .. controls (0.5,-0.52) and (0.42,0) .. (0,0);\end{scope}
  \node[ld vtx] at (0,0) {};
  \node[ld order,fill=white,fill opacity=0.85,text opacity=1,rounded corners=1pt] at (0,0.13) {$(1)$};
  \node[font=\footnotesize,anchor=north] at (0,-1) {(b)\ create $a$};
\end{scope}
\begin{scope}[shift={(11.2,0)}]
  \draw[ld ad,ld mark at=0.68] (0,0) .. controls (-0.42,0) and (-0.5,0.52) .. (-0.92,0.52) -- (-2,0.52);
  \node[ld label,anchor=east,inner sep=0.5pt] at (-2.05,0.52) {\scriptsize $m{-}1$};
  \begin{scope}[color=blue!65!black]\draw[ld ad,ld mark at=0.5] (0,0) -- (-2,0);\end{scope}
  \draw[ld a,ld mark at=0.68] (0,0) .. controls (-0.42,0) and (-0.5,-0.52) .. (-0.92,-0.52) -- (-2,-0.52);
  \node[ld label,anchor=east,inner sep=0.5pt] at (-2.05,-0.52) {\scriptsize $n$};
  \draw[ld ad,ld mark at=0.32] (2,0.26) -- (0.92,0.26) .. controls (0.5,0.26) and (0.42,0) .. (0,0);
  \node[ld label,anchor=west,inner sep=0.5pt] at (2.05,0.26) {\scriptsize $m{-}1$};
  \draw[ld a,ld mark at=0.32] (2,-0.26) -- (0.92,-0.26) .. controls (0.5,-0.26) and (0.42,0) .. (0,0);
  \node[ld label,anchor=west,inner sep=0.5pt] at (2.05,-0.26) {\scriptsize $n$};
  \node[ld vtx] at (0,0) {};
  \node[ld order,fill=white,fill opacity=0.85,text opacity=1,rounded corners=1pt] at (0,0.13) {$(1)$};
  \node[font=\footnotesize,anchor=north] at (0,-1) {(c)\ absorb an $\ad$ line};
\end{scope}
\begin{scope}[shift={(16.8,0)}]
  \draw[ld ad,ld mark at=0.68] (0,0) .. controls (-0.42,0) and (-0.5,0.52) .. (-0.92,0.52) -- (-2,0.52);
  \node[ld label,anchor=east,inner sep=0.5pt] at (-2.05,0.52) {\scriptsize $m$};
  \draw[ld a,ld mark at=0.5] (0,0) -- (-2,0);
  \node[ld label,anchor=east,inner sep=0.5pt] at (-2.05,0) {\scriptsize $n{-}1$};
  \begin{scope}[color=blue!65!black]\draw[ld a,ld mark at=0.68] (0,0) .. controls (-0.42,0) and (-0.5,-0.52) .. (-0.92,-0.52) -- (-2,-0.52);\end{scope}
  \draw[ld ad,ld mark at=0.32] (2,0.26) -- (0.92,0.26) .. controls (0.5,0.26) and (0.42,0) .. (0,0);
  \node[ld label,anchor=west,inner sep=0.5pt] at (2.05,0.26) {\scriptsize $m$};
  \draw[ld a,ld mark at=0.32] (2,-0.26) -- (0.92,-0.26) .. controls (0.5,-0.26) and (0.42,0) .. (0,0);
  \node[ld label,anchor=west,inner sep=0.5pt] at (2.05,-0.26) {\scriptsize $n{-}1$};
  \node[ld vtx] at (0,0) {};
  \node[ld order,fill=white,fill opacity=0.85,text opacity=1,rounded corners=1pt] at (0,0.13) {$(1)$};
  \node[font=\footnotesize,anchor=north] at (0,-1) {(d)\ absorb an $a$ line};
\end{scope}
\end{tikzpicture}}
\caption{The four elementary moves, drawn as first-order diagrams. Left arrows
are factors of $\ad$ (or $\beta^\dagger$), right arrows factors of $a$ (or
$\beta$); the labels give the number of lines in each bundle, and the line
that is created or absorbed is shown in colour. (a),(b)~A mode operator
creates its own line, with weight one. (c),(d)~It absorbs one existing line.
The weight of an absorption is the steady-state contraction of the incoming
operator $A$ with the absorbed one $X$, taken in the order in which they stand:
$\contr{A}{X}$ when $A$ multiplies from the left and $\contr{X}{A}$ when it
multiplies from the right, times the number of lines of that kind;
for one mode these are $N$, $N+1$, $M$, and $M^{*}$,
Eqs.~\eqref{eq:gmove1}--\eqref{eq:gmove4}. In vacuum only $\contr{a}{\ad}=1$ is
nonzero, so only $a$ multiplying from the left absorbs an $\ad$ line [panel
(c), Eq.~\eqref{eq:move3}] and only $\ad$ multiplying from the right absorbs an
$a$ line [panel (d), Eq.~\eqref{eq:move4}]. A thermal steady state
($M=0$, $N=\bar n$) adds absorptions of the other kind, and a squeezed steady
state ($M\neq0$) adds absorption of a line of the same orientation. For many
modes the lines carry a mode label and the weights are the entries of the
covariance $V_{\mu\nu}$, Eqs.~\eqref{eq:mmove1}--\eqref{eq:mmove2}.}
\label{fig:vertices}
\end{figure*}

\emph{The general single-mode rule.} Multiplying a Wick monomial by one linear
operator $A$ obeys Wick's identity,
\begin{align}
  A\,\normord{X} &= \normord{AX}
    + \sum_{i} \contr{A}{X_i}\;\normord{X/X_i}, \nonumber\\
  \normord{X}\,A &= \normord{XA}
    + \sum_{i} \contr{X_i}{A}\;\normord{X/X_i},
  \label{eq:wickmult}
\end{align}
where the sum runs over the factors $X_i$ of $X$, $X/X_i$ omits one of them,
and the contractions are the ordered steady-state second moments of
$\Lop_0$, the same ones that define the Wick ordering in
Eq.~\eqref{eq:contraction}. For a damped mode $c$ with
$O_{mn}\equiv O^{(N,M)}_{mn}$ the contractions are $\contr{c\dg}{c}=N$,
$\contr{c}{c}=M$, $\contr{c\dg}{c\dg}=M^{*}$, and $\contr{c}{c\dg}=N+1$, and
Eq.~\eqref{eq:wickmult} gives the four general moves
\begin{align}
  c\dg O_{mn} &= O_{(m+1)n} + mM^{*}\,O_{(m-1)n} + nN\,O_{m(n-1)}, \label{eq:gmove1}\\
  O_{mn}\,c &= O_{m(n+1)} + mN\,O_{(m-1)n} + nM\,O_{m(n-1)}, \label{eq:gmove2}\\
  c\,O_{mn} &= O_{m(n+1)} + m(N+1)\,O_{(m-1)n} \nonumber\\ &\quad + nM\,O_{m(n-1)}, \label{eq:gmove3}\\
  O_{mn}\,c\dg &= O_{(m+1)n} + mM^{*}\,O_{(m-1)n} \nonumber\\ &\quad + n(N+1)\,O_{m(n-1)}. \label{eq:gmove4}
\end{align}
In words: \emph{a mode operator entering a vertex either creates its own
line, with weight one, or absorbs one existing line of either orientation,
with weight equal to the steady-state contraction of the two operators in the
order in which they stand, times the number of such lines.} Figure~\ref{fig:vertices}
shows the moves. Because the weights are steady-state moments rather than
commutators, they carry the temperature and squeezing dependence of $\Lop_0$
into the vertices rather than into the propagators, which remain
temperature independent (Sec.~\ref{sec:singlemode}).

\emph{Vacuum.} For $N=M=0$ only $\contr{c}{c\dg}=1$ survives and
Eqs.~\eqref{eq:gmove1}--\eqref{eq:gmove4} reduce to
\begin{align}
  \ad\,O_{mn} &= O_{(m+1)n}, \label{eq:move1}\\
  O_{mn}\,a   &= O_{m(n+1)}, \label{eq:move2}\\
  a\,O_{mn}   &= O_{m(n+1)} + m\,O_{(m-1)n}, \label{eq:move3}\\
  O_{mn}\,\ad &= O_{(m+1)n} + n\,O_{m(n-1)}, \label{eq:move4}
\end{align}
which are the four shift relations
\begin{align}
  \ad O_{mn} &= O_{(m+1)n}, & [a,O_{mn}] &= m\,O_{(m-1)n}, \nonumber \\
  O_{mn}a &= O_{m(n+1)}, & [O_{mn},\ad] &= n\,O_{m(n-1)}
  \label{eq:shift}
\end{align}
in another arrangement: a creation operator on the left or an annihilation
operator on the right creates a line and nothing else, while the other two
placements have no direct rule, and commuting the operator across produces the
creation term plus an absorption of one line of the opposite orientation with
weight $m$ or $n$. Those absorption terms are the non-derivation correction of
Eq.~\eqref{eq:leibniz}; in vacuum they are the only absorptions.

\emph{Thermal bath.} For $M=0$ and $N=\bar n$ the vacuum absorptions acquire
the weight $\bar n+1$ and two channels open that vacuum forbids: $\ad$ from the
left absorbs an $a$ line and $a$ from the right absorbs an $\ad$ line, each
with weight $\bar n$. The vertex table acquires stimulated as well as
spontaneous terms, matching the rates $\gamma(\bar n+1)$ and $\gamma\bar n$
of the thermal dissipator, but at the vertices rather than in the
propagators.

\emph{Squeezed bath.} For $M\neq0$ the remaining channel opens: an operator
absorbs a line of its own orientation, $c\dg$ absorbing a $c\dg$ line with
weight $M^{*}$ and $c$ absorbing a $c$ line with weight $M$. These are the
anomalous contractions of Eq.~\eqref{eq:NMsqueezed}, and they are what lets a
squeezed reservoir change the photon statistics at first order in the
interaction.

\emph{Many modes without anomalous terms.} For the particle-conserving
multimode problem of Sec.~\ref{sec:multimode} the eigenoperators are the
monomials $O_{\bvec m,\bvec n}$ in the left modes $a_{k,L}$, and a line carries
the mode label $k$ together with its arrow. The interaction is first written
in the left modes through the completeness relation
$\sum_k\bvec{v}_{k,R}\bvec{v}_{k,L}\dg=1$,
\begin{equation}
  \bvec a = \sum_k \bvec{v}_{k,R}\,a_{k,L},
  \qquad
  \bvec a\dg = \sum_k \bvec{v}_{k,R}^{*}\,a_{k,L}\dg,
  \label{eq:leftexpand}
\end{equation}
so a physical operator $a_j$ entering a vertex is a superposition of the
left modes with the right eigenvector as coefficients. Each $a_{k,L}$ then acts
by the vacuum rule, except that the absorption weight is no longer a Kronecker
delta: the left modes are not orthonormal among themselves, and
$\contr{a_{k,L}}{a_{k',L}\dg}=[a_{k,L},a_{k',L}\dg]=\bvec v_{k,L}\dg\bvec v_{k',L}\equiv G_{kk'}$.
The moves are
\begin{align}
  a_{k,L}\,O_{\bvec m,\bvec n} &= O_{\bvec m,\bvec n+\bvec e_k}
     + \sum_{k'} m_{k'}G_{kk'}\,O_{\bvec m-\bvec e_{k'},\bvec n}, \label{eq:mmove1}\\
  O_{\bvec m,\bvec n}\,a_{k,L}\dg &= O_{\bvec m+\bvec e_k,\bvec n}
     + \sum_{k'} n_{k'}G_{k'k}\,O_{\bvec m,\bvec n-\bvec e_{k'}}, \label{eq:mmove2}
\end{align}
with $a_{k,L}\dg$ from the left and $a_{k,L}$ from the right creating a line
$k$ and nothing else. An incoming line $k$ can therefore be absorbed as any
line $k'$ that overlaps with it, which is how correlated decay enters the
vertices; the propagators stay diagonal. When $\Gamma$ is normal the left
modes are orthonormal, $G_{kk'}=\delta_{kk'}$, absorption conserves the mode
label, and the rule is the vacuum rule with a label attached. This is the
case of the Kerr ring of Sec.~\ref{sec:ring}, where the label is momentum and
Eqs.~\eqref{eq:mmove1}--\eqref{eq:mmove2} become momentum conservation. A
thermal multimode bath adds, exactly as for one mode, the absorptions
$\contr{a_{k,L}\dg}{a_{k',L}}=N_{kk'}$ of an $a$ line by $a\dg$ from the left
and of an $\ad$ line by $a$ from the right, and changes the vacuum weights to
$G_{kk'}+N_{kk'}$.

\emph{Many modes with anomalous terms.} When the jump operators contain
creation operators, or $H_0$ contains parametric terms, the eigenoperators are
the Wick monomials $O_{\bvec n}$ of Eq.~\eqref{eq:generaleigen} in the $2K$
damped normal modes $\beta_\mu$, which run over both $\beta_{k,L}$ and
$\beta_{k,L}\dg$, and a line carries the label $\mu$. The interaction is
rewritten in these modes through the inverse of the eigendecomposition of
$\mathcal G$,
\begin{equation}
  \bvec A = \sum_{\mu=1}^{2K} \bvec w_{\mu,R}\,\beta_\mu ,
  \label{eq:modeinverse}
\end{equation}
so every monomial in $a_j$ and $a_j\dg$ becomes a polynomial in the
$\beta_\mu$, and Eq.~\eqref{eq:wickmult} applies with the full covariance,
\begin{align}
  \beta_\mu\,O_{\bvec n} &= O_{\bvec n+\bvec e_\mu}
     + \sum_{\nu} n_{\nu}V_{\mu\nu}\,O_{\bvec n-\bvec e_\nu}, \nonumber\\
  O_{\bvec n}\,\beta_\mu &= O_{\bvec n+\bvec e_\mu}
     + \sum_{\nu} n_{\nu}V_{\nu\mu}\,O_{\bvec n-\bvec e_\nu},
  \label{eq:mmove3}
\end{align}
where $V_{\mu\nu}=\contr{\beta_\mu}{\beta_\nu}$ is the ordered covariance of
Eq.~\eqref{eq:contraction}. The single-mode moves
Eqs.~\eqref{eq:gmove1}--\eqref{eq:gmove4} are the $K=1$ case with
$(\beta_1,\beta_2)=(c\dg,c)$ and $V$ built from $N$, $M$, $M^{*}$, and $N+1$;
the particle-conserving moves Eqs.~\eqref{eq:mmove1}--\eqref{eq:mmove2} are
the case in which $V$ is block diagonal, with $\contr{a_{k,L}}{a_{k',L}\dg}=G_{kk'}$
in vacuum. What the anomalous terms add is the same channel as for one
squeezed mode, now between any two lines: $\beta_\mu$ may absorb a line $\nu$
of the same orientation whenever $V_{\mu\nu}\neq0$, which is the diagrammatic
signature of two-mode squeezing.

\emph{Emitters.} For a two-level emitter whose steady state is diagonal (the
undriven emitter of Sec.~\ref{sec:tls} at zero or finite temperature) the
eigenoperators are $\tilde O_{00}=I$, $\tilde O_{01}=\sigm$,
$\tilde O_{10}=\sigp$, and $\tilde O_{11}=\sigp\sigm-p_e$, with
$p_e=\avg{\sigp\sigm}_{\mathrm{ss}}$ equal to zero at zero temperature and to
$N/(2N+1)$ in a thermal bath~\cite{SM}. The shift relations
become
\begin{align}
  \sigp\tilde O_{mn} &= \delta_{m0}\big[\tilde O_{1n} + \delta_{n1}\,p_e\,\tilde O_{00}\big]
      \nonumber\\ &\quad - \delta_{m1}\delta_{n1}\,p_e\,\tilde O_{10}, \nonumber\\
  \tilde O_{mn}\sigm &= \delta_{n0}\big[\tilde O_{m1} + \delta_{m1}\,p_e\,\tilde O_{00}\big]
      \nonumber\\ &\quad - \delta_{m1}\delta_{n1}\,p_e\,\tilde O_{01}, \nonumber\\
  \sigm\tilde O_{mn} &= \delta_{m0}\delta_{n0}\,\tilde O_{01}
      + \delta_{m1}\delta_{n1}\,\bar p_e\,\tilde O_{01} \nonumber\\
      &\quad + \delta_{m1}\delta_{n0}\big(\bar p_e\,\tilde O_{00}-\tilde O_{11}\big), \nonumber\\
  \tilde O_{mn}\sigp &= \delta_{m0}\delta_{n0}\,\tilde O_{10}
      + \delta_{m1}\delta_{n1}\,\bar p_e\,\tilde O_{10} \nonumber\\
      &\quad + \delta_{m0}\delta_{n1}\big(\bar p_e\,\tilde O_{00}-\tilde O_{11}\big),
  \label{eq:emittermoves}
\end{align}
with $\bar p_e = 1-p_e$. They obey the same rule with two changes. The
absorption weights are again steady-state expectation values of the product,
$\avg{\sigp\sigm}_{\mathrm{ss}}=p_e$ and $\avg{\sigm\sigp}_{\mathrm{ss}}=\bar p_e$;
and exclusion forbids a second line of the same orientation, $(\sigma^\pm)^2=0$,
which is why an operator acting on the occupied label $\tilde O_{11}$ can only
return a single line, with weight $-p_e$ or $\bar p_e$. At $p_e=0$
Eq.~\eqref{eq:emittermoves} is the zero-temperature table,
$\sigm O_{1n}=O_{0n}-O_{1(n+1)}$ and its companions, in which
$\sigm\sigp=1-\sigp\sigm$ gives the creation term a minus sign.

For a driven emitter the steady state has coherences,
$\avg{\sigma^\mp}_{\mathrm{ss}}\neq0$, and the eigenoperators are the dressed
modes $X_k$ of Sec.~\ref{sec:drivenatom}. The rule keeps its form:
\begin{align}
  \sigma^\pm X_k &= \avg{\sigma^\pm X_k}_{\mathrm{ss}}\,I + \sum_j F^{\pm}_{kj}X_j,
  \nonumber\\
  X_k\sigma^\pm &= \avg{X_k\sigma^\pm}_{\mathrm{ss}}\,I + \sum_j \tilde F^{\pm}_{kj}X_j,
  \nonumber\\
  \sigma^\pm I &= \avg{\sigma^\pm}_{\mathrm{ss}}\,I + \sum_j c^\pm_j X_j ,
  \label{eq:dressedmoves}
\end{align}
where the identity components are once more steady-state expectation values,
now including the one-point functions, which let an emitter operator enter a
vertex and leave no line at all. The coefficients $F^\pm$, $\tilde F^\pm$ and
$c^\pm$ are the structure constants of the four-dimensional emitter algebra in
the dressed basis, a finite table computed once for given $\Omega$ and
$\Delta_a$. Because the algebra is four dimensional, an emitter never carries
more than one line.

\emph{Interaction vertices in general.} Nothing so far assumed a particular
interaction. Any $\Lint\dg$ is a sum of terms of the form $O\mapsto AOB$ with
$A$ and $B$ products of mode and emitter operators: a Hamiltonian interaction
gives $i[H_{\mathrm{int}},O]=iH_{\mathrm{int}}O-iOH_{\mathrm{int}}$, with one of
the two factors trivial in each term, and a dissipative one gives the
two-sided $L\dg O L$ of Eq.~\eqref{eq:Ddagcompact}. Each term is
evaluated by applying the elementary moves once per operator factor, those in
$A$ from the left and those in $B$ from the right. The vertex table for any
interaction is therefore assembled from the same moves; what follows records
the assembly for the cases that occur in practice.

\emph{One mode, Hamiltonian, vacuum.} For a normal-ordered monomial the
composition can be done once and for all. Applying Eqs.~\eqref{eq:move1}--\eqref{eq:move4}
$p+q$ times on each side of the commutator and collecting terms gives the
closed form
\begin{align}
  i\big[(\ad)^pa^q,\,O_{mn}\big] = {}& i\sum_{k\geq0}\left[\binom{q}{k}m^{\underline{k}}
     - \binom{p}{k}n^{\underline{k}}\right] \nonumber\\
   &\times O_{(m+p-k)(n+q-k)},
  \label{eq:generalvertex}
\end{align}
with $x^{\underline{k}}=x!/(x-k)!$ the falling factorial. The index $k$ counts
how many of the $p+q$ operators absorb a line instead of creating one; the two
terms are the two orders in which the monomial can stand, and they cancel
whenever $p=q$ and $m=n$, which is why $\avg{\bd b}$ has no first-order Kerr
correction in Sec.~\ref{sec:kerr}. Every entry of Table~\ref{tab:vertices} is
Eq.~\eqref{eq:generalvertex} evaluated for $p+q\leq4$, and a general
$H_{\mathrm{int}}=\sum_{pq}h_{pq}(\ad)^pa^q$ is the corresponding sum, with the
$\alpha$-dependent coefficients of a displaced frame carried along unchanged.

\emph{One mode, arbitrary Gaussian $\Lop_0$.} Away from vacuum nothing changes
except which moves are used: the factors of $H_{\mathrm{int}}$ are applied one
at a time by Eqs.~\eqref{eq:gmove1}--\eqref{eq:gmove4} instead of
Eqs.~\eqref{eq:move1}--\eqref{eq:move4}, reading the monomial inwards from
each side of the commutator. No preliminary Wick ordering of
$H_{\mathrm{int}}$ is needed. A line created by one factor of the vertex is
available to be absorbed by a later factor of the same vertex, and those terms
are exactly the contractions between operators of $H_{\mathrm{int}}$; the rules
for $c$ and $c\dg$ generate them automatically. For instance, applying
Eq.~\eqref{eq:gmove3} and then Eq.~\eqref{eq:gmove1} to the number operator
gives
\begin{align}
  c\dg c\,O_{mn} = {}& O_{(m+1)(n+1)} + mM^{*}O_{(m-1)(n+1)} \nonumber\\
   &+ nM\,O_{(m+1)(n-1)} \nonumber\\
   &+ \big[(n+1)N + m(N+1)\big]O_{mn} + \ldots ,
  \label{eq:selfcontract}
\end{align}
where the $(n+1)N$ piece is the line just created by $c$ being absorbed again
by $c\dg$, the self-contraction $\contr{c\dg}{c}=N$ of the vertex with
itself, and the ellipsis stands for the terms of lower degree, in which both
factors absorb.
In Eq.~\eqref{eq:generalvertex} the effect is to replace the falling factorials
by sums of products of $N$, $N+1$, $M$, and $M^{*}$ over which lines are
absorbed and which factors contract with each other.

\emph{Many modes.} For several modes the monomial carries mode labels,
$\prod_j(a_j\dg)^{p_j}a_j^{q_j}$; the interaction is first written in the
normal modes by Eq.~\eqref{eq:leftexpand} or Eq.~\eqref{eq:modeinverse}, and
each factor then acts by Eqs.~\eqref{eq:mmove1}--\eqref{eq:mmove3}. The
structure of Eq.~\eqref{eq:generalvertex} survives with the absorption weights
$G_{kk'}$ or $V_{\mu\nu}$ in place of the falling factorials; the Kerr ring of
Sec.~\ref{sec:ring}, whose vertex is Eq.~\eqref{eq:ringvertex}, is this case
with momentum labels.

\emph{Field--emitter interactions.} An interaction that couples a field to an
emitter is a product $H_{\mathrm{int}}=h\,F\,A$ of a field monomial $F$ and an
emitter operator $A$, acting on a product label $O_{mn}\tilde O_{\mu\nu}$.
Because the two sectors commute,
\begin{equation}
  i\big[h\,FA,\;O_{mn}\tilde O\big]
   = ih\Big[(F\,O_{mn})(A\,\tilde O) - (O_{mn}\,F)(\tilde O\,A)\Big],
  \label{eq:productvertex}
\end{equation}
so each term of the vertex is a field move times an emitter move, and the
vertex factor is their product. It is \emph{not} a product of two commutators:
that is the failure of the Leibniz rule, Eq.~\eqref{eq:leibniz}, seen at the
level of the vertices. For the exchange coupling $h\ad\sigm$ of a
Jaynes--Cummings interaction, Eq.~\eqref{eq:productvertex} with
Eqs.~\eqref{eq:move1}, \eqref{eq:move4} gives the two entries
\begin{align}
  (j,j') = (+1,0):&\quad ih\,[\sigm,\tilde O], \nonumber\\
  (j,j') = (0,-1):&\quad -ih\,n\,\tilde O\sigm ,
  \label{eq:jcvertex}
\end{align}
and the conjugate monomial $h^{*}a\sigp$ gives the Hermitian-conjugate table.
The first entry creates a photon line and acts on the emitter label with the
commutator; the second absorbs one of the $n$ annihilation-type photon lines
and multiplies the emitter label from the right. Both are collected, with the
other interactions used in this paper, in Table~\ref{tab:interactions}.

\begin{table}[t]
\caption{Vertex tables for common interactions, in vacuum and for one mode
unless stated. Entries are the coefficients $\Vint_{j,j'}$ of
Eq.~\eqref{eq:vertexdef} acting on the label $O_{mn}$ (times $\tilde O$ for the
emitter), for the monomial shown; the Hermitian-conjugate monomial gives the
transposed table. All follow from Eq.~\eqref{eq:generalvertex} for the field
factor and Eq.~\eqref{eq:productvertex} for the product with an emitter. The
three-wave monomial $g(\ad)^2a$ is the degenerate single-mode form of the
three-wave interaction, with all three modes identified.}
\label{tab:interactions}
\begin{ruledtabular}
\begin{tabular}{llll}
Interaction & monomial & $(j,j')$ & $\Vint_{j,j'}$ \\ \hline
drive            & $\eta\ad$          & $(0,-1)$  & $-i\eta\,n$ \\
parametric drive & $\lambda(\ad)^2$   & $(1,-1)$  & $-2i\lambda\,n$ \\
                 &                    & $(0,-2)$  & $-i\lambda\,n(n-1)$ \\
three-wave       & $g(\ad)^2a$        & $(1,0)$   & $ig(m-2n)$ \\
                 &                    & $(0,-1)$  & $-ig\,n(n-1)$ \\
Kerr             & $\tfrac{U}{2}(\ad)^2a^2$ & $(1,1)$ & $iU(m-n)$ \\
                 &                    & $(0,0)$   & $\tfrac{i}{2}U(m-n)(m+n-1)$ \\
exchange (JC)    & $g\ad\sigm$        & $(1,0)$   & $ig\,[\sigm,\tilde O]$ \\
                 &                    & $(0,-1)$  & $-ig\,n\,\tilde O\sigm$ \\
dispersive       & $\chi\ad a\,\sigma_z$ & $(1,1)$ & $i\chi\,[\sigma_z,\tilde O]$ \\
                 &                    & $(0,0)$   & $i\chi\big(m\,\sigma_z\tilde O - n\,\tilde O\sigma_z\big)$ \\
\end{tabular}
\end{ruledtabular}
\end{table}

\emph{Dissipative interactions.} If $\Lint$ contains a dissipator rather than a
commutator, the recycling term $L\dg OL$ puts operator factors on both sides of
the label at the same vertex, and the moves are applied on both sides
accordingly. Two consequences follow. The vertex can change the label in ways a
Hamiltonian vertex cannot, and the pruning rule Eq.~\eqref{eq:killrule}, which
rests on the vertex being a commutator, no longer applies; the enumeration must
then be done without it, as noted for dephasing placed in $\Lint$.

\begin{table}[t]
\caption{Vertex table in vacuum: the coefficients $\Vint_{j,j'}$ of
Eq.~\eqref{eq:vertexdef} for $\Lint\dg O_{mn}=i[h\,(\ad)^pa^q+\mathrm{h.c.},O_{mn}]$,
for the monomials of a $\chi^{(2)}$ ($p+q=3$) and $\chi^{(3)}$ ($p+q=4$)
interaction, i.e.\ Eq.~\eqref{eq:generalvertex} evaluated term by term. Entries
are the factor multiplying $O_{(m+j)(n+j')}$, omitting the overall $ih$ (and
$ih^{*}$ for the conjugate monomial, which gives the transposed table). Displacing to a driven frame multiplies each entry by the
powers of $\alpha$, $\alpha^{*}$ left over, as in Eq.~\eqref{eq:kerrb1table};
a nonvacuum steady state adds the contraction channels of
Eqs.~\eqref{eq:gmove1}--\eqref{eq:gmove4}.}
\label{tab:vertices}
\begin{ruledtabular}
\begin{tabular}{lll}
$(\ad)^pa^q$ & $(j,j')$ & $\Vint_{j,j'}$ \\ \hline
$\ad a$        & $(0,0)$   & $m-n$ \\
$(\ad)^2a$     & $(1,0)$   & $m-2n$ \\
                & $(0,-1)$  & $-n(n-1)$ \\
$\ad a^2$      & $(0,1)$   & $2m-n$ \\
                & $(-1,0)$  & $m(m-1)$ \\
$(\ad)^2a^2$   & $(1,1)$   & $2(m-n)$ \\
                & $(0,0)$   & $(m-n)(m+n-1)$ \\
$(\ad)^3a$     & $(2,0)$   & $m-3n$ \\
                & $(1,-1)$  & $-3n(n-1)$ \\
                & $(0,-2)$  & $-n(n-1)(n-2)$ \\
$\ad a^3$      & $(0,2)$   & $3m-n$ \\
                & $(-1,1)$  & $3m(m-1)$ \\
                & $(-2,0)$  & $m(m-1)(m-2)$ \\
\end{tabular}
\end{ruledtabular}
\end{table}

\subsection{The propagator}
\label{sec:propagator}

The expansion Eq.~\eqref{eq:Nth} needs two inputs: the vertex map
Eq.~\eqref{eq:vertexdef}, supplied by the index algebra above, and the free
propagator, the matrix element between two labels of
\begin{equation}
  \int_0^{\infty}\!\! e^{\Lop_0\tau}e^{-s\tau}\,d\tau = \left(s-\Lop_0\right)^{-1},
  \label{eq:resolvent}
\end{equation}
with $s\to0$ giving the steady-state limit of Eq.~\eqref{eq:ss}.

The diagrams are drawn in the labels $O_{mn}$, which are eigenoperators of
$\Lop_0$, so the resolvent is diagonal on them and every line of every diagram
carries the scalar
\begin{equation}
  \left[(s-\Lop_0\dg)^{-1}\right]_{mn} = \frac{1}{s-\Lambda_{mn}} .
  \label{eq:scalarprop}
\end{equation}
There is no matrix to invert and no ladder to sum. Only the dependence of
$\Lambda_{mn}$ on the label varies from problem to problem, and two cases
occur.

\emph{(i) Additive.} $\Lambda_{\bvec m,\bvec n} = -\sum_k(\zeta_kn_k+\zeta_k^{*}m_k)$,
so the propagator is read off the picture line by line: each line contributes
its own rate to the denominator. This is the Gaussian case of
Secs.~\ref{sec:gaussmode}--\ref{sec:general}, and it makes the rules of
Sec.~\ref{sec:ssrules} local in the lines.

\emph{(ii) Diagonal but not additive.} $\Lambda$ is still a number attached to
the label, so Eq.~\eqref{eq:scalarprop} and every rule in
Sec.~\ref{sec:ssrules} hold; only the line-by-line assembly of the denominator
is lost, and $\Lambda$ is evaluated on the whole label. Pure dephasing, for
which $\Lambda_{mn}$ acquires $-\tfrac{\gamma_\phi}{2}(m-n)^2$~\cite{SM}, is
of this type; the non-additive piece is pairwise and may be regarded as a
static line--line interaction carried along the propagator.

\emph{Operators that are not labels.} If an operator of interest (an
observable, an initial state, or an operator left behind at a vertex) is not
one of the $O_{mn}$, it is expanded,
\begin{equation}
  X = \sum_{pq} c_{pq}\,O_{pq},
  \qquad
  c_{pq} \ \text{fixed by Eq.~\eqref{eq:dual}},
  \label{eq:expandX}
\end{equation}
and each term starts its own diagram, with the weight $c_{pq}$ and the scalar
propagator of its own label. This is the step used in Sec.~\ref{sec:recombine}
to assemble $\avg{(\ad)^ma^n}$ from eigenoperator moments and in
Sec.~\ref{sec:emitter} to expand $\sigma^{\mp}$ in the dressed modes of a
driven emitter: one diagram per eigenoperator, not one dressed line per
operator. A finite expansion costs a finite number of diagrams. It replaces
the two constructions one might otherwise use, a propagator dressed by rungs
along a ladder or a matrix inverse over a block, both of which amount to
writing $\Lop_0$ in a basis that is not its eigenbasis. For a thermal bath, in
the plain monomials $\Lop_0\dg$ is triangular and its resolvent is a path sum
along the rungs, whereas in the Wick monomials of Sec.~\ref{sec:singlemode},
which are what Eq.~\eqref{eq:expandX} produces, it is
Eq.~\eqref{eq:scalarprop} again. The driven emitter of
Sec.~\ref{sec:drivenatom} is the same statement with a $4\times4$ block.

Two requirements remain. The eigenoperators of $\Lop_0$ must be available in
closed form, which Sec.~\ref{sec:gaussian} supplies for Gaussian generators
and for emitters. And the expansion Eq.~\eqref{eq:expandX} must be short for
the operators of interest, which Sec.~\ref{sec:recombine} shows it is for
low-order moments. When either fails, the offending part of $\Lop$ is moved
into $\Lint$, where it becomes vertices. The split $\Lop = \Lop_0 + \Lint$ is
thus a choice of how much to diagonalize and how much to expand in. The same
elementary move algebra is used for whichever terms remain in $\Lint$, while
the actual vertex table changes with that choice~\cite{SM}.

\subsection{Diagram elements}
\label{sec:elements}

The examples of Secs.~\ref{sec:kerr}--\ref{sec:emitter} use several kinds of
line; Table~\ref{tab:elements} fixes the vocabulary. The orientation of the arrow carries the operator type: a right
arrow is an annihilation-type factor and propagates with the rate $z$ (or
$z_a$, or $-\lambda_k$), a left arrow is a creation-type factor and propagates
with the conjugate rate. The incoming propagator of a diagram is the inverse
of the sum of the rates of its incoming lines whenever $\Lop_0$ is additive
across them, and $(s-\Lambda)^{-1}$ evaluated on the whole label otherwise,
Sec.~\ref{sec:propagator}. Lines are drawn from the initial state on
the right to the observable on the left.

\begin{table}[t]
\caption{Diagram elements. ``Rate'' is the contribution of a line to the
denominator of the incoming propagator. $z=\kappa/2+i\Delta_c$ for a cavity,
$z_a=\gamma/2+i\Delta_a$ for an emitter, and $\lambda_k$ are the eigenvalues of
the driven-emitter generator of Sec.~\ref{sec:drivenatom}.}
\label{tab:elements}
\begin{ruledtabular}
\begin{tabular}{clcc}
Glyph & Element & Arrow & Rate \\ \hline
\ldglyph{ld a}  & bare mode $a$ & right & $z$ \\
\ldglyph{ld ad} & bare mode $\ad$ & left & $z^{*}$ \\
\ldglyph{ld b}  & fluctuation $b=a-\alpha$ & right & $z$ \\
\ldglyph{ld bd} & fluctuation $\bd$ & left & $z^{*}$ \\
\ldglyph{ld sm} & emitter $\sigm$ & right & $z_a$ \\
\ldglyph{ld sp} & emitter $\sigp$ & left & $z_a^{*}$ \\
\ldglyph{ld d1} & dressed mode $X_1=X_1^{\dagger}$ & circle & $-\lambda_1$ \\
\ldglyph{ld d2} & dressed mode $X_2$ & right & $-\lambda_2$ \\
\ldglyph{ld d3} & dressed mode $X_3=X_2^{\dagger}$ & left & $-\lambda_2^{*}$ \\[2pt]
\tikz[baseline=-0.6ex]\node[ld vtx]{}; & vertex, marked with its order $(N)$ & --- & --- \\
\end{tabular}
\end{ruledtabular}
\end{table}

\emph{Bare and fluctuation lines.} A wavy line is a factor of $a$ or $\ad$ of
an undriven mode. For a coherently driven mode we work in the displaced frame
of Sec.~\ref{sec:drive} and draw the fluctuation $b=a-\alpha$ as a coil. The
c-number factors $\alpha$ and $\alpha^{*}$ that the displacement leaves in an
interaction are part of the vertex factor and are not drawn.

\emph{Emitter lines.} For the undriven emitter of Sec.~\ref{sec:tls} the
labels $\tilde O_{mn}$, $m,n\in\{0,1\}$, of Eq.~\eqref{eq:emittermoves} are
eigenoperators with $\Lambda_{mn}=-(z_an+z_a^{*}m)$ at zero temperature, which
is additive, so a $\sigm$ line carries a right arrow and a $\sigp$ line a left
arrow, exactly as for a mode.

\emph{Dressed emitter lines.} When the drive is kept inside $\Lop_0$ the
emitter's Heisenberg generator has, besides the identity, three eigenoperators
$X_k$ with rates $\lambda_k$ and $\Tr[X_k\rho_{\mathrm{ss}}]=0$. Because
$\Lop_a\dg$ preserves Hermiticity, the $\lambda_k$ are real or come in
conjugate pairs. Generically there is one real eigenvalue $\lambda_1$ with
$X_1$ Hermitian and one pair $\lambda_3=\lambda_2^{*}$, $X_3=X_2^{\dagger}$; off
resonance these connect continuously to the population $\sigp\sigm-p_e$, to
$\sigm$, and to $\sigp$ as $\Omega\to0$, and we label $X_2$ the mode with
$\mathrm{Im}\,\lambda_2<0$. We draw them as double lines, the doubling marking
that the drive has been resummed into them: $X_2$ with a right arrow, $X_3$
with a left arrow, and the self-conjugate $X_1$ with an open circle. On
resonance the labelling is ambiguous below the exceptional point
$\Omega=\gamma/4$, where both $\lambda_\pm$ of Sec.~\ref{sec:drivenatom} are
real; the propagator Eq.~\eqref{eq:scalarprop} is unaffected,
and only the pictures need a convention. The Schr\"odinger-picture duals
$\rho_k$, with $\Tr[X_j\rho_k]=\delta_{jk}$, label the initial-state terms on
which dressed lines close.

\subsection{Feynman rules}
\label{sec:ssrules}

The rules are stated for a general matrix element between a label of the
observable and a label of the state, $\Tr[O_{mn}e^{\Lop\Delta t}r_{m'n'}]$, which
is the object Eq.~\eqref{eq:Nth} expands. Reading that expansion from the
observable towards the state, each factor $\Lint$ moves the label and each
stretch between two factors is a propagator, Eq.~\eqref{eq:scalarprop}. We
collect a stretch and the move that follows it into a first-order diagram and
build every order by gluing such diagrams.

\begin{enumerate}
\item Draw the observable label $O_{mn}$ as $m$ lines with a left arrow and $n$
lines with a right arrow entering a single vertex from the left.
\item Mark the vertex with the order $N$, and draw the label $O_{pq}$ reached
after $N$ applications of $\Lint\dg$ as $p$ left-arrow and $q$ right-arrow lines
leaving to the right. They close on the state label $r_{qp}$, with the index
transposition of Eq.~\eqref{eq:dual}.
\item In Laplace space, $\tilde f(s)=\int_0^\infty e^{-s\Delta t}f(\Delta t)\,d\Delta t$,
a first-order diagram has the value
\begin{equation}
  D^{(1)}_{mn\to pq}(s) = \frac{\left[\Lint\dg O_{mn}\right]_{pq}}{s+zn+z^{*}m},
  \label{eq:D1}
\end{equation}
the incoming propagator times the vertex factor of Sec.~\ref{sec:vertices}.
\item Higher orders follow by gluing the outgoing lines of one diagram to the
incoming lines of the next,
\begin{equation}
  D^{(N)}_{mn\to pq} = \sum_{(k,l)} D^{(1)}_{mn\to kl}\,D^{(N-1)}_{kl\to pq}.
  \label{eq:gluing}
\end{equation}
\item Closing the diagram on the state supplies one further propagator for the
outgoing label, so that
\begin{equation}
  \left(\widetilde{\Tr\!\left[O_{mn}e^{\Lop\Delta t}r_{m'n'}\right]}\right)^{(N)}
   = \frac{D^{(N)}_{mn\to n'm'}(s)}{s+zm'+z^{*}n'},
  \label{eq:laplace}
\end{equation}
with $D^{(0)}_{mn\to pq}=\delta_{mp}\delta_{nq}$.
\item Intermediate labels $(k,l)=(0,0)$ contribute nothing.
\end{enumerate}

\emph{The steady state as a special case.} Take $m'=n'=0$, so that
$\Lambda_{00}=0$ and the closing propagator in Eq.~\eqref{eq:laplace} is $1/s$;
the steady-state value is $\lim_{s\to0}s\,\widetilde{(\cdots)}$, which keeps only
the term with no outgoing lines and sets $s=0$ everywhere else. Equivalently,
in the time domain, reversing each integration variable in Eq.~\eqref{eq:Nth}
via $x\to x_{\max}-x$ and using
$e^{\Lop_0(\Delta t-\sum_i\tau_i)}r_{00}=r_{00}$, the limit $t_0\to-\infty$
decouples the upper limits entirely,
\begin{widetext}
\begin{equation}
  \lim_{t_0\to-\infty}\!\left(\Tr\!\left[O_{mn}e^{\Lop(t-t_0)}r_{00}\right]\right)^{(N)}
   = \int_0^{\infty}\!\!\!\cdots\!\int_0^{\infty}
     \Tr\!\left[O_{mn}e^{\Lop_0\tau_1}\Lint e^{\Lop_0\tau_2}\cdots\Lint r_{00}\right] d\tau^N,
  \label{eq:ss}
\end{equation}
\end{widetext}
and since $\int_0^{\infty}e^{\Lambda\tau}d\tau=-1/\Lambda$ every propagator
collapses to $-1/\Lambda$, the $s\to0$ limit of Eq.~\eqref{eq:D1}. The rules
then reduce to their steady-state form: only diagrams with no outgoing lines
survive, and
\begin{equation}
  \avg{O_{mn}}^{(N)} = D^{(N)}_{mn\to00}\big|_{s=0} .
  \label{eq:ssvalue}
\end{equation}
This is the form used in Secs.~\ref{sec:examples} and \ref{sec:ring}. At finite
time the poles of Eq.~\eqref{eq:laplace} invert to sums of exponentials; that
calculation, and a worked switch-on of the Kerr cavity, are given in the
Supplemental Material~\cite{SM}.

\emph{A general state expands in the labels.} Nothing above assumed that the
state is one of the $r_{m'n'}$, just as nothing assumed that the observable is
one of the $O_{mn}$. An arbitrary initial state is expanded in the
Schr\"odinger labels exactly as an arbitrary operator is expanded in the
Heisenberg ones, Eq.~\eqref{eq:expandX},
\begin{equation}
  \rho_0 = \sum_{p,q} \Tr\!\left[O_{pq}\rho_0\right] r_{qp},
  \label{eq:stateexpand}
\end{equation}
the coefficients again supplied by biorthogonality, Eq.~\eqref{eq:dual}. Each
term starts its own set of diagrams, weighted by $\Tr[O_{pq}\rho_0]$, and the
answer is their sum. There is therefore one set of diagrams for each order,
each observable label, and each state label; the steady state is the case in
which a single state label, $r_{00}$, survives.

The diagrams of a given order therefore form a table indexed by the observable
label and the state label. In practice one builds them from the first-order
table, and the pruning rule Eq.~\eqref{eq:killrule} derived in
Sec.~\ref{sec:kerr} removes intermediate labels that can never be closed;
Sec.~\ref{sec:recombine} then says how to assemble a physical moment out of
such diagram sets.

The rules are stated for the single damped mode, but nothing in them is
specific to it. Rules~1, 2, 4, 5, and 6 belong to the index algebra and never
change. Rule~3 contains the propagator, which is the scalar
Eq.~\eqref{eq:scalarprop} of the incoming label, and the vertex factor, which
follows the moves of Sec.~\ref{sec:vertices}. For a general Gaussian $\Lop_0$,
replace $zn+z^{*}m$ by $-\Lambda_{\bvec{m},\bvec{n}}$ from
Eq.~\eqref{eq:generaleigen}, one rate $\zeta_k$ or $\zeta_k^{*}$ per line with
the line labelled by the normal mode it carries, and use the contraction moves
of Eqs.~\eqref{eq:gmove1}--\eqref{eq:gmove4}. For a non-additive $\Lambda$,
evaluate it on the whole label rather than line by line.

Rule~6 is justified as follows. Rule~3 would give a divergent propagator in
the steady-state limit for a first-order diagram entering with the label
$(0,0)$, since $\Lambda_{00}=0$, and such intermediate labels do occur with
nonzero amplitude: every lower-order steady-state contribution is one. They
contribute nothing because the next vertex annihilates them: $\Lint$ is trace
preserving, so $\Lint\dg O_{00}=\Lint\dg I=0$. Equivalently, the
$t_0\to-\infty$ limit of Eq.~\eqref{eq:ss} acts on the traceless complement of
the unperturbed steady state, on which $\Lop_0$ has no zero eigenvalue. Rule~6
is the counterpart of the cancellation of vacuum bubbles in field theory, and
it is what makes the steady-state series well defined at every order. The
resulting steady-state expansion is equivalent to the Drazin-inverse expansion
of the perturbed steady state~\cite{lenarcic2018perturbative,nagib2025exact}.

\subsection{From diagrams to physical moments}
\label{sec:recombine}

The diagrams compute matrix elements between eigenoperators of $\Lop_0$. What
one usually wants is a moment of the physical operators: a quadrature, a photon
number, a two-photon correlation $\avg{(\ad)^m a^n}$, or an emitter observable
such as $\avg{\sigma^{\mp}}$ or an excited-state population. Two steps, both
fixed by $\Lop_0$ and not by the interaction, take one to the other; they
decide which diagrams to evaluate and with what weight.

\emph{Step 1: undo the displacement.} If $\Lop_0$ was diagonalized in the
displaced frame $b=a-\alpha$ of Sec.~\ref{sec:drive}, expand the physical
operators binomially,
\begin{equation}
  (\ad)^m a^n = \sum_{j=0}^{m}\sum_{k=0}^{n}
    \binom{m}{j}\binom{n}{k}\,
    (\alpha^{*})^{m-j}\alpha^{\,n-k}\;(\bd)^j b^k .
  \label{eq:undisplace}
\end{equation}
The coherent amplitude $\alpha$ is the unperturbed mean field, not the exact
one; the difference is supplied order by order by the diagrams for $\avg{b}$.

\emph{Step 2: invert the Wick ordering.} The eigenoperators
$O^{(N,M)}_{mn}=\normord{(c\dg)^mc^n}$ of Eq.~\eqref{eq:Omn} are the monomials
with all contractions subtracted, so the monomials are the eigenoperators with
all contractions added back. Reading Eq.~\eqref{eq:wickgen} the other way,
$e^{sc\dg}e^{tc}=e^{\frac12M^{*}s^2+Nst+\frac12Mt^2}\,\mathcal{W}(s,t)$, which
gives the inverse of Eq.~\eqref{eq:Omn} with the three signs reversed,
\begin{widetext}
\begin{equation}
  (c\dg)^m c^n = \sum_{\substack{p,q,k\geq0\\ 2p+k\leq m,\ 2q+k\leq n}}
    \frac{m!\,n!\,(M^{*})^p M^q N^k}{2^{p+q}\,p!\,q!\,k!\,(m-2p-k)!\,(n-2q-k)!}
    \;O^{(N,M)}_{(m-2p-k)(n-2q-k)} .
  \label{eq:unwick}
\end{equation}
\end{widetext}
For a vacuum steady state, $N=M=0$, this is the identity
$(c\dg)^mc^n=O_{mn}$ and there is nothing to do. For a thermal bath, $M=0$ and
$N=\bar n$, it is a single sum, for instance
\begin{align}
  \ad a &= O^{\mathrm{th}}_{11} + \bar n, \nonumber\\
  (\ad)^2a &= O^{\mathrm{th}}_{21} + 2\bar n\,O^{\mathrm{th}}_{10}, \nonumber\\
  (\ad)^2a^2 &= O^{\mathrm{th}}_{22} + 4\bar n\,O^{\mathrm{th}}_{11} + 2\bar n^2 .
  \label{eq:unwickthermal}
\end{align}
For a squeezed steady state the anomalous contractions appear as well,
\begin{equation}
  (\ad)^2a^2 = O_{22} + 4N\,O_{11} + M\,O_{20} + M^{*}O_{02}
     + 2N^2 + |M|^2 .
\end{equation}
Because $\avg{O_{pq}}_{\mathrm{ss}}=0$ for $p+q>0$, the constant term of
Eq.~\eqref{eq:unwick} is exactly the unperturbed moment: the inversion
separates $\avg{(\ad)^ma^n}$ into its Gaussian part and the corrections the
diagrams compute.

\emph{Putting it together.} Combining the two steps, a physical moment is a
finite, $\Lop_0$-dependent linear combination of eigenoperator moments,
\begin{equation}
  \avg{(\ad)^m a^n} = \sum_{p,q} w^{mn}_{pq}\;\avg{O_{pq}},
  \qquad
  \avg{O_{pq}} = \sum_{N\geq0}D^{(N)}_{pq\to00},
  \label{eq:moment}
\end{equation}
with $w^{mn}_{pq}$ the product of the binomial factors of
Eq.~\eqref{eq:undisplace} and the contraction factors of
Eq.~\eqref{eq:unwick}, and each $\avg{O_{pq}}$ the sum of the diagrams with
that external operator. The labels needed are bounded by $p\leq m$ and
$q\leq n$, so a quadrature needs $O_{01}$ and $O_{10}$, a photon number
$O_{11}$ as well, and a two-photon correlation $O_{22}$ and everything below
it: a few diagram sets, computed once and reused for every moment of the same
order.

\emph{Emitters.} The same two steps apply, with the roles of the displacement
and the contractions played by the steady-state one-point functions. For an
undriven emitter the labels of Eq.~\eqref{eq:emittermoves} are
$\tilde O_{01}=\sigm$, $\tilde O_{10}=\sigp$, and
$\tilde O_{11}=\sigp\sigm-p_e$, so the operators one measures are
\begin{align}
  \sigma^{\mp} &= \tilde O_{01},\ \tilde O_{10}, \nonumber\\
  \sigp\sigm &= \tilde O_{11} + p_e, \nonumber\\
  \sigm\sigp &= -\tilde O_{11} + (1-p_e), \nonumber\\
  \sigma_z &= 2\tilde O_{11} + (2p_e-1),
  \label{eq:emitterinvert}
\end{align}
with $p_e=\avg{\sigp\sigm}_{\mathrm{ss}}$ zero at zero temperature and
$\bar n/(2\bar n+1)$ in a thermal bath; the constant is again the unperturbed
value, and the diagrams supply the correction. For a driven emitter the labels
are the dressed modes $X_k$ of Sec.~\ref{sec:drivenatom} and the inversion
is the last line of Eq.~\eqref{eq:dressedmoves} read as an expansion of the
physical operator,
\begin{equation}
  \sigma^{\mp} = \avg{\sigma^{\mp}}_{\mathrm{ss}}\,I + \sum_{k=1}^{3} c^{\mp}_k X_k,
  \qquad
  c^{\mp}_k = \Tr\!\left[\sigma^{\mp}\rho_k\right],
  \label{eq:dressedinvert}
\end{equation}
with $\rho_k$ the Schr\"odinger duals, $\Tr[X_j\rho_k]=\delta_{jk}$, and the
same formula for $\sigp\sigm$ and $\sigma_z$ with their own means and
coefficients. Each term of Eq.~\eqref{eq:dressedinvert} starts its own diagram,
which is how the three Mollow components of Sec.~\ref{sec:emitter} arise from
one measured operator. Because the emitter algebra is four dimensional, the
expansion always terminates after the identity and three labels.

For several modes the covariance $V_{\mu\nu}$ replaces $N$ and $M$ in
Eq.~\eqref{eq:unwick}, and the displacement of Eq.~\eqref{eq:undisplace} is
applied mode by mode.

Every construction of Secs.~\ref{sec:gaussian} and \ref{sec:diagrams}, and
every diagram sum reported in Secs.~\ref{sec:examples} and \ref{sec:ring},
has been checked symbolically or numerically against an independent
calculation; the checks, references, and levels of agreement are tabulated
in the Supplemental Material~\cite{SM} (section ``Numerical checks'').

\section{Worked examples}
\label{sec:examples}

\subsection{The driven Kerr cavity}
\label{sec:kerr}

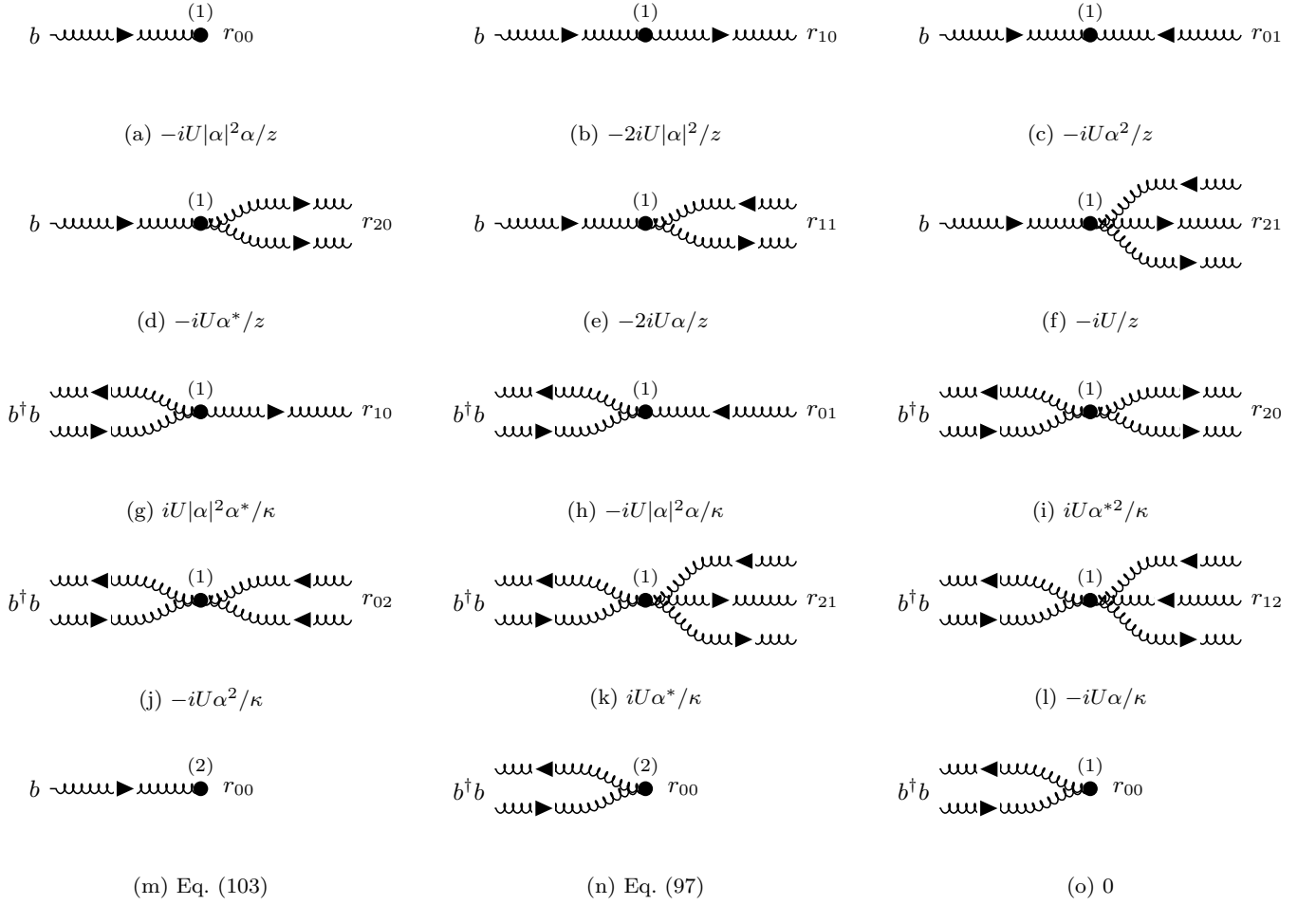
\begin{figure*}[t]
\centering
\resizebox{\textwidth}{!}{%
\begin{tikzpicture}[x=1cm,y=1cm]
\begin{scope}[shift={(0,0)}]
  \draw[ld b,ld mark at=0.5] (0,0) -- (-2,0);
  \node[ld vtx] at (0,0) {};
  \node[ld order,fill=white,fill opacity=0.85,text opacity=1,rounded corners=1pt] at (0,0.13) {$(1)$};
  \node[ld ext,anchor=east] at (-2.06,0) {$b$};
  \node[ld ext,anchor=west] at (0.22,0) {$r_{00}$};
  \node[font=\footnotesize,anchor=north] at (0,-1.05) {(a)\ $-iU|\alpha|^2\alpha/z$};
\end{scope}
\begin{scope}[shift={(5.9,0)}]
  \draw[ld b,ld mark at=0.5] (0,0) -- (-2,0);
  \draw[ld b,ld mark at=0.5] (2,0) -- (0,0);
  \node[ld vtx] at (0,0) {};
  \node[ld order,fill=white,fill opacity=0.85,text opacity=1,rounded corners=1pt] at (0,0.13) {$(1)$};
  \node[ld ext,anchor=east] at (-2.06,0) {$b$};
  \node[ld ext,anchor=west] at (2.06,0) {$r_{10}$};
  \node[font=\footnotesize,anchor=north] at (0,-1.05) {(b)\ $-2iU|\alpha|^2/z$};
\end{scope}
\begin{scope}[shift={(11.8,0)}]
  \draw[ld b,ld mark at=0.5] (0,0) -- (-2,0);
  \draw[ld bd,ld mark at=0.5] (2,0) -- (0,0);
  \node[ld vtx] at (0,0) {};
  \node[ld order,fill=white,fill opacity=0.85,text opacity=1,rounded corners=1pt] at (0,0.13) {$(1)$};
  \node[ld ext,anchor=east] at (-2.06,0) {$b$};
  \node[ld ext,anchor=west] at (2.06,0) {$r_{01}$};
  \node[font=\footnotesize,anchor=north] at (0,-1.05) {(c)\ $-iU\alpha^2/z$};
\end{scope}
\begin{scope}[shift={(0,-2.5)}]
  \draw[ld b,ld mark at=0.5] (0,0) -- (-2,0);
  \draw[ld b,ld mark at=0.32] (2,0.26) -- (0.92,0.26) .. controls (0.5,0.26) and (0.42,0) .. (0,0);
  \draw[ld b,ld mark at=0.32] (2,-0.26) -- (0.92,-0.26) .. controls (0.5,-0.26) and (0.42,0) .. (0,0);
  \node[ld vtx] at (0,0) {};
  \node[ld order,fill=white,fill opacity=0.85,text opacity=1,rounded corners=1pt] at (0,0.13) {$(1)$};
  \node[ld ext,anchor=east] at (-2.06,0) {$b$};
  \node[ld ext,anchor=west] at (2.06,0) {$r_{20}$};
  \node[font=\footnotesize,anchor=north] at (0,-1.05) {(d)\ $-iU\alpha^{*}/z$};
\end{scope}
\begin{scope}[shift={(5.9,-2.5)}]
  \draw[ld b,ld mark at=0.5] (0,0) -- (-2,0);
  \draw[ld bd,ld mark at=0.32] (2,0.26) -- (0.92,0.26) .. controls (0.5,0.26) and (0.42,0) .. (0,0);
  \draw[ld b,ld mark at=0.32] (2,-0.26) -- (0.92,-0.26) .. controls (0.5,-0.26) and (0.42,0) .. (0,0);
  \node[ld vtx] at (0,0) {};
  \node[ld order,fill=white,fill opacity=0.85,text opacity=1,rounded corners=1pt] at (0,0.13) {$(1)$};
  \node[ld ext,anchor=east] at (-2.06,0) {$b$};
  \node[ld ext,anchor=west] at (2.06,0) {$r_{11}$};
  \node[font=\footnotesize,anchor=north] at (0,-1.05) {(e)\ $-2iU\alpha/z$};
\end{scope}
\begin{scope}[shift={(11.8,-2.5)}]
  \draw[ld b,ld mark at=0.5] (0,0) -- (-2,0);
  \draw[ld bd,ld mark at=0.32] (2,0.52) -- (0.92,0.52) .. controls (0.5,0.52) and (0.42,0) .. (0,0);
  \draw[ld b,ld mark at=0.5] (2,0) -- (0,0);
  \draw[ld b,ld mark at=0.32] (2,-0.52) -- (0.92,-0.52) .. controls (0.5,-0.52) and (0.42,0) .. (0,0);
  \node[ld vtx] at (0,0) {};
  \node[ld order,fill=white,fill opacity=0.85,text opacity=1,rounded corners=1pt] at (0,0.13) {$(1)$};
  \node[ld ext,anchor=east] at (-2.06,0) {$b$};
  \node[ld ext,anchor=west] at (2.06,0) {$r_{21}$};
  \node[font=\footnotesize,anchor=north] at (0,-1.05) {(f)\ $-iU/z$};
\end{scope}
\begin{scope}[shift={(0,-5)}]
  \draw[ld bd,ld mark at=0.68] (0,0) .. controls (-0.42,0) and (-0.5,0.26) .. (-0.92,0.26) -- (-2,0.26);
  \draw[ld b,ld mark at=0.68] (0,0) .. controls (-0.42,0) and (-0.5,-0.26) .. (-0.92,-0.26) -- (-2,-0.26);
  \draw[ld b,ld mark at=0.5] (2,0) -- (0,0);
  \node[ld vtx] at (0,0) {};
  \node[ld order,fill=white,fill opacity=0.85,text opacity=1,rounded corners=1pt] at (0,0.13) {$(1)$};
  \node[ld ext,anchor=east] at (-2.06,0) {$\bd b$};
  \node[ld ext,anchor=west] at (2.06,0) {$r_{10}$};
  \node[font=\footnotesize,anchor=north] at (0,-1.05) {(g)\ $iU|\alpha|^2\alpha^{*}/\kappa$};
\end{scope}
\begin{scope}[shift={(5.9,-5)}]
  \draw[ld bd,ld mark at=0.68] (0,0) .. controls (-0.42,0) and (-0.5,0.26) .. (-0.92,0.26) -- (-2,0.26);
  \draw[ld b,ld mark at=0.68] (0,0) .. controls (-0.42,0) and (-0.5,-0.26) .. (-0.92,-0.26) -- (-2,-0.26);
  \draw[ld bd,ld mark at=0.5] (2,0) -- (0,0);
  \node[ld vtx] at (0,0) {};
  \node[ld order,fill=white,fill opacity=0.85,text opacity=1,rounded corners=1pt] at (0,0.13) {$(1)$};
  \node[ld ext,anchor=east] at (-2.06,0) {$\bd b$};
  \node[ld ext,anchor=west] at (2.06,0) {$r_{01}$};
  \node[font=\footnotesize,anchor=north] at (0,-1.05) {(h)\ $-iU|\alpha|^2\alpha/\kappa$};
\end{scope}
\begin{scope}[shift={(11.8,-5)}]
  \draw[ld bd,ld mark at=0.68] (0,0) .. controls (-0.42,0) and (-0.5,0.26) .. (-0.92,0.26) -- (-2,0.26);
  \draw[ld b,ld mark at=0.68] (0,0) .. controls (-0.42,0) and (-0.5,-0.26) .. (-0.92,-0.26) -- (-2,-0.26);
  \draw[ld b,ld mark at=0.32] (2,0.26) -- (0.92,0.26) .. controls (0.5,0.26) and (0.42,0) .. (0,0);
  \draw[ld b,ld mark at=0.32] (2,-0.26) -- (0.92,-0.26) .. controls (0.5,-0.26) and (0.42,0) .. (0,0);
  \node[ld vtx] at (0,0) {};
  \node[ld order,fill=white,fill opacity=0.85,text opacity=1,rounded corners=1pt] at (0,0.13) {$(1)$};
  \node[ld ext,anchor=east] at (-2.06,0) {$\bd b$};
  \node[ld ext,anchor=west] at (2.06,0) {$r_{20}$};
  \node[font=\footnotesize,anchor=north] at (0,-1.05) {(i)\ $iU\alpha^{*2}/\kappa$};
\end{scope}
\begin{scope}[shift={(0,-7.5)}]
  \draw[ld bd,ld mark at=0.68] (0,0) .. controls (-0.42,0) and (-0.5,0.26) .. (-0.92,0.26) -- (-2,0.26);
  \draw[ld b,ld mark at=0.68] (0,0) .. controls (-0.42,0) and (-0.5,-0.26) .. (-0.92,-0.26) -- (-2,-0.26);
  \draw[ld bd,ld mark at=0.32] (2,0.26) -- (0.92,0.26) .. controls (0.5,0.26) and (0.42,0) .. (0,0);
  \draw[ld bd,ld mark at=0.32] (2,-0.26) -- (0.92,-0.26) .. controls (0.5,-0.26) and (0.42,0) .. (0,0);
  \node[ld vtx] at (0,0) {};
  \node[ld order,fill=white,fill opacity=0.85,text opacity=1,rounded corners=1pt] at (0,0.13) {$(1)$};
  \node[ld ext,anchor=east] at (-2.06,0) {$\bd b$};
  \node[ld ext,anchor=west] at (2.06,0) {$r_{02}$};
  \node[font=\footnotesize,anchor=north] at (0,-1.05) {(j)\ $-iU\alpha^{2}/\kappa$};
\end{scope}
\begin{scope}[shift={(5.9,-7.5)}]
  \draw[ld bd,ld mark at=0.68] (0,0) .. controls (-0.42,0) and (-0.5,0.26) .. (-0.92,0.26) -- (-2,0.26);
  \draw[ld b,ld mark at=0.68] (0,0) .. controls (-0.42,0) and (-0.5,-0.26) .. (-0.92,-0.26) -- (-2,-0.26);
  \draw[ld bd,ld mark at=0.32] (2,0.52) -- (0.92,0.52) .. controls (0.5,0.52) and (0.42,0) .. (0,0);
  \draw[ld b,ld mark at=0.5] (2,0) -- (0,0);
  \draw[ld b,ld mark at=0.32] (2,-0.52) -- (0.92,-0.52) .. controls (0.5,-0.52) and (0.42,0) .. (0,0);
  \node[ld vtx] at (0,0) {};
  \node[ld order,fill=white,fill opacity=0.85,text opacity=1,rounded corners=1pt] at (0,0.13) {$(1)$};
  \node[ld ext,anchor=east] at (-2.06,0) {$\bd b$};
  \node[ld ext,anchor=west] at (2.06,0) {$r_{21}$};
  \node[font=\footnotesize,anchor=north] at (0,-1.05) {(k)\ $iU\alpha^{*}/\kappa$};
\end{scope}
\begin{scope}[shift={(11.8,-7.5)}]
  \draw[ld bd,ld mark at=0.68] (0,0) .. controls (-0.42,0) and (-0.5,0.26) .. (-0.92,0.26) -- (-2,0.26);
  \draw[ld b,ld mark at=0.68] (0,0) .. controls (-0.42,0) and (-0.5,-0.26) .. (-0.92,-0.26) -- (-2,-0.26);
  \draw[ld bd,ld mark at=0.32] (2,0.52) -- (0.92,0.52) .. controls (0.5,0.52) and (0.42,0) .. (0,0);
  \draw[ld bd,ld mark at=0.5] (2,0) -- (0,0);
  \draw[ld b,ld mark at=0.32] (2,-0.52) -- (0.92,-0.52) .. controls (0.5,-0.52) and (0.42,0) .. (0,0);
  \node[ld vtx] at (0,0) {};
  \node[ld order,fill=white,fill opacity=0.85,text opacity=1,rounded corners=1pt] at (0,0.13) {$(1)$};
  \node[ld ext,anchor=east] at (-2.06,0) {$\bd b$};
  \node[ld ext,anchor=west] at (2.06,0) {$r_{12}$};
  \node[font=\footnotesize,anchor=north] at (0,-1.05) {(l)\ $-iU\alpha/\kappa$};
\end{scope}
\begin{scope}[shift={(0,-10)}]
  \draw[ld b,ld mark at=0.5] (0,0) -- (-2,0);
  \node[ld vtx] at (0,0) {};
  \node[ld order,fill=white,fill opacity=0.85,text opacity=1,rounded corners=1pt] at (0,0.13) {$(2)$};
  \node[ld ext,anchor=east] at (-2.06,0) {$b$};
  \node[ld ext,anchor=west] at (0.22,0) {$r_{00}$};
  \node[font=\footnotesize,anchor=north] at (0,-1.05) {(m)\ Eq.~\eqref{eq:kerrb2}};
\end{scope}
\begin{scope}[shift={(5.9,-10)}]
  \draw[ld bd,ld mark at=0.68] (0,0) .. controls (-0.42,0) and (-0.5,0.26) .. (-0.92,0.26) -- (-2,0.26);
  \draw[ld b,ld mark at=0.68] (0,0) .. controls (-0.42,0) and (-0.5,-0.26) .. (-0.92,-0.26) -- (-2,-0.26);
  \node[ld vtx] at (0,0) {};
  \node[ld order,fill=white,fill opacity=0.85,text opacity=1,rounded corners=1pt] at (0,0.13) {$(2)$};
  \node[ld ext,anchor=east] at (-2.06,0) {$\bd b$};
  \node[ld ext,anchor=west] at (0.22,0) {$r_{00}$};
  \node[font=\footnotesize,anchor=north] at (0,-1.05) {(n)\ Eq.~\eqref{eq:kerr2}};
\end{scope}
\begin{scope}[shift={(11.8,-10)}]
  \draw[ld bd,ld mark at=0.68] (0,0) .. controls (-0.42,0) and (-0.5,0.26) .. (-0.92,0.26) -- (-2,0.26);
  \draw[ld b,ld mark at=0.68] (0,0) .. controls (-0.42,0) and (-0.5,-0.26) .. (-0.92,-0.26) -- (-2,-0.26);
  \node[ld vtx] at (0,0) {};
  \node[ld order,fill=white,fill opacity=0.85,text opacity=1,rounded corners=1pt] at (0,0.13) {$(1)$};
  \node[ld ext,anchor=east] at (-2.06,0) {$\bd b$};
  \node[ld ext,anchor=west] at (0.22,0) {$r_{00}$};
  \node[font=\footnotesize,anchor=north] at (0,-1.05) {(o)\ $0$};
\end{scope}
\end{tikzpicture}}
\caption{Diagrams for the driven Kerr cavity, each labelled by its value and
by the initial-state term $r_{qp}$ its outgoing lines close on.
(a)--(f)~The complete first-order table for the observable $b$, from
Eq.~\eqref{eq:kerrb1table}. (g)--(l)~The complete first-order table for
$\bd b$, Eq.~\eqref{eq:kerrV1}. (m),(n)~Second-order steady-state diagrams for
$\avg{b}$ and $\avg{\bd b}$, obtained by gluing the first-order tables,
Eqs.~\eqref{eq:kerrb2} and \eqref{eq:kerr2}. (o)~The first-order steady-state
diagram for $\avg{\bd b}$ vanishes. In the steady state only (a), (m), (n), and
(o) contribute; at finite time every panel does, weighted by
$\Tr[O_{pq}\rho_0]$.}
\label{fig:kerrdiagrams}
\end{figure*}

We first check the rules where the answer is known: the exact steady state of
this model is due to Drummond and Walls~\cite{drummond1980quantum}, later
extended to two-photon driving and loss and to its hidden time-reversal
symmetry~\cite{bartolo2016exact,roberts2020driven,roberts2021hidden}, and we
compare with direct numerical solution. Take a coherently driven, damped cavity with a Kerr nonlinearity,
\begin{equation}
  H = \Delta_c \ad a + i(\eta\ad - \eta^{*}a) + \frac{U}{2}(\ad)^2a^2,
  \qquad \kappa\Dop[a].
\end{equation}
Following Sec.~\ref{sec:drive}, displace to $b = a-\alpha$ with
$\alpha=\eta/z_c$ and $z_c = \tfrac{\kappa}{2}+i\Delta_c$. The unperturbed
generator is then the damped mode in the $b$ operators, with
$\Lop_0\dg O_{mn} = -(z_c n + z_c^{*}m)O_{mn}$ for $O_{mn}=(\bd)^m b^n$ and
unperturbed steady state $r_{00}=\ketbra{0_b}{0_b}$. The interaction is the
Kerr term, $\Lint\dg O = i[H_K,O]$ with $H_K=\tfrac{U}{2}(\ad)^2a^2$, which in
the displaced frame is a polynomial of degree two in $\bd$ and $b$ on each
side. We abbreviate $z\equiv z_c$ below. Because the drive has been absorbed
into the displacement, every line in this section is a fluctuation line. The
diagrams used below are collected in Fig.~\ref{fig:kerrdiagrams}; the same
expansion drawn with one vertex per order, in which the single-vertex tables
appear as the label-resolved content of a chain, is shown in the Supplemental
Material~\cite{SM}.

\subsubsection{First order: the mean-field Kerr shift}

The field amplitude is $\avg{b}=\Tr[O_{01}e^{\Lop t}r_{00}]$. A single
application of the interaction gives
\begin{align}
  \Lint\dg O_{01} &= i[H_K,b] = -iU\ad a^2
   = -iU(\bd+\alpha^{*})(b+\alpha)^2 \nonumber\\
  &= -iU\big(O_{12} + 2\alpha\,O_{11} + \alpha^{*}O_{02} + \alpha^2 O_{10}
     \nonumber\\ &\qquad\;\; + 2|\alpha|^2 O_{01} + |\alpha|^2\alpha\,O_{00}\big),
  \label{eq:kerrb1table}
\end{align}
and with the incoming propagator $(z\cdot 1 + z^{*}\cdot 0)^{-1}=1/z$ its six
terms are the six first-order diagrams of Fig.~\ref{fig:kerrdiagrams}(a)--(f).
In the steady state only the diagram without outgoing lines survives, panel
(a), so
\begin{equation}
  \avg{b}^{(1)} = -\frac{iU|\alpha|^2\alpha}{z}.
  \label{eq:kerr1}
\end{equation}
This is the first-order solution of the classical driven Kerr equation
$z\bar\alpha + iU|\bar\alpha|^2\bar\alpha = \eta$: the diagram reproduces the
mean-field Kerr shift with no mean-field assumption. The other five diagrams
do not contribute here; they are the building blocks of second order, and at
finite time they contribute directly.

\subsubsection{Second order: the photon number}

The intracavity fluctuation number is
$\avg{\bd b}=\Tr[O_{11}e^{\Lop t}r_{00}]$. After expanding
$i[H_K,\bd b] = iU(\alpha^{*}\ad a^2 - \alpha(\ad)^2a)$ in the displaced
operators,
\begin{align}
  \Lint\dg O_{11} = iU\Big(&\alpha^{*}O_{12} - \alpha O_{21}
    + \alpha^{*2}O_{02} - \alpha^{2}O_{20} \nonumber\\
   &+ |\alpha|^2\alpha^{*}O_{01} - |\alpha|^2\alpha\,O_{10}\Big),
  \label{eq:kerrV1}
\end{align}
which with the incoming propagator $1/(z+z^{*})=1/\kappa$ is the first-order
table of Fig.~\ref{fig:kerrdiagrams}(g)--(l). The $O_{11}$ terms cancel, and so
does the $O_{00}$ term, which is why $\avg{\bd b}$ has no first-order correction,
panel (o).

Second order glues each output of this table to a first-order diagram that
closes it on $r_{00}$. There is a general shortcut for which outputs can be
closed. The $O_{00}$ coefficient of a normal-ordered polynomial in $b,\bd$ is
its expectation value in the $b$ vacuum, and for a Hamiltonian interaction
\begin{equation}
  \bra{0_b}[H,O_{m'n'}]\ket{0_b} = 0
  \qquad \text{whenever } m'\geq1 \text{ and } n'\geq1,
  \label{eq:killrule}
\end{equation}
since $O_{m'n'}\ket{0_b}=0$ and $\bra{0_b}O_{m'n'}=0$ simultaneously: a
Hamiltonian interaction cannot close a label that still has lines of both
orientations. The rule uses a vacuum steady state and does not survive
$N,M\neq0$, where the $O_{00}$ coefficient is a steady-state moment rather than
a vacuum expectation value; with a thermal or squeezed $\Lop_0$ the
enumeration must be done without it. Panels (k) and (l) therefore drop out, and four routes remain,
through $O_{01}$, $O_{10}$, $O_{02}$, and $O_{20}$. The closing vertex factors
are, computing $[H_K,\cdot]$ in the displaced frame and taking the c-number
part,
\begin{align}
  O_{01}&:\ -iU|\alpha|^2\alpha, &
  O_{10}&:\ +iU|\alpha|^2\alpha^{*}, \nonumber\\
  O_{02}&:\ -iU\alpha^{2}, &
  O_{20}&:\ +iU\alpha^{*2}.
\end{align}
The two-line results require care: $[H_K,b^2]=[H_K,a^2]-2\alpha[H_K,a]$,
and the second piece cancels the $2|\alpha|^2\alpha^2$ term that
$[H_K,a^2]$ alone would contribute, leaving only $-U\alpha^2$. The
displacement of the composite operator must not be omitted.

With the propagators $1/z$, $1/z^{*}$, $1/2z$, and $1/2z^{*}$ of the four
intermediate labels, Eq.~\eqref{eq:gluing} gives
\begin{align}
  \avg{\bd b}^{(2)} = \frac{1}{\kappa}\Bigg[&
    \frac{U^2|\alpha|^4\,\alpha^{*}\alpha}{z}
   + \frac{U^2|\alpha|^4\,\alpha\alpha^{*}}{z^{*}} \nonumber\\
  &+ \frac{U^2\alpha^{*2}\alpha^{2}}{2z}
   + \frac{U^2\alpha^{2}\alpha^{*2}}{2z^{*}}\Bigg].
\end{align}
Using $1/z+1/z^{*}=\kappa/|z|^2$, this collapses to the value of
Fig.~\ref{fig:kerrdiagrams}(n),
\begin{equation}
  \boxed{\;
  \avg{\bd b}^{(2)} = \frac{U^2|\alpha|^4}{|z|^2}\left(|\alpha|^2+\frac{1}{2}\right).
  \;}
  \label{eq:kerr2}
\end{equation}
The factor $|\alpha|^2+\tfrac12$ is exactly the linearized (Bogoliubov) result
about the mean field: $|\alpha|^2$ is the shift of the mean field and
$\tfrac12$ the Bogoliubov fluctuation number~\cite{SM}, so second order
reproduces linearized theory and higher orders go beyond it.

\subsubsection{The physical photon number}
\label{sec:kerrphysical}

Equations~\eqref{eq:kerr1} and \eqref{eq:kerr2} are moments of the
fluctuation operator. The measured photon number follows from
Sec.~\ref{sec:recombine}. Here $\Lop_0$ has a vacuum steady state in the
displaced frame, so $N=M=0$, Eq.~\eqref{eq:unwick} is the identity, and only
the displacement of Eq.~\eqref{eq:undisplace} is needed:
\begin{equation}
  \ad a = \bd b + \alpha^{*}b + \alpha\bd + |\alpha|^2 .
  \label{eq:kerrundisplace}
\end{equation}
Taking expectation values term by term, with $\avg{\bd}=\avg{b}^{*}$,
\begin{equation}
  \avg{\ad a} = |\alpha|^2 + 2\,\mathrm{Re}\!\left[\alpha^{*}\avg{b}\right]
     + \avg{\bd b} ,
  \label{eq:kerrphoton}
\end{equation}
so the $O_{01}$ and $O_{10}$ diagrams enter the photon number weighted by the
coherent amplitude, and the $O_{11}$ diagrams enter directly. Through second
order this gives
\begin{equation}
  \avg{\ad a} = |\alpha|^2
    + 2\,\mathrm{Re}\!\left[-\frac{iU|\alpha|^4}{z}\right]
    + O(U^2),
\end{equation}
whose $U^2$ terms combine Eqs.~\eqref{eq:kerrb2} and \eqref{eq:kerr2}. The same
two diagram sets, with the weights of Eq.~\eqref{eq:undisplace} for $m=n=2$,
give the two-photon correlation $\avg{(\ad)^2a^2}$ and hence
$g^{(2)}(0)$; no new diagrams are needed, only new weights.

\subsubsection{Loops and the semiclassical expansion}
\label{sec:loops}

The agreement with linearized theory has a reading in terms of the routes glued into
Fig.~\ref{fig:kerrdiagrams}(n). The routes through $O_{01}$ and $O_{10}$ create
no line along the way: the first vertex absorbs one of the two external lines
and the second absorbs the other. They give the displacement term. The routes
through $O_{02}$ and $O_{20}$ create one new line at the first vertex, which is
absorbed at the second together with the remaining external line. They give
the Bogoliubov fluctuations.

The distinction is general. Along a route of $N$ first-order steps starting
from $E=m+n$ external lines, let $I$ be the number of lines created. Every
line is eventually absorbed and every step absorbs at least one line, so
$E+I\geq N$, and
\begin{equation}
  L \equiv E+I-N \geq 0 .
\end{equation}
If the route were drawn with one vertex per step, $L$ would be the number of
closed loops, and we keep that name. A vertex of degree $d$ in $a$ and $\ad$
has $d$ operator slots; the $E+2I$ line ends fill some and factors of $\alpha$
or $\alpha^{*}$ fill the rest, so a route carries $dN-E-2I=(d-2)N+E-2L$ such
factors. For the Kerr vertex, $d=4$ and
\begin{equation}
  \avg{O}^{(N)}\Big|_{L}\ \propto\ U^N|\alpha|^{2N+E-2L}.
  \label{eq:loopcount}
\end{equation}
At fixed order each loop costs a factor $1/|\alpha|^2$, the inverse of the
coherent photon number: the loop expansion is the semiclassical expansion.

The field amplitude at second order, Fig.~\ref{fig:kerrdiagrams}(m), shows the
same structure. Gluing the outputs (b), (c), and (d) of the first-order table to
the diagrams that close them gives ($O_{11}$ and $O_{12}$ cannot be closed, by
Eq.~\eqref{eq:killrule})
\begin{equation}
  \avg{b}^{(2)} = -\frac{2U^2|\alpha|^4\alpha}{z^2}
     + \frac{U^2|\alpha|^4\alpha}{|z|^2}
     - \frac{U^2|\alpha|^2\alpha}{2z^2}.
  \label{eq:kerrb2}
\end{equation}
The first two terms come from routes with $L=0$ (the frequency shift of panel
(b) and the conversion of panel (c)) and are exactly the second-order terms of
the classical equation $z\bar\alpha+iU|\bar\alpha|^2\bar\alpha=\eta$. The third
comes from the route through panel (d), which creates two lines and has $L=1$;
it is a quantum correction that no mean-field treatment contains. As in field
theory, the loop-free contributions solve the classical equation of motion at
every order.

\subsubsection{Higher orders: an asymptotic series}
\label{sec:kerrhigh}

Beyond second order the gluing of Eq.~\eqref{eq:gluing} is best done by
machine. Rules~3 and 5 define a linear map $\mathcal{G}_0$ on the label
lattice, which multiplies the $O_{mn}$ component by $(zn+z^{*}m)^{-1}$ and sets
the $O_{00}$ component to zero, and the vertex factors are the map $\Lint\dg$.
The order-$N$ steady-state diagram is then
\begin{equation}
  \avg{O}^{(N)} = D^{(N)}_{O\to00} = \Big[\big(\Lint\dg\,\mathcal{G}_0\big)^N O\Big]_{00},
  \label{eq:recursion}
\end{equation}
where $[\,\cdot\,]_{00}$ denotes the $O_{00}$ coefficient. Each vertex changes
the total degree $m+n$ by a bounded amount, so only finitely many labels can
still reach $O_{00}$, and Eq.~\eqref{eq:recursion} is a finite computation at
every order~\cite{SM}.

\begin{figure}[t]
\centering
\includegraphics[width=\columnwidth]{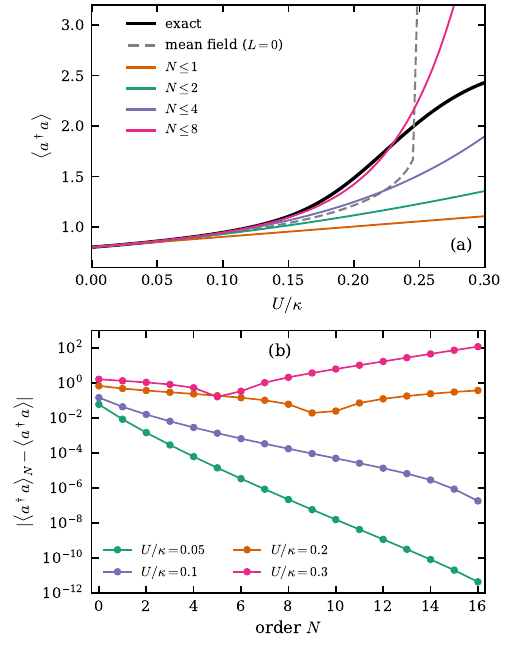}
\caption{Driven Kerr cavity beyond second order, for $\Delta_c=-\kappa$ and
$\eta=\kappa$, so that $|\alpha|^2=0.8$. (a)~Intracavity photon number versus
$U$: exact steady state (black), mean-field theory, which is the sum of all
$L=0$ contributions (grey dashed), and partial sums of the diagrammatic series
through order $N$. (b)~Absolute error of the partial sums versus order $N$.
The series is asymptotic: the error decreases to a minimum at an order that
shrinks as $U$ grows and increases beyond it. At $U=0.05\kappa$ and
$0.1\kappa$ the error is still decreasing at $N=16$.}
\label{fig:kerrconv}
\end{figure}

Figure~\ref{fig:kerrconv} shows the photon number for $\Delta_c=-\kappa$ and
$\eta=\kappa$. Two features are visible. First, the loops matter. On this side
of resonance the Kerr shift pulls the cavity towards the drive, and mean-field
theory, which is the sum of all loop-free contributions, responds steeply as it
approaches the onset of bistability: at $U=0.2\kappa$ it gives
$\avg{\ad a}=1.215$ against the exact $1.485$, and at $U=0.3\kappa$ it gives
$3.76$ against $2.43$. The partial sums, which contain the loops, follow the
exact curve through this region.

Second, the series is asymptotic rather than convergent. Writing
$\avg{\ad a}=\sum_N c_NU^N$, the root-test estimate $|c_N|^{-1/N}$ of the
radius of convergence falls from $0.27\kappa$ at $N=10$ to $0.15\kappa$ at
$N=40$ without levelling off, and for $\Delta_c=\eta=\kappa/2$ the ratio
$|c_{N+1}/c_N|$ of the coefficients of $\avg{a}$ grows linearly through
$N=40$. Correspondingly the partial sums improve up to a most accurate order and
deteriorate beyond it, Fig.~\ref{fig:kerrconv}(b). At $U=0.05\kappa$ the error is below $10^{-11}$ at
$N=16$ and still falling; at $U=0.2\kappa$ it reaches its minimum, $0.019$ or
$1.3\%$, at $N=9$; and at $U=0.3\kappa$ its minimum is $0.16$ at $N=5$, against
an error of $1.3$ for mean-field theory. A Dyson-type argument~\cite{Dyson1952,SM} makes this plausible: continued
to imaginary $U$ the Kerr shift outgrows the damping for arbitrarily large
coherences, so no neighbourhood of $U=0$ is free of singularities. The same
diagrams also give the ring-up after the drive is switched on~\cite{SM}.

Because the series is asymptotic, a partial sum is only as good as the order
at which it is stopped, and no single order is best at every coupling: the
order-$16$ partial sum is accurate to $4\times10^{-12}$ at $U=0.05\kappa$ and
$2\times10^{-7}$ at $U=0.1\kappa$, but at $U=0.2\kappa$ the eighth-order
sum ($0.061$) is closer to the exact value than the sixteenth ($0.37$). In
this paper every diagrammatic result is therefore quoted at a stated, fixed
order, and wherever an exact or independently converged reference exists its
error is given against that reference.

\subsection{A driven two-level atom in a cavity}
\label{sec:emitter}

The emitter moves of Eq.~\eqref{eq:emittermoves} differ from the bosonic
moves of Eqs.~\eqref{eq:move1}--\eqref{eq:move4} in two ways: exclusion
forbids a second line of the same orientation, and $\sigm\sigp=1-\sigp\sigm$
gives the creation term the opposite sign. Everything that distinguishes a
two-level atom from a classical, harmonic dipole with the same rates appears
through those two changes. For an undriven atom seen through a weakly driven
cavity the consequence is the antibunching $g^{(2)}(0)=0.89$ of
cavity-enhanced absorption~\cite{Rice1988}, which the diagrams reproduce to
$10^{-5}$ while the dipole gives exactly one~\cite{SM}. Here we take the case where the
difference is largest and the choice of $\Lop_0$ matters most: an atom driven
directly and strongly.

\begin{figure*}[t]
\centering
\resizebox{\textwidth}{!}{%
\begin{tikzpicture}[x=1cm,y=1cm]
\begin{scope}[shift={(0,0)}]
  \draw[ld sp,ld mark at=0.68] (0,0) .. controls (-0.42,0) and (-0.5,0.26) .. (-0.92,0.26) -- (-2,0.26);
  \draw[ld sm,ld mark at=0.68] (0,0) .. controls (-0.42,0) and (-0.5,-0.26) .. (-0.92,-0.26) -- (-2,-0.26);
  \node[ld vtx] at (0,0) {};
  \node[ld order,fill=white,fill opacity=0.85,text opacity=1,rounded corners=1pt] at (0,0.13) {$(2)$};
  \node[ld ext,anchor=east] at (-2.06,0) {$\sigp\sigm$};
  \node[ld ext,anchor=west] at (0.22,0) {$\ketbra{g}{g}$};
  \node[font=\footnotesize,anchor=north] at (0,-1.05) {(a)\ $\Omega^2/4D$};
\end{scope}
\begin{scope}[shift={(6.3,0)}]
  \draw[ld sp,ld mark at=0.68] (0,0) .. controls (-0.42,0) and (-0.5,0.26) .. (-0.92,0.26) -- (-2,0.26);
  \draw[ld sm,ld mark at=0.68] (0,0) .. controls (-0.42,0) and (-0.5,-0.26) .. (-0.92,-0.26) -- (-2,-0.26);
  \node[ld vtx] at (0,0) {};
  \node[ld order,fill=white,fill opacity=0.85,text opacity=1,rounded corners=1pt] at (0,0.13) {$(4)$};
  \node[ld ext,anchor=east] at (-2.06,0) {$\sigp\sigm$};
  \node[ld ext,anchor=west] at (0.22,0) {$\ketbra{g}{g}$};
  \node[font=\footnotesize,anchor=north] at (0,-1.05) {(b)\ $-\Omega^4/8D^2$ (dipole: $0$)};
\end{scope}
\begin{scope}[shift={(12.6,0)}]
  \draw[ld sp,ld mark at=0.68] (0,0) .. controls (-0.42,0) and (-0.5,0.26) .. (-0.92,0.26) -- (-2,0.26);
  \draw[ld sm,ld mark at=0.68] (0,0) .. controls (-0.42,0) and (-0.5,-0.26) .. (-0.92,-0.26) -- (-2,-0.26);
  \node[ld vtx] at (0,0) {};
  \node[ld order,fill=white,fill opacity=0.85,text opacity=1,rounded corners=1pt] at (0,0.13) {$(2r)$};
  \node[ld ext,anchor=east] at (-2.06,0) {$\sigp\sigm$};
  \node[ld ext,anchor=west] at (0.22,0) {$\ketbra{g}{g}$};
  \node[font=\footnotesize,anchor=north] at (0,-1.05) {(c)\ $\frac{\Omega^2}{4D}\big(-\frac{\Omega^2}{2D}\big)^{r-1}$ (dipole: $0$)};
\end{scope}
\begin{scope}[shift={(0,-2.55)}]
  \draw[ld ad,ld mark at=0.68] (0,0) .. controls (-0.42,0) and (-0.5,0.26) .. (-0.92,0.26) -- (-2,0.26);
  \draw[ld a,ld mark at=0.68] (0,0) .. controls (-0.42,0) and (-0.5,-0.26) .. (-0.92,-0.26) -- (-2,-0.26);
  \node[ld vtx] at (0,0) {};
  \node[ld order,fill=white,fill opacity=0.85,text opacity=1,rounded corners=1pt] at (0,0.13) {$(2)$};
  \node[ld ext,anchor=east] at (-2.06,0) {$\ad a$};
  \node[ld ext,anchor=west] at (0.22,0) {$\rho^{(0)}_{\mathrm{ss}}$};
  \node[font=\footnotesize,anchor=north] at (0,-1.05) {(d)\ Eq.~\eqref{eq:emitter2}};
\end{scope}
\begin{scope}[shift={(6.3,-2.55)}]
  \draw[ld a,ld mark at=0.5] (0,0) -- (-2,0);
  \node[ld vtx] at (0,0) {};
  \node[ld order,fill=white,fill opacity=0.85,text opacity=1,rounded corners=1pt] at (0,0.13) {$(1)$};
  \node[ld ext,anchor=east] at (-2.06,0) {$a$};
  \node[ld ext,anchor=west] at (0.22,0) {$\rho^{(0)}_{\mathrm{ss}}$};
  \node[font=\footnotesize,anchor=north] at (0,-1.05) {(e)\ $-ig\avg{\sigm}_{\mathrm{ss}}/z$};
\end{scope}
\begin{scope}[shift={(12.6,-2.55)}]
  \draw[ld a,ld mark at=0.5] (0,0) -- (-2,0);
  \draw[ld d1,ld mark at=0.5] (2,0) -- (0,0);
  \node[ld vtx] at (0,0) {};
  \node[ld order,fill=white,fill opacity=0.85,text opacity=1,rounded corners=1pt] at (0,0.13) {$(1)$};
  \node[ld ext,anchor=east] at (-2.06,0) {$a$};
  \node[ld ext,anchor=west] at (2.06,0) {$\rho_1$};
  \node[font=\footnotesize,anchor=north] at (0,-1.05) {(f)\ $-igc^{-}_1/z$};
\end{scope}
\begin{scope}[shift={(0,-5.1)}]
  \draw[ld a,ld mark at=0.5] (0,0) -- (-2,0);
  \draw[ld d2,ld mark at=0.5] (2,0) -- (0,0);
  \node[ld vtx] at (0,0) {};
  \node[ld order,fill=white,fill opacity=0.85,text opacity=1,rounded corners=1pt] at (0,0.13) {$(1)$};
  \node[ld ext,anchor=east] at (-2.06,0) {$a$};
  \node[ld ext,anchor=west] at (2.06,0) {$\rho_2$};
  \node[font=\footnotesize,anchor=north] at (0,-1.05) {(g)\ $-igc^{-}_2/z$};
\end{scope}
\begin{scope}[shift={(6.3,-5.1)}]
  \draw[ld a,ld mark at=0.5] (0,0) -- (-2,0);
  \draw[ld d3,ld mark at=0.5] (2,0) -- (0,0);
  \node[ld vtx] at (0,0) {};
  \node[ld order,fill=white,fill opacity=0.85,text opacity=1,rounded corners=1pt] at (0,0.13) {$(1)$};
  \node[ld ext,anchor=east] at (-2.06,0) {$a$};
  \node[ld ext,anchor=west] at (2.06,0) {$\rho_3$};
  \node[font=\footnotesize,anchor=north] at (0,-1.05) {(h)\ $-igc^{-}_3/z$};
\end{scope}
\begin{scope}[shift={(12.6,-5.1)}]
  \draw[ld ad,ld mark at=0.68] (0,0) .. controls (-0.42,0) and (-0.5,0.26) .. (-0.92,0.26) -- (-2,0.26);
  \draw[ld a,ld mark at=0.68] (0,0) .. controls (-0.42,0) and (-0.5,-0.26) .. (-0.92,-0.26) -- (-2,-0.26);
  \draw[ld ad,ld mark at=0.5] (2,0) -- (0,0);
  \node[ld vtx] at (0,0) {};
  \node[ld order,fill=white,fill opacity=0.85,text opacity=1,rounded corners=1pt] at (0,0.13) {$(1)$};
  \node[ld ext,anchor=east] at (-2.06,0) {$\ad a$};
  \node[ld ext,anchor=west] at (2.06,0) {$r_{01}$};
  \node[font=\footnotesize,anchor=north] at (0,-1.05) {(i)\ $-ig\avg{\sigm}_{\mathrm{ss}}/\kappa$};
\end{scope}
\begin{scope}[shift={(0,-7.65)}]
  \draw[ld ad,ld mark at=0.68] (0,0) .. controls (-0.42,0) and (-0.5,0.26) .. (-0.92,0.26) -- (-2,0.26);
  \draw[ld a,ld mark at=0.68] (0,0) .. controls (-0.42,0) and (-0.5,-0.26) .. (-0.92,-0.26) -- (-2,-0.26);
  \draw[ld ad,ld mark at=0.32] (2,0.26) -- (0.92,0.26) .. controls (0.5,0.26) and (0.42,0) .. (0,0);
  \draw[ld d1,ld mark at=0.32] (2,-0.26) -- (0.92,-0.26) .. controls (0.5,-0.26) and (0.42,0) .. (0,0);
  \node[ld vtx] at (0,0) {};
  \node[ld order,fill=white,fill opacity=0.85,text opacity=1,rounded corners=1pt] at (0,0.13) {$(1)$};
  \node[ld ext,anchor=east] at (-2.06,0) {$\ad a$};
  \node[ld ext,anchor=west] at (2.06,0) {$r_{01}\rho_1$};
  \node[font=\footnotesize,anchor=north] at (0,-1.05) {(j)\ $-igc^{-}_1/\kappa$};
\end{scope}
\begin{scope}[shift={(6.3,-7.65)}]
  \draw[ld ad,ld mark at=0.68] (0,0) .. controls (-0.42,0) and (-0.5,0.26) .. (-0.92,0.26) -- (-2,0.26);
  \draw[ld a,ld mark at=0.68] (0,0) .. controls (-0.42,0) and (-0.5,-0.26) .. (-0.92,-0.26) -- (-2,-0.26);
  \draw[ld ad,ld mark at=0.32] (2,0.26) -- (0.92,0.26) .. controls (0.5,0.26) and (0.42,0) .. (0,0);
  \draw[ld d2,ld mark at=0.32] (2,-0.26) -- (0.92,-0.26) .. controls (0.5,-0.26) and (0.42,0) .. (0,0);
  \node[ld vtx] at (0,0) {};
  \node[ld order,fill=white,fill opacity=0.85,text opacity=1,rounded corners=1pt] at (0,0.13) {$(1)$};
  \node[ld ext,anchor=east] at (-2.06,0) {$\ad a$};
  \node[ld ext,anchor=west] at (2.06,0) {$r_{01}\rho_2$};
  \node[font=\footnotesize,anchor=north] at (0,-1.05) {(k)\ $-igc^{-}_2/\kappa$};
\end{scope}
\begin{scope}[shift={(12.6,-7.65)}]
  \draw[ld ad,ld mark at=0.68] (0,0) .. controls (-0.42,0) and (-0.5,0.26) .. (-0.92,0.26) -- (-2,0.26);
  \draw[ld a,ld mark at=0.68] (0,0) .. controls (-0.42,0) and (-0.5,-0.26) .. (-0.92,-0.26) -- (-2,-0.26);
  \draw[ld ad,ld mark at=0.32] (2,0.26) -- (0.92,0.26) .. controls (0.5,0.26) and (0.42,0) .. (0,0);
  \draw[ld d3,ld mark at=0.32] (2,-0.26) -- (0.92,-0.26) .. controls (0.5,-0.26) and (0.42,0) .. (0,0);
  \node[ld vtx] at (0,0) {};
  \node[ld order,fill=white,fill opacity=0.85,text opacity=1,rounded corners=1pt] at (0,0.13) {$(1)$};
  \node[ld ext,anchor=east] at (-2.06,0) {$\ad a$};
  \node[ld ext,anchor=west] at (2.06,0) {$r_{01}\rho_3$};
  \node[font=\footnotesize,anchor=north] at (0,-1.05) {(l)\ $-igc^{-}_3/\kappa$};
\end{scope}
\end{tikzpicture}}
\caption{Diagrams for a driven emitter in a cavity, Eq.~\eqref{eq:emitterH}.
(a)--(c)~Bare split: steady-state diagrams for the emitter population at orders
$2$, $4$, and $2r$, which form the geometric series of Eq.~\eqref{eq:pebare}; for a classical dipole
all diagrams beyond second order vanish.
(d)~Dressed split: the second-order steady-state diagram for the photon number,
Eq.~\eqref{eq:emitter2}. (e)--(h)~First-order table for the cavity field $a$:
the emitter operator left at the vertex is either replaced by its steady-state
mean (e) or starts one of the three dressed lines (f)--(h), which close on the
duals $\rho_k$. (i)--(l)~First-order table for $\ad a$, routes through $\sigm$;
the routes through $\sigp$ are their Hermitian conjugates. Gluing (i) to the
diagram that closes $\ad$ gives coherent scattering ($L=0$), and gluing
(j)--(l) gives the three components of the Mollow triplet ($L=1$) in
Fig.~\ref{fig:emitterplot}(a).}
\label{fig:emitterdiagrams}
\end{figure*}
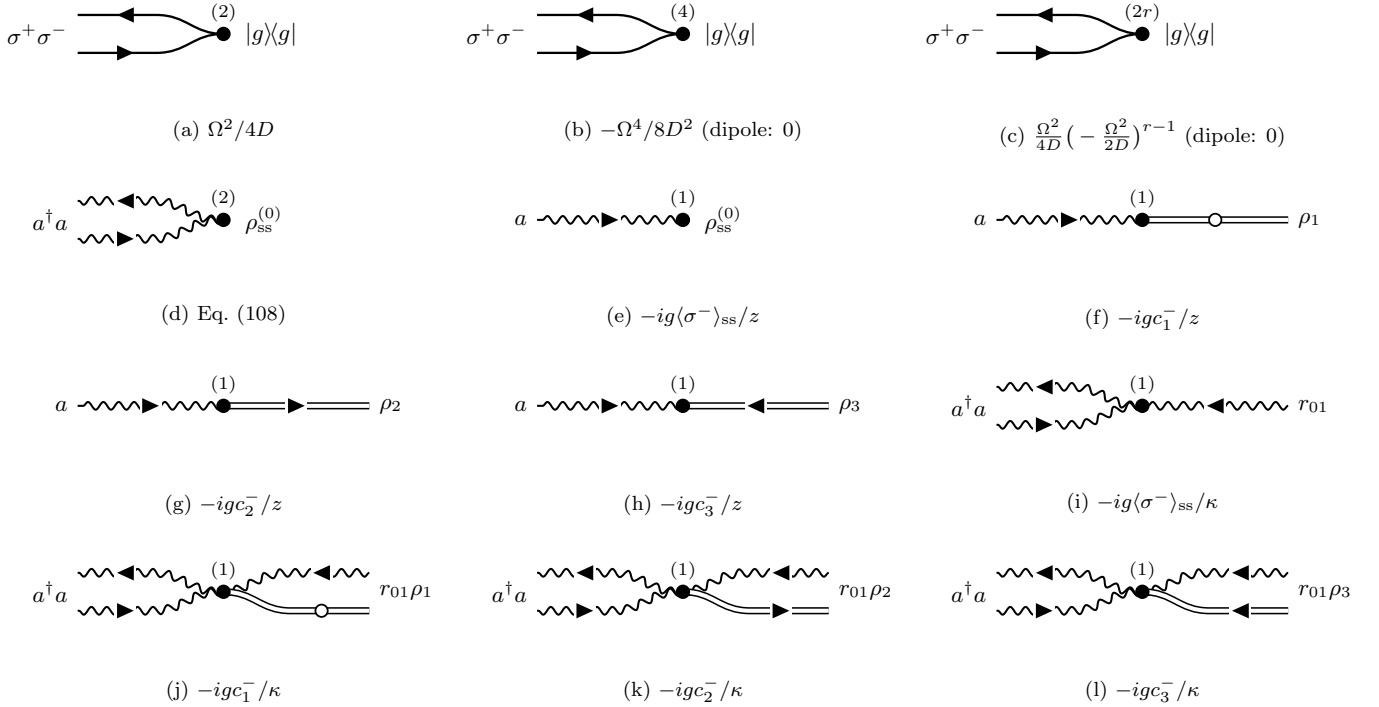

A two-level atom is driven by a laser of Rabi frequency $\Omega$ and coupled
to an undriven cavity. In the frame of the laser,
\begin{align}
  H &= \Delta_c\ad a + \frac{\Delta_a}{2}\sigma_z + \frac{\Omega}{2}\sigma_x
       + g\left(\ad\sigm + a\sigp\right), \nonumber\\
  \Lop\rho &= -i[H,\rho] + \kappa\Dop[a]\rho + \gamma\Dop[\sigm]\rho .
  \label{eq:emitterH}
\end{align}
The cavity has no coherent drive, so its lines are bare wavy lines, and every
jump operator acts on a single sector, so any $\Lop_0$ built from the cavity
and emitter parts separately is additive across them
(Sec.~\ref{sec:drivenatom}). Two splits suggest themselves. In the
\emph{bare} split $\Lop_0$ contains the undriven emitter of
Sec.~\ref{sec:tls} and both the drive $\Omega$ and the exchange $g$ are part of
$\Lint$. In the \emph{dressed} split $\Lop_0$ contains the driven emitter of
Sec.~\ref{sec:drivenatom} and only the exchange is perturbative,
$\Lint\dg O = ig[\ad\sigm + a\sigp, O]$. Diagrams for both are collected in
Fig.~\ref{fig:emitterdiagrams}.

\begin{figure}[t]
\centering
\includegraphics[width=\columnwidth]{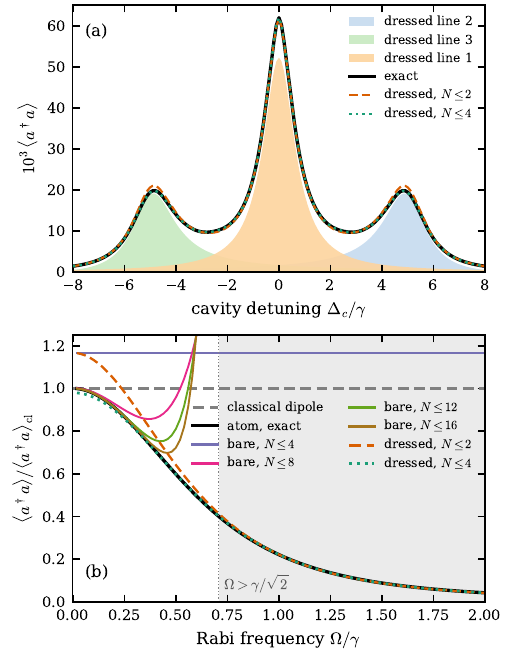}
\caption{Emitter in a cavity, $\gamma=1$, $\kappa=\gamma/2$, $\Delta_a=0$.
(a)~Cavity photon number versus cavity detuning for $\Omega=5\gamma$ and
$g=\gamma/5$: exact steady state, dressed expansion through second and fourth
order, and the contributions of the three dressed lines in
Eq.~\eqref{eq:emitter2} (shaded). (b)~Photon number at $\Delta_c=0$ versus
$\Omega$ for $g=\gamma/10$, divided by that of a classical dipole with the same
rates (grey dashed line at one): exact atom, bare partial sums through order $N$,
which diverge as $\Omega$ approaches $\gamma/\sqrt2$ (shaded region beyond), and
the dressed expansion, which remains accurate throughout.}
\label{fig:emitterplot}
\end{figure}

\subsubsection{Bare lines and the saturation radius}

Start with the emitter alone, $g=0$, and the population $\avg{\sigp\sigm}$. The
drive vertex $i(\Omega/2)[\sigma_x,\cdot\,]$ acts through the emitter moves of
Eq.~\eqref{eq:emittermoves} with $p_e=0$: on $O_{11}=\sigp\sigm$ it absorbs
either line with factor $\pm i\Omega/2$, and on a single line it either absorbs
it with factor $\mp i\Omega/2$ or creates the missing line of the opposite
orientation with factor $\pm i\Omega$. The second-order diagram,
Fig.~\ref{fig:emitterdiagrams}(a), has two routes, through $\sigm$ and through
$\sigp$, which give $(\Omega^2/4\gamma)(1/z_a+1/z_a^{*})=\Omega^2/4D$, with
$D=\Delta_a^2+\gamma^2/4$ as in Eq.~\eqref{eq:atomss}. At fourth order,
Fig.~\ref{fig:emitterdiagrams}(b), the third step recreates the absorbed line
and the pattern repeats. Each repetition, summed over its two routes,
contributes $(i\Omega)(i\Omega/2)\gamma^{-1}(1/z_a+1/z_a^{*})=-\Omega^2/2D$, so
the diagram of order $2r$, Fig.~\ref{fig:emitterdiagrams}(c), is a geometric
term and
\begin{equation}
  \avg{\sigp\sigm} = \frac{\Omega^2}{4D}\sum_{r=0}^{\infty}
     \left(-\frac{\Omega^2}{2D}\right)^{r}
   = \frac{\Omega^2/4}{D+\Omega^2/2},
  \label{eq:pebare}
\end{equation}
which reproduces Eq.~\eqref{eq:atomss} for $\Omega^2<2D$ and diverges beyond.
The singularity is that of the exact result at $\Omega^2=-2D$, and it is
inherited by every bare expansion of this model: with the cavity included, the
bare partial sums for the photon number diverge as $\Omega$ approaches
$\sqrt{2D}=\gamma/\sqrt2$ on resonance, Fig.~\ref{fig:emitterplot}(b).
Saturation is a nonperturbative effect of the drive, and no finite order of
the bare expansion captures it.

The classical dipole shows where the ladder comes from. For $c$ the drive vertex
$i(\Omega/2)[c+c\dg,\,\cdot\,]$ acting on a single line only absorbs it: the
creation terms produced by multiplying from the left and from the right cancel.
For the atom they add, because the creation term that accompanies an
absorption has the opposite sign, and that is the factor $i\Omega$ in each rung. For the dipole every diagram beyond second order therefore
vanishes, $\avg{c\dg c}=\Omega^2/4D$ exactly, and the scattered intensity
grows without bound. The geometric ladder, the saturation of
Eq.~\eqref{eq:pebare}, and the radius of convergence $\Omega^2<2D$ are the
anticommutator of the atom made diagrammatic. Drawn with one vertex per step,
the bare expansion is a ladder alternating between the population and
coherence labels; the picture is given in the Supplemental
Material~\cite{SM}.

\subsubsection{Dressed lines and the Mollow triplet}

The dressed split removes the problem by construction: the drive is inside
the lines, and the expansion parameter is the exchange $g$ alone. At first
order, Fig.~\ref{fig:emitterdiagrams}(e)--(h), the vertex acting on $a$ leaves
$-ig\sigm$, which Eq.~\eqref{eq:dressedexpand} splits into its steady-state
mean and the three dressed lines $X_1$, $X_+$, $X_-$ with the coefficients
$c^{-}_k$ computed there. In the steady state only the mean survives, and the
cavity field is driven by the emitter's mean dipole,
\begin{equation}
  \avg{a}^{(1)} = -\frac{ig\avg{\sigm}_{\mathrm{ss}}}{z},
  \label{eq:emitter1}
\end{equation}
with $z=\kappa/2+i\Delta_c$ and $\avg{\sigm}_{\mathrm{ss}}$ from
Eq.~\eqref{eq:atomss} at arbitrary $\Omega$. At second order the first step on
$\ad a$ absorbs the $a$ line and leaves $\sigm$ in the same way,
Fig.~\ref{fig:emitterdiagrams}(i)--(l). Gluing output (i) to the step that
absorbs the $\ad$ line with the mean $\avg{\sigp}_{\mathrm{ss}}$ creates no line
and gives coherent scattering. Gluing output (j), (k), or (l) to the step that
absorbs the $\ad$ line and the dressed line $k$ together, with the weight
$\avg{\sigp X_k}_{\mathrm{ss}}$ of Eq.~\eqref{eq:dressedmoves} and the propagator
$1/(z^{*}-\lambda_k)$, has $L=1$. Adding the Hermitian-conjugate routes,
\begin{align}
  \avg{\ad a}^{(2)} = {}&\frac{g^2|\avg{\sigm}_{\mathrm{ss}}|^2}{|z|^2}
   \nonumber\\
   &+ \frac{2g^2}{\kappa}\sum_{k=1}^{3}\mathrm{Re}\,
     \frac{c^{-}_k\avg{\sigp X_k}_{\mathrm{ss}}}{z^{*}-\lambda_k},
  \label{eq:emitter2}
\end{align}
the value of Fig.~\ref{fig:emitterdiagrams}(d). The first term is
$|\avg{a}^{(1)}|^2$, the coherently scattered light. In the second, the quantum
regression theorem identifies
$\sum_kc^{-}_ke^{\lambda_k\tau}\avg{\sigp X_k}_{\mathrm{ss}}$ with
$\avg{\delta\sigp(0)\delta\sigm(\tau)}_{\mathrm{ss}}$, where
$\delta\sigma^\pm=\sigma^\pm-\avg{\sigma^\pm}_{\mathrm{ss}}$, so the $L=1$ routes
sum to
\begin{equation}
  \avg{\ad a}^{(2)}_{\mathrm{inc}} = \frac{2g^2}{\kappa}\,\mathrm{Re}\!\int_0^\infty
    \!d\tau\, e^{-(\kappa/2-i\Delta_c)\tau}
    \avg{\delta\sigp(0)\delta\sigm(\tau)}_{\mathrm{ss}},
\end{equation}
the incoherent resonance-fluorescence spectrum seen through a Lorentzian
filter of width $\kappa$ centred on the cavity~\cite{Eberly1977}. Each dressed
line contributes one Lorentzian, centred at $\Delta_c=-\mathrm{Im}\,\lambda_k$
with half-width $\kappa/2-\mathrm{Re}\,\lambda_k$, and the explicit
eigenoperators of Sec.~\ref{sec:drivenatom} fix all three. On resonance,
$X_1=\delta\sigma_x$ with $\lambda_1=-\gamma/2$ [Eq.~\eqref{eq:X1}] gives a peak
at $\Delta_c=0$ of half-width $\kappa/2+\gamma/2$, and
$X_\pm$ with $\lambda_\pm=-3\gamma/4\pm i(\Omega^2-\gamma^2/16)^{1/2}$
[Eqs.~\eqref{eq:lampm} and \eqref{eq:Xpm}] give the sidebands at
$\Delta_c=\mp(\Omega^2-\gamma^2/16)^{1/2}$ of half-width $\kappa/2+3\gamma/4$.
That is the Mollow triplet~\cite{mollow1969power}, one dressed line per component,
with the weight of each peak fixed by $c^{-}_k$ of
Eq.~\eqref{eq:dressedexpand} times $\avg{\sigp X_k}_{\mathrm{ss}}$: for
$\Omega=2\gamma$, $c^{-}_1=\tfrac12$ and $c^{-}_\pm=0.250\mp0.032i$, so the
sidebands carry equal weight and the central line half of the total. A classical dipole has no
counterpart. Its unperturbed steady state is coherent, all $L=1$ routes vanish,
and the cavity sees a single coherent Lorentzian at $\Delta_c=0$; for the
parameters below it would hold $9.2$ photons there, $150$ times the atom's, and
none at the sidebands.

\subsubsection{Comparison with the exact steady state}

Figure~\ref{fig:emitterplot}(a) shows the photon number for $\Omega=5\gamma$,
$\kappa=\gamma/2$, and $g=\gamma/5$. The second-order result
Eq.~\eqref{eq:emitter2} is within about $6\%$ of the exact steady state at
every detuning, and fourth order within $0.4\%$; the shaded areas are the three
dressed-line contributions, which account for the triplet line by line.
Figure~\ref{fig:emitterplot}(b) fixes $\Delta_c=0$ and $g=\gamma/10$, scans
the drive, and divides by the classical dipole. At weak drive the atom scatters
like the dipole; by $\Omega=2\gamma$ it scatters only $4\%$ as much. The bare
partial sums through $N=16$ follow the exact result at weak drive and diverge
before $\Omega=\gamma/\sqrt2$, as Eq.~\eqref{eq:pebare} requires, while the
dressed expansion at fourth order is within about $2\%$ of the exact result
over the whole range, including deep saturation. The two
expansions differ only in which part of $\Lop$ was resummed into the lines
(Sec.~\ref{sec:propagator}): the split is a choice about what to put in the
propagator, and here that choice converts a divergent series into a rapidly
convergent one.

\section{A Kerr ring beyond direct diagonalization}
\label{sec:ring}

The examples so far were chosen because their answers are independently
known, and their Hilbert spaces are small enough to diagonalize. A
diagrammatic method is intended for the opposite regime. Here the unperturbed
generator is a chain of coupled cavities: its Liouvillian is Gaussian and
diagonalized once for any number of sites, while the exact treatment costs the
square of a Hilbert-space dimension that grows exponentially with that number.

Take $K$ single-mode cavities on a ring, uniformly driven and damped, with
nearest-neighbour hopping $J$ and an on-site Kerr term, the driven-dissipative
Bose-Hubbard model of coupled nonlinear
resonators~\cite{Carusotto2013,leboite2013steady,jin2016cluster,rota2017critical,fink2017observation,vaneph2018observation}:
\begin{align}
  H &= \sum_{j=0}^{K-1}\Big[\Delta\, a_j\dg a_j
      + \frac{U}{2}(a_j\dg)^2a_j^2
      + i\big(\eta a_j\dg - \eta^{*}a_j\big)\Big] \nonumber\\
    &\quad - J\sum_{j}\left(a_j\dg a_{j+1} + \mathrm{h.c.}\right),
  \qquad \kappa\,\Dop[a_j] .
  \label{eq:ringH}
\end{align}
Everything except the Kerr term is quadratic in the mode operators with linear
jump operators, so it belongs to $\Lop_0$ by Sec.~\ref{sec:general}. Two
reductions make the unperturbed problem trivial. The uniform drive is removed
by a uniform displacement, $b_j = a_j - \alpha$ with
$\alpha = \eta/(\kappa/2+i(\Delta-2J))$, as in Sec.~\ref{sec:drive}; and
translation invariance diagonalizes the drift matrix by Fourier transform, so
the damped normal modes of Eq.~\eqref{eq:betamodes} are the momentum modes
\begin{equation}
  \beta_q = \frac{1}{\sqrt K}\sum_j e^{-iqj}\,b_j,
  \qquad
  z_q = \frac{\kappa}{2} + i\left(\Delta - 2J\cos q\right),
  \label{eq:ringmodes}
\end{equation}
with $q = 2\pi\ell/K$. The steady state of $\Lop_0$ is the displaced vacuum, so
$N=M=0$ and the eigenoperators are plain normal-ordered monomials in the
$\beta_q$, one line per momentum, with the rates of
Eq.~\eqref{eq:ringmodes}.

This is the following case of the general construction; the hopping, which
is the part that makes the Hilbert space exponential, is treated exactly, and
only the on-site nonlinearity is expanded. The jump operators
conserve particle number and the loss is uniform, so the drift
$\Gamma=\kappa/2+i\Omega$ is normal: the left and right modes coincide and are
orthonormal, $G_{kk'}=\delta_{kk'}$ in Eq.~\eqref{eq:mmove1}, and absorption at
a vertex conserves the mode label. With translation invariance that label is
momentum, and the moves of Sec.~\ref{sec:vertices} become momentum
conservation. Two generalizations are already covered by the theory and simply
not needed here. Site-dependent loss rates or non-reciprocal hopping make
$\Gamma$ non-normal; the left modes are then not orthonormal, absorption no
longer conserves the label, and the Kronecker delta is replaced by the Gram
matrix $G_{kk'}=\bvec v_{k,L}\dg\bvec v_{k',L}$, which is how correlated decay
enters the vertices. Parametric driving or a squeezed reservoir adds anomalous
terms; the eigenoperators are then the Wick monomials of
Eq.~\eqref{eq:generaleigen} in the $2K$ modes and the weights are the full
covariance $V_{\mu\nu}$, Eq.~\eqref{eq:mmove3}. In either case the propagator
stays diagonal and the rules of Sec.~\ref{sec:ssrules} are unchanged; only the
vertex weights differ.

The Kerr term is the perturbation. Written in the momentum modes it is a
momentum-conserving vertex,
\begin{widetext}
\begin{align}
  \frac{2}{U}H_{\mathrm{int}} = {}& K|\alpha|^4
   + 2\sqrt K\left(\alpha^{*}\alpha^2\,\beta_0\dg + \mathrm{h.c.}\right)
   + \sum_q\left[4|\alpha|^2\beta_q\dg\beta_q
     + \left(\alpha^2\,\beta_q\dg\beta_{-q}\dg + \mathrm{h.c.}\right)\right]
   \nonumber\\
   &+ \frac{2}{\sqrt K}\sum_{q_1q_2}\left(\alpha\,
       \beta_{q_1}\dg\beta_{q_2}\dg\beta_{q_1+q_2} + \mathrm{h.c.}\right)
   + \frac{1}{K}\sum_{q_1q_2q_3}
       \beta_{q_1}\dg\beta_{q_2}\dg\beta_{q_3}\beta_{q_1+q_2-q_3},
  \label{eq:ringvertex}
\end{align}
\end{widetext}
where each term is one entry of the vertex table of Sec.~\ref{sec:vertices},
and the factors of $\alpha$ are the c-numbers left by the displacement. The
diagrams are those of Sec.~\ref{sec:kerr} with a momentum label on every line
and momentum conservation at the vertex. Physical moments are assembled by
Sec.~\ref{sec:recombine}: a site operator is
$a_j = \alpha + K^{-1/2}\sum_q e^{iqj}\beta_q$, and on-site and intersite
correlators follow from the same diagram sets with different weights.

\begin{figure}[t]
\centering
\includegraphics[width=\columnwidth]{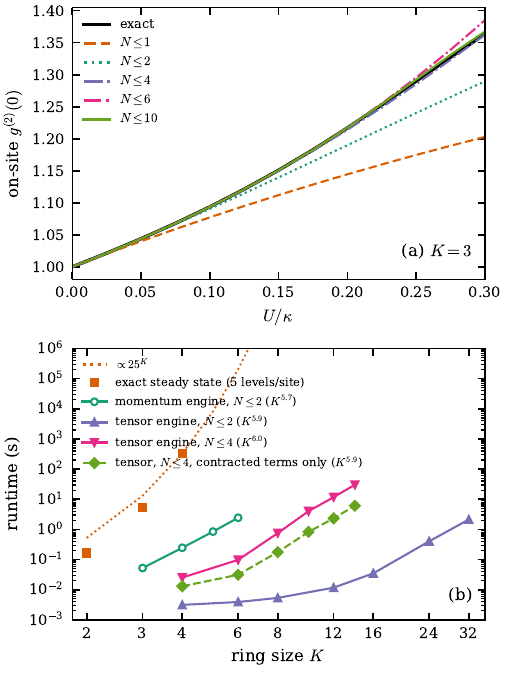}
\caption{Driven-dissipative Kerr ring, Eq.~\eqref{eq:ringH}, in units of
$\kappa$, with $\Delta=-\kappa$, $J=0.4\kappa$, and $\eta=\kappa$.
(a)~On-site $g^{(2)}(0)$ at $K=3$ versus the Kerr strength: exact steady state
(displaced-frame Fock basis, converged in the cutoff) and partial sums
through order $N$. (b)~Runtime on one core versus ring size for the
exact steady state by sparse time evolution with five levels per site
(squares; the dotted line $\propto25^{K}$, anchored at the $K=4$ point, is
the growth $n_{\max}^{2}=25$ per site of the Liouvillian dimension at five
levels, whereas the measured runtime grows by $45$ per site), for the
momentum-space engine at second order, and for the tensor engine at second
and fourth order, with the fitted exponents in the legend. The fourth-order
tensor engine is also shown with the commutator restricted to contracted
terms. The exact calculation is affordable at $K\leq3$ and out of reach
beyond $K=5$; at fixed perturbative order the diagrammatic implementations
scale polynomially with $K$ throughout the range shown.}
\label{fig:ring}
\end{figure}

\emph{Cost.} A label is a normal-ordered monomial in the $2K$ operators
$\beta_q$ and $\beta_q\dg$. A Kerr vertex followed by the propagator changes
the degree of a label by at most two, and only labels of degree zero
contribute to the steady state, so an observable of degree $d_0$ carried to
order $N$ never needs labels of degree above
$D_{\max}=2\lfloor N/2+d_0/4\rfloor$. The number of labels is bounded by the
number of monomials of degree at most $D_{\max}$ in $2K$ operators,
$\binom{2K+D_{\max}}{D_{\max}}$. For the on-site $g^{(2)}$ ($d_0=4$) it grows
as $K^4$ at $N=2$ and $3$ and as $K^6$ at $N=4$ and $5$, with one power of $K$
less on the translation-invariant ring, where only labels of zero net momentum
can reach the steady state~\cite{SM}. The exact steady state, by contrast,
requires the null vector of a Liouvillian of dimension $n_{\max}^{2K}$ with
$n_{\max}$ the retained photon number per site. Figure~\ref{fig:ring}(b) shows
runtimes measured on one core. The exact steady state grows by a factor of
about $45$ per added site with five levels per site ($328$~s at $K=4$). The
momentum-space implementation at second order grows as $K^{5.7}$ ($2.4$~s at
$K=6$). The tensor implementation described below takes $2.1$~s at $K=32$ at
second order, growing as $K^{5.9}$ above $K=16$, where construction of the
vertex tensor dominates (the series evaluation alone grows as $K^{4.0}$), and
$30$~s at $K=14$ at fourth order, growing as $K^{6.0}$. Restricting the
commutator to terms with at least one contraction, which removes labels that
carry only roundoff, reduces the fourth-order time to $6.1$~s. Timings on the
shared node varied by up to a factor of five with its load; the values shown
come from a single pass at fixed load. The recursion Eq.~\eqref{eq:recursion}
is linear, so the map ``vertex after propagator'' can also be stored once as
a sparse matrix over the reachable zero-momentum labels and the series
becomes one sparse matrix-vector product per order; since the vertex is
proportional to $U$, the $N$th-order term is $c_NU^N$ and one run serves
every coupling. At $K=3$ this gives orders $0$ to $10$ of $n$,
$\avg{(a_0\dg)^2a_0^2}$, and $\avg{(a_0\dg)^3a_0^3}$ at all couplings in
$4$~s on one core~\cite{SM}; the construction of the matrix grows faster
with $K$ than the tensor implementation, which is kept for large arrays.

\emph{Accuracy.} At $K=3$ the exact steady state is still available, and
Fig.~\ref{fig:ring}(a) compares it with the expansion. For the per-site photon
number the second-order error is $0.01\%$ at $U=0.05\kappa$, $0.3\%$ at
$U=0.15\kappa$ and $3\%$ at $U=0.3\kappa$, between two and twelve times smaller
than the first-order error. The same diagram sets give the correlators that
are the reason for studying such arrays. At $U=0.15\kappa$ the exact values
are $\avg{a_0\dg a_0}=0.3029$ and on-site $g^{(2)}(0)=1.151$; second order
gives $0.3019$ and $1.140$, fourth order gives $1.150$, and tenth order gives
$1.15089$ against the exact $1.15087$, an error of $1.7\times10^{-5}$. The
ring is bunched on site and antibunched between neighbours, and the expansion
reproduces both. The error of the tenth-order $g^{(2)}$ grows from
$2\times10^{-7}$ at $U=0.1\kappa$ to $4\times10^{-3}$ at $U=0.3\kappa$. From
$U\approx0.2\kappa$ upward a lower order is sometimes more accurate: at
$U=0.3\kappa$ the fourth-order sum (error $1\times10^{-3}$) is closer to the
exact value than the tenth.

\emph{Disorder.} Translation invariance is a convenience, not a requirement. For
site-dependent detunings the drift $\Gamma=\kappa/2+i\Omega$ with $\Omega$ real
symmetric is still normal, so the momentum modes are replaced by the
numerically diagonalized normal modes, the displacement becomes site dependent,
$\bvec\alpha=\Gamma^{-1}\bvec\eta$, and the momentum delta of
Eq.~\eqref{eq:ringvertex} becomes a dense tensor
$\sum_j v^{*}_{q_1j}v^{*}_{q_2j}v_{q_3j}v_{q_4j}$. Evaluating the diagrams with
operators stored as symmetric tensors by degree sector, and discarding at each
vertex every sector the remaining vertices cannot bring back to the steady
state, costs the same with and without disorder: the timings of
Fig.~\ref{fig:ring}(b), $2.1$~s for second-order site-resolved $g^{(2)}$ at
$K=32$ and $30$~s at fourth order for $K=14$, apply to a disordered ring as
well. The disordered chain of Sec.~\ref{sec:fcs} uses this implementation.

\subsection{Comparison with cumulant closures}
\label{sec:cumulant}

Exact diagonalization is not the competitor at large $K$. The standard tools
there are cumulant
closures~\cite{kira2008cluster,casteels2016truncated,weimer2021simulation,plankensteiner2022quantumcumulants},
which keep all moments up to a chosen order $N_c$ and close the hierarchy by
setting every joint cumulant of higher order to zero. At $N_c=2$ this is
time-dependent Hartree--Fock--Bogoliubov theory, with an effective on-site
quadratic Hamiltonian
$2U(|\alpha|^2+n)\,\delta a\dg\delta a + \tfrac{U}{2}[(\alpha^2+m)\delta a\dg{}^2+\mathrm{H.c.}]$
determined self-consistently by the moments $\alpha$, $n$, and $m$; it is a
Gaussian theory and evaluates every higher moment by Wick's theorem. At
$N_c=3$ and $4$ the closures retain part of the non-Gaussian
content~\cite{plankensteiner2022quantumcumulants}. For the three-site ring
they contain $27$, $83$, and $209$ complex moment variables and cost $0.13$,
$0.6$, and $5$~s per evaluation on one core, against $0.21$~s for the
diagrams through second order and $1.1$~s through fourth order with the
pruning of Sec.~\ref{sec:ring}, or $0.03$ and $0.11$~s for all couplings at
once in the sparse-matrix form~\cite{SM}. At fixed truncation order all are
polynomial in $K$, so the
comparison is one of accuracy at a given cost.
Truncated-cumulant trajectories~\cite{verstraelen2023quantum} and
positive-$P$ sampling~\cite{deuar2021fully} are further polynomial-cost
alternatives, with a sampling error in place of a truncation error; they are
not compared here.

\begin{figure}[t]
\centering
\includegraphics[width=\columnwidth]{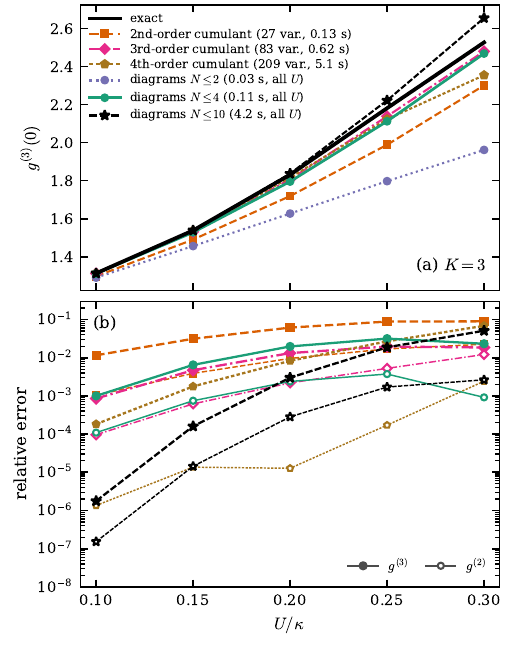}
\caption{Photon statistics of the $K=3$ Kerr ring of Sec.~\ref{sec:ring}
against the converged exact steady state. (a)~Three-photon correlation
$g^{(3)}(0)$ from the cumulant closures at second, third, and fourth order,
and from the diagrams through second, fourth, and tenth order. The legend gives the number of moment variables of
each closure and the time on one core: per evaluation for the closures,
and for all couplings at once for the diagrams in the sparse-matrix form
of the Supplemental Material~\cite{SM}. (b)~Relative errors on
$g^{(3)}$ (filled symbols) and $g^{(2)}$ (open symbols). The second-order
closure is a Gaussian theory and its error grows with cumulant order; the
third- and fourth-order closures and the fourth-order diagrams are comparable
at moderate coupling, and the tenth-order diagrams are the most
accurate on $g^{(3)}$ for $U\leq0.25\kappa$.}
\label{fig:compare}
\end{figure}

The comparison is made on the three-photon correlation
$g^{(3)}(0)=\avg{(a_0\dg)^3a_0^3}/\avg{a_0\dg a_0}^3$, where the Gaussian
closure cannot be right even in principle: for it the non-Gaussian part of any
correlator is zero by construction, and the diagrams compute exactly that
part. Figure~\ref{fig:compare} shows the closures and the diagrams against the
exact steady state, computed in the displaced frame and converged in the
cutoff~\cite{SM}. Against the second-order closure the diagrams through fourth
order reduce the error on $g^{(3)}$ by a factor of $12$ at $U=0.1\kappa$
(exact value $1.3126$; errors of $1.2\%$ and $0.10\%$) and by factors of
$2.8$ to $4.8$ at $U=0.15$--$0.3\kappa$. Against the third- and fourth-order
closures the comparison reverses. The better of the two has the smaller
$g^{(3)}$ error at every coupling of Fig.~\ref{fig:compare}, by a factor of
$5.5$ at $U=0.1\kappa$ falling to $1.25$ at $U=0.3\kappa$, at a cost
between $0.6$ and $5$ times that of the diagrams, and on $g^{(2)}$ the
fourth-order closure is more accurate still. The diagrams regain the lead when carried to
tenth order: their relative $g^{(3)}$ error is then $1.8\times10^{-6}$ at
$U=0.1\kappa$, $1.6\times10^{-4}$ at $0.15\kappa$, $3.0\times10^{-3}$ at
$0.2\kappa$, and $1.9\times10^{-2}$ at $0.25\kappa$, smaller than that of the
best closure by factors of $103$, $11$, $2.8$, and $1.05$. They are worse by
factors of $1.75$ and $2.8$ at $U=0.275\kappa$ and $0.3\kappa$, where the
asymptotic series has passed its most accurate order
(Sec.~\ref{sec:kerrhigh}). This lead is specific to $g^{(3)}$:
on $g^{(2)}$ and on the photon number the tenth-order diagrams are
ahead of the fourth-order closure only up to $U=0.125\kappa$ and
$0.1\kappa$ (Supplemental Material~\cite{SM}, Table~SI). Its cost is
$4$~s on one core for orders $0$ to $10$ of the three moments at all
thirteen couplings in the sparse-matrix form ($54$~s per coupling with the
symbolic recursion), less than one fourth-order closure evaluation. The same diagram sets that gave
$g^{(2)}$ give $g^{(3)}$; only the weights of Sec.~\ref{sec:recombine}
change.

Two conclusions follow for this small ring. The diagrams improve
systematically on the Gaussian closure and recover the non-Gaussian part it
lacks, and carried to tenth order they give the smallest $g^{(3)}$ error of
the methods compared for $U\leq0.25\kappa$. On $g^{(2)}$ the fourth-order
closure is more accurate from $U\approx0.15\kappa$ and on the photon number
from $U\approx0.125\kappa$, and at moderate coupling the third- and
fourth-order closures are more accurate than the fourth-order diagrams on
every correlator, at between $0.6$ and $5$ times their cost. Closures
beyond fourth order, truncated-cumulant trajectories, and positive-$P$
sampling were not run; the statement is about the methods compared here. Photon-counting experiments on these arrays measure exactly these
correlators. A finite-order perturbative correlator is not guaranteed to respect
the positivity bounds that an exact correlator obeys, such as $g^{(2)}\geq0$;
no correlator plotted in Figs.~\ref{fig:ring} and \ref{fig:compare} violates
them, all values lying above one.

\subsection{Counting statistics of the emitted light}
\label{sec:fcs}

The observables considered so far are steady-state moments. The statistics of
the photons that leave the array is described instead by the
counting-field-tilted
generator~\cite{esposito2009nonequilibrium,bagrets2003full,flindt2008counting,landi2024current},
\begin{equation}
  \Lop_\chi\rho = \Lop\rho + \left(e^{\chi}-1\right)\kappa\,a_c\rho a_c\dg ,
  \label{eq:tilted}
\end{equation}
whose leading eigenvalue $\theta(\chi)$ generates the cumulants of the number
of photons emitted from site $c$: $c_1=\theta'(0)$ is the emission rate,
$c_2/c_1$ the Fano factor, $c_3/c_1$ the skewness ratio. This is a harder
object than a steady state: one needs an eigenvalue of a non-trace-preserving
generator of the same exponential dimension. Three reference points bracket
the problem. Mean-field theory emits a coherent field and gives exactly
Poissonian statistics,
$\theta_{\mathrm{MF}}(\chi)=\kappa|\alpha_c|^2(e^{\chi}-1)$ and
$c_2/c_1=c_3/c_1=1$; for a single Kerr resonator the exact cumulants and their
divergence at the dissipative critical point are known~\cite{kewming2022diverging}. The self-consistent Gaussian
(Hartree--Fock--Bogoliubov) approximation is \emph{not} Poissonian: its
anomalous moment $m$ makes the fluctuations squeezed, and its counting
statistics, obtained from the tilted quadratic generator, already carry
nontrivial cumulants. It cannot produce the cumulants generated by the interaction beyond Gaussian
order; those are what the diagrams add. We
compute it for arrays from the tilted moment equations: with
$\xi=\delta a-m$ the fluctuation about the tilted mean, $\mu=\bar\alpha+m$ the
total mean, $n_{ij}=\avg{\xi_i\dg\xi_j}$, $q_{ij}=\avg{\xi_i\xi_j}$ and
$\varepsilon=\kappa_c(e^{\chi}-1)$, the tilted state stays Gaussian and its
moments obey a closed Riccati system whose tilt terms are
\begin{align}
  \dot m_j &\supset \varepsilon\big[\mu_c^{*}q_{jc} + \mu_c n_{cj}\big],
    \nonumber\\
  \dot n_{ij} &\supset \varepsilon\big[q^{*}_{ic}q_{jc}
     + n_{ic}n_{cj}\big], \nonumber\\
  \dot q_{ij} &\supset \varepsilon\big[n_{ci}q_{jc} + q_{ic}n_{cj}\big],
  \label{eq:riccati}
\end{align}
with $\theta_{\mathrm{G}}(\chi)=\varepsilon(|\mu_c|^2+n_{cc})$ at stationarity.
The cost is $O(K^2)$ variables, so the Gaussian baseline is available at every
array size. It reproduces the Fock-space diagonalization of the same quadratic
model to five digits at $K=1$ and, at $K=2$, to the cutoff accuracy of that
reference~\cite{SM}.

\begin{figure}[t]
\centering
\includegraphics[width=\columnwidth]{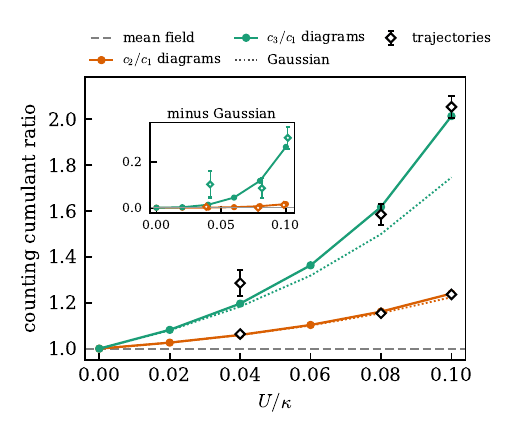}
\caption{Counting statistics of the light emitted from site $0$ of a
disordered eight-site Kerr chain ($\Delta_j=-\kappa+0.3\kappa\,\xi_j$,
$J=0.4\kappa$; the drive acts on site $0$ only, which is also the counted
site; the eight detunings are listed in the Supplemental
Material~\cite{SM}): Fano factor $c_2/c_1$ and skewness ratio $c_3/c_1$
versus $U$. Mean-field theory gives the dashed line at one; the Gaussian
(HFB) baseline is Eq.~\eqref{eq:riccati}; the diagrams are the full recursion
Eq.~\eqref{eq:thetaN} through sixth order, the highest computed. The inset shows
the diagrammatic values minus the Gaussian ones. Diamonds with error bars:
quantum-jump trajectories at $U=0.04$, $0.08$, and $0.1\kappa$ ($1216$,
$3072$, and $3552$ trajectories; total counting times $1.2$, $3.1$, and
$3.6\times10^{6}/\kappa$), one standard error. The tilted
Liouvillian of this chain has dimension $n_{\max}^{16}$ and cannot be
diagonalized directly.}
\label{fig:fcs}
\end{figure}

The tilt fits the construction of Sec.~\ref{sec:diagrams} exactly. In the
displaced frame the extra term of Eq.~\eqref{eq:tilted} reads
\begin{align}
  &\varepsilon\left[\,\bd_c X b_c + \alpha_c^{*}X b_c + \alpha_c \bd_c X
      + |\alpha_c|^2 X\,\right], \nonumber\\
  &\qquad \varepsilon = \kappa\left(e^{\chi}-1\right),
  \label{eq:tiltdisplaced}
\end{align}
which \emph{only raises} the label. The tilted generator is therefore upper
triangular on the label lattice, and three consequences follow at once. Its
spectrum is $\Lambda^{\chi}_{mn}=\Lambda_{mn}+\varepsilon|\alpha_c|^2$, so the
leading eigenvalue of the Gaussian problem is the Poissonian
$\theta_0(\chi)=\kappa|\alpha_c|^2(e^{\chi}-1)$. The propagator is unchanged,
$\theta_0-\Lambda^{\chi}_{mn}=zn+z^{*}m$. And the left eigenoperator is the
resummation of the raising ladder, which for one mode is the normal-ordered
Gaussian
\begin{equation}
  Y_{00} = \;\normord{\exp\!\left[\frac{\varepsilon}{\kappa}\bd b
     + \frac{\varepsilon\alpha^{*}}{z}\bd
     + \frac{\varepsilon\alpha}{z^{*}}b\right]}\;.
  \label{eq:Ychi}
\end{equation}
The interacting corrections are Rayleigh--Schr\"odinger perturbation theory
for the leading eigenvalue, with $Y_{00}$ as the unperturbed eigenoperator and
the same vertices as before. Writing $\theta=\theta_0+\sum_{N\geq1}\theta^{(N)}$
for the eigenvalue and $\psi=\sum_{N\geq0}\psi_N$ with $\psi_0=Y_{00}$ for the
eigenoperator,
\begin{align}
  \psi_N &= \mathcal G_\chi\Big[\Lint\dg\psi_{N-1}
      - \sum_{k=1}^{N-1}\theta^{(k)}\psi_{N-k}\Big], \nonumber\\
  \theta^{(N)} &= \big[\Lint\dg\psi_{N-1}\big]_{Y},
  \label{eq:thetaN}
\end{align}
where $[\,\cdot\,]_{Y}$ is the component along $Y_{00}$ and $\mathcal G_\chi$
is the resolvent $(\theta_0-\Lop_{0,\chi}\dg)^{-1}$ of the Gaussian part of
the tilted generator on the complement of $Y_{00}$: it removes the $Y_{00}$
direction, not the $O_{00}$ label, before propagating. The renormalization
terms $\theta^{(k)}\psi_{N-k}$ vanish for the untilted steady state, where
every $\theta^{(k)}=0$, which is why they are absent from
Eq.~\eqref{eq:recursion}; in the tilted problem they first contribute at
fourth order for one mode and at third order for several, and they shift
$c_2$ and $c_3$ but not $c_1$. The cumulants are obtained from $\theta(\chi)$
by a Cauchy integral on a circle $|\chi|=r$ with $r\leq0.1$, for the exact,
Gaussian, and diagrammatic $\theta$ alike; the radius is limited by the exact
single cavity, whose leading tilted eigenvalue nearly crosses the next one at
$\chi\approx0.2$ for $U=0.1\kappa$~\cite{SM}.

Table~\ref{tab:fcs} compares the expansion with the exact tilted eigenvalue
for one and three cavities. The ordering is the same in both: mean-field
theory fails first, the Gaussian approximation captures most of the effect at
weak nonlinearity, and the diagrams improve on it order by order. For the
single cavity at $U=0.05\kappa$ the diagrams through eighth order reproduce
the Fano factor to $8\times10^{-6}$ and $c_3/c_1$ to $1.5\times10^{-4}$. At
$U=0.1\kappa$ the Gaussian error on $c_3/c_1$ is $25\%$ and the eighth-order
diagrammatic error $7\%$. For the three-site ring the error on $c_3/c_1$
falls through every order computed, from $0.087$ at $N\leq3$ and $0.040$ at
$N\leq4$ to $7\times10^{-3}$ at $N\leq6$, where the Fano factor is accurate to
$3.4\times10^{-4}$.

\begin{table*}[t]
\caption{Counting cumulants against the exact tilted Liouvillian. Upper
block: single driven Kerr cavity, $\Delta_c=-\kappa$, $\eta=\kappa$. Lower
block: a three-site ring, $\Delta_j=-\kappa$, $J=0.4\kappa$, drive and counting
on one site, $U=0.05\kappa$. All cumulants are derivatives of $\theta(\chi)$
obtained by a Cauchy integral on a circle in the complex $\chi$ plane. Exact
values are computed in the displaced frame $a=\alpha+b$ and converged in the
fluctuation cutoff. The diagrams use the full recursion Eq.~\eqref{eq:thetaN};
each row gives the sum through the stated order. Mean field is the self-consistent classical
solution, which emits Poissonian light (for the ring, the $U=0$ displacement,
as marked); the Gaussian entry is the tilted quadratic model frozen at the
self-consistent $\alpha$, $n$, $m$ of the Hartree--Fock--Bogoliubov steady
state.}
\label{tab:fcs}
\begin{ruledtabular}
\begin{tabular}{llccc}
 & method & $c_1/\kappa$ & $c_2/c_1$ & $c_3/c_1$ \\ \hline
$K=1$, $U=0.05\kappa$   & exact                             & $0.8596$ & $1.0749$ & $1.2511$ \\
                        & mean field                        & $0.8576$ & $1.0000$ & $1.0000$ \\
                        & Gaussian FCS                      & $0.8596$ & $1.0730$ & $1.2296$ \\
                        & diagrams $N\leq8$                 & $0.8596$ & $1.0749$ & $1.2510$ \\[2pt]
$K=1$, $U=0.10\kappa$   & exact                             & $0.9460$ & $1.2702$ & $2.3585$ \\
                        & mean field                        & $0.9327$ & $1.0000$ & $1.0000$ \\
                        & Gaussian FCS                      & $0.9453$ & $1.2292$ & $1.7759$ \\
                        & diagrams $N\leq8$                 & $0.9458$ & $1.2629$ & $2.1848$ \\[2pt]
$K=3$, $U=0.05\kappa$   & exact                             & $1.1458$ & $1.1653$ & $1.6393$ \\
                        & mean field ($U=0$ displacement)   & $1.0381$ & $1.0000$ & $1.0000$ \\
                        & diagrams $N\leq2$                 & $1.1422$ & $1.1385$ & $1.4572$ \\
                        & diagrams $N\leq3$                 & $1.1450$ & $1.1558$ & $1.5520$ \\
                        & diagrams $N\leq4$                 & $1.1456$ & $1.1620$ & $1.5997$ \\
                        & diagrams $N\leq6$                 & $1.1458$ & $1.1649$ & $1.6321$ \\
\end{tabular}
\end{ruledtabular}
\end{table*}

Figure~\ref{fig:fcs} applies this to a disordered eight-site chain, where the
tilted Liouvillian cannot be diagonalized directly and the polynomial
alternatives are approximations of a different kind. The drive acts on site
$0$ only, which is also the counted site, and the detunings
$\Delta_j=-\kappa+0.3\kappa\,\xi_j$ of the single disorder realization are
listed in the Supplemental Material~\cite{SM}. Three statements follow. The
interaction-induced deviation from Poissonian statistics is large: $24\%$ on
the Fano factor and $101\%$ on $c_3/c_1$ at $U=0.1\kappa$. Most of it is
Gaussian: the squeezed HFB state accounts for $93\%$ of the deviation on
$c_2/c_1$ and $74\%$ on $c_3/c_1$ at $U=0.1\kappa$, and for more at smaller
$U$, so a Gaussian treatment is a strong competitor here and not, as for
$g^{(3)}$ in Sec.~\ref{sec:cumulant}, structurally blind. What is left over is
the non-Gaussian part, and the higher cumulant carries it: the difference
between the diagrams and the Gaussian theory grows from $0.002$ to $0.27$ on
$c_3/c_1$ between $U=0.02\kappa$ and $0.1\kappa$, seventeen to forty times the
corresponding difference on $c_2/c_1$. No exact result exists for this
chain; whether the sixth-order diagrams resolve the non-Gaussian part is
tested against quantum-jump trajectories.

Quantum-jump trajectories provide an independent benchmark, since the
recorded jumps of $a_0$ are the photon counts whose cumulants are
plotted~\cite{SM}. Runs at $U=0.04$, $0.08$, and $0.1\kappa$, with $1216$,
$3072$, and $3552$ trajectories and total counting times of $1.2$, $3.1$,
and $3.6\times10^{6}/\kappa$, use Fock cutoffs $(10,6,5,4,3,3,2,2)$,
$(12,7,5,4,3,3,2,2)$, and $(14,7,5,4,3,3,2,2)$, set by the convergence of
$c_3/c_1$ itself~\cite{SM}; the standard errors come from a bootstrap over
trajectories, and the leading $1/T$ finite-window bias of the counting
cumulants is removed by a two-window difference; the residual finite-window
bias is checked explicitly in the Supplemental Material. At $U=0.1\kappa$ the trajectories give
$c_1=0.9309(6)\kappa$, $c_2/c_1=1.2363(49)$, and $c_3/c_1=2.054(49)$. The
sixth-order diagrams give $0.9306\kappa$, $1.2401$, and $2.013$, within one
standard error on every cumulant. Third order gives
$0.9282\kappa$, $1.2075$, and $1.749$, low by $4.6$, $5.8$, and $6.2$
standard errors. The Gaussian theory gives $0.9305\kappa$, $1.2241$, and
$1.746$: $2.5$ standard errors low on the Fano factor and $6.3$ standard
errors low on $c_3/c_1$. The non-Gaussian part of the third cumulant, $0.31(5)$
by the trajectories and $0.27$ by the diagrams, is therefore resolved and
correctly given. At $U=0.08\kappa$ the trajectories give $0.9009(6)\kappa$,
$1.1541(49)$, and $1.585(45)$, the diagrams $0.9002\kappa$, $1.1613$,
and $1.617$, within $1.5$ standard errors, and the Gaussian theory
$0.9002\kappa$, $1.1545$, and $1.499$, $1.9$ standard errors low on
$c_3/c_1$. At $U=0.04\kappa$ the non-Gaussian part is $0.013$ on $c_3/c_1$,
below the trajectory error of $0.058$; the trajectories give
$0.8507(9)\kappa$, $1.0635(71)$, and $1.286(58)$, within $1.6$ standard
errors of the diagrams ($0.8497\kappa$, $1.0597$, $1.196$) and within $1.8$
of the Gaussian theory. The sixth order is the last one computed; at
$U=0.1\kappa$ the increments of $c_3/c_1$ still shrink by a factor of about
$0.65$ per order, so the series has not yet reached its smallest term there,
and the geometric tail would add about $0.1$, which stays within $1.2$
standard errors of the trajectories. Over the three couplings the
trajectories and the sixth-order diagrams differ by at most $0.6\%$
on the Fano factor and by $2\%$ on $c_3/c_1$ at $U=0.08$ and $0.1\kappa$;
at $U=0.04\kappa$, where the trajectory error on $c_3/c_1$ is $4.5\%$, the
difference is $7\%$, or $1.6$ standard errors. The benchmark therefore
confirms the diagrams within its own precision across the range of
Fig.~\ref{fig:fcs}, and it separates them from the Gaussian theory where
the two differ.

Two limits should be stated. The expansion is in $U$ as before, so the
accessible range is the weakly nonlinear one; and at fixed $U$ the expansion
in the counting field is reliable only for small $\chi$. At larger $\chi$ the
exact $\theta(\chi)$ of the single cavity jumps sharply (at $U=0.1\kappa$ from
$0.101$ at $\chi=0.1$ to $0.998$ at $\chi=0.3$) as the dominant trajectories
switch to the bright mean-field branch, a dynamical transition in the counting
field that the perturbative branch does not follow~\cite{SM}. The low
cumulants, which are what experiments measure, lie well inside the reliable
region.

\section{Discussion}
\label{sec:discussion}

The construction above is a set of Feynman rules for Lindbladians whose
solvable part is an open Gaussian system. Its scope and its limits follow from
the same structural facts.

\emph{Essential property.} The construction rests on the closure of the label
lattice under a small, fixed set of elementary moves. Their algebra is fixed by
the operator basis (with Wick-basis weights set by the unperturbed steady
covariance), while the vertex table for a specified interaction is assembled
mechanically from those moves. It is this assembly rule, rather than one
interaction-independent table of numerical coefficients, that is reused across
problems.
Additivity of the spectrum is a property of one family of propagators, the
Gaussian one; it allows the propagator to be assembled line by line rather
than evaluated on the whole label, and it is a convenience rather than a
condition. In the eigenoperator labels the propagator is always a scalar
(Sec.~\ref{sec:propagator}); what varies is whether it factorizes over the
lines and how long an expansion a given observable needs.

Within the Gaussian family, Secs.~\ref{sec:gaussmode}--\ref{sec:general} show
how far additivity reaches: correlated multimode decay, thermal occupation,
squeezed reservoirs and parametric drives, continuum reservoirs, and an
emitter driven to arbitrary strength. The price is that the eigenoperators
become Wick ordered rather than normal ordered, with contractions supplied by
the steady-state moments $N$ and $M$ or, for many modes, the covariance $V$.
This relocates all temperature and squeezing dependence from the propagators
into the vertices, and can be avoided by dressing the line instead. Two
failures remain, both noted above: a non-diagonalizable drift matrix
(Sec.~\ref{sec:general}), for which there is no eigenbasis, and a jump
operator acting on two sectors at once (Sec.~\ref{sec:drivenatom}), for which
the correction term in Eq.~\eqref{eq:leibniz} is not a c-number and the Wick
construction fails. Neither is a matter of additivity.

\emph{Limits.} The method requires short expansions of both the observable and
the initial state in the respective eigenbases. For the observables usually
wanted this is automatic: Sec.~\ref{sec:recombine} turns $\avg{(\ad)^ma^n}$
into a finite combination of eigenoperator moments with $p\leq m$ and
$q\leq n$, whatever the steady state of $\Lop_0$. The requirement is
restrictive for states far from the unperturbed steady state. The natural
domain is therefore small deviations from a steady state, which is also where
the diagram count is smallest.

\emph{Behaviour of the series.} The dressed emitter series converges rapidly;
the bare one has a finite radius set by saturation, exposed by the diagrams as
a geometric ladder; the Kerr series is asymptotic. Every Kerr series in this
paper is therefore quoted at a stated order and compared with an exact result
wherever one exists. For the single cavity the most accurate order lies
beyond sixteen at $U\leq0.1\kappa$ and falls to nine at $U=0.2\kappa$ and five
at $U=0.3\kappa$; on the three-site ring a lower order is more accurate than
the tenth for $g^{(3)}$ from $U\approx0.175\kappa$. Where no exact result
exists, as for the eight-site chain, the adequacy of a given order has to be
established independently, as the trajectories do in Sec.~\ref{sec:fcs}.
Borel or Pad\'e resummation are the next tools at stronger coupling.

\emph{Steady-state selection.} Taking $t_0\to-\infty$ presumes that the
perturbed system has the same number of steady states as the unperturbed one.
If $\Lint$ splits a degenerate steady-state manifold, or creates additional
stationary solutions, the expansion under this limit diverges and the correct
branch must be selected separately. The intermediate returns to the $(0,0)$
sector disposed of by rule~6 are the benign version of this issue; the
divergent version is a breakdown of the expansion.

\emph{Domain of application.} The examples locate the method's domain. The
diagrams are appropriate when four conditions hold: the generator splits into
a Gaussian part and a weak nonlinearity, with $U|\alpha|^2$ small against the
linewidth; the Gaussian part involves many modes, so that the density matrix
is out of reach; the quantity wanted is a correlator beyond the second
moments, for which a Gaussian closure is wrong by construction; and the
coupling is weak enough that the series is still improving at the order
computed.
Coupled-cavity arrays measured by photon counting are the standard instance.
Outside that domain, at strong nonlinearity, blockade-regime numerics are
appropriate; for intensities alone a cumulant closure is adequate and simpler;
and on a small ring at moderate coupling, third- and fourth-order cumulant
closures are more accurate than the fourth-order diagrams at comparable cost
(Sec.~\ref{sec:cumulant}). Where the diagrams are not required they still add
legibility: each correction is a picture, the loop count identifies the terms
mean-field theory misses, the vertex table is assembled once for the chosen
interaction and unperturbed basis, and one set of diagrams serves every moment
of the same order.

\emph{Outlook.} Next targets are waveguide QED with several emitters, where the
dressed lines of Sec.~\ref{sec:emitter} combine with continuum lines and the
bosonization $\sigma_z\approx-I$ can be replaced by a controlled expansion,
and two-time correlation functions, which the finite-time rules in Laplace
space reach directly and which would extend the method to spectra.

\begin{acknowledgments}
D.S. acknowledges support from the U.S. Department of Energy, Office of
Science, under Award No.~DE-SC0025910, and from the National Science
Foundation under Award No.~2529700. The funders played no role in study
design, data collection, analysis and interpretation of data, or the writing
of this manuscript. OpenAI ChatGPT (GPT-5.6 Sol) was used under author-directed
prompts in the final manuscript review to assist with consistency checks of the
text, code, numerical reproducibility, and limited wording revisions. The
authors independently inspected and, where applicable, tested every
AI-assisted output and retain full responsibility for the content and results.
\end{acknowledgments}

\section*{Author contributions}
P.E. conceived the diagrammatic approach and developed its initial
formulation, including the Gaussian eigenoperators, the vertex shift
relations, and the diagram rules for a damped mode. P.H.N. generalized the
formulation into the form presented here, carried out the calculations
and numerical checks, and wrote the manuscript. D.S. supervised the project.
All authors discussed the results and contributed to the final manuscript.

\section*{Data availability}
The source code, processed benchmark outputs, numerical result files, and
notebooks and scripts used to generate the figures and tables in this article
and the Supplemental Material are publicly available in the associated
repository~\cite{deposit}. The repository contains the
symbolic vertex machinery, the single-mode and multimode diagram engines, the
tensor implementation used for the disordered arrays, the tilted engines,
exact and Gaussian reference calculations, cumulant closures, the quantum-jump
solver and estimators, and the processed results used in the comparisons. 

\bibliography{refs}

\clearpage
\setcounter{section}{0}
\setcounter{subsection}{0}
\setcounter{equation}{0}
\setcounter{figure}{0}
\setcounter{table}{0}
\renewcommand{\thesection}{S\Roman{section}}
\renewcommand{\theequation}{S\arabic{equation}}
\renewcommand{\thefigure}{S\arabic{figure}}
\renewcommand{\thetable}{S\Roman{table}}
\makeatletter
\renewcommand{\theHsection}{supp.\arabic{section}}
\renewcommand{\theHsubsection}{supp.\arabic{section}.\arabic{subsection}}
\renewcommand{\theHequation}{supp.\arabic{equation}}
\renewcommand{\theHfigure}{supp.\arabic{figure}}
\renewcommand{\theHtable}{supp.\arabic{table}}
\makeatother
\newcommand{\meqn}[1]{(\ref{#1})}
\newcommand{\msec}[1]{\ref{#1}}
\newcommand{\mfig}[1]{\ref{#1}}
\newcommand{\mtab}[1]{\ref{#1}}

\onecolumngrid
\begin{center}
{\large\bfseries Supplemental Material for ``Operator-language Feynman rules for driven-dissipative quantum systems:\\ from mean field to non-Gaussian photon correlations''\par}
\vspace{0.65em}
Peter Ehlers, Phi Hung Nguyen, and Daniel Soh\par
\vspace{0.25em}
James C.\ Wyant College of Optical Sciences, University of Arizona, Tucson, Arizona 85721, USA\par
\vspace{0.25em}
{\small Corresponding author: danielsoh@arizona.edu\quad (Dated: \today)}
\end{center}
\vspace{0.6em}
\twocolumngrid

\noindent Equation, figure, section, and table numbers without an S prefix
refer to the main text; the notation is that of the main text throughout.

\section{Further solvable starting points}

\subsection{Continuum limit: waveguide and cavity QED}
\label{sec:continuum}

The multimode results carry over to a continuum of modes by replacing dot
products with integrals: $\Omega\to\omega(k)\delta(k-k')$, $X\to X(k,k')$, and
$\bvec{a}\to a(k)$ with $[a(k),a\dg(k')]=\delta(k-k')$. The eigenoperators
become
\begin{equation}
  O_{k_1\ldots k_m,k'_1\ldots k'_n}
    = \prod_{i=1}^{m}a_L\dg(k_i)\prod_{j=1}^{n}a_L(k'_j),
\end{equation}
with eigenvalues
$-\sum_{j}\gamma(k'_j)-\sum_i\gamma^{*}(k_i)$. Each diagram line then carries a
continuous momentum label and each vertex an integral over it, recovering the
familiar structure of a field-theoretic expansion. This is the form relevant
to waveguide QED and to structured-reservoir problems.

\subsection{Pure dephasing}
\label{sec:nonlinear}

Everything above assumes jump operators linear in the mode operators. That
assumption buys eigenoperators with additive eigenvalues, but additivity is
not what the expansion needs; what it needs is a resolvent one can write down,
and the classification of those is deferred to Sec.~\msec{sec:propagator}. The
one nonlinear dissipator we treat is pure dephasing, $\gamma_\phi\Dop[\ad a]$,
which keeps the monomials as eigenoperators.

Using Eq.~\meqn{eq:Ddagcompact} with $L=\ad a$ and
$[\ad a,O_{mn}]=(m-n)O_{mn}$,
\begin{equation}
  \Dop\dg[\ad a]\,O_{mn} = -\tfrac{1}{2}(m-n)^2\,O_{mn},
  \label{eq:dephasing}
\end{equation}
so the monomials remain exact eigenoperators and the damped mode with dephasing
is diagonalized in closed form,
\begin{equation}
  \Lambda_{mn} = -(zn+z^{*}m) - \tfrac{\gamma_\phi}{2}(m-n)^2,
\end{equation}
correctly leaving the photon number untouched ($m=n$) while broadening every
coherence. The spectrum is quadratic rather than linear in the labels. Nothing
in the diagram rules changes on that account: the propagator is still the
scalar $-1/\Lambda$. What one gives up is only the ability to assemble
$\Lambda$ one line at a time.

One caution concerns $\Lint$ rather than $\Lop_0$. If the dephasing is placed
in $\Lint$ instead, the pruning rule Eq.~\meqn{eq:killrule}, which was derived
for a commutator vertex, no longer applies to it, because $\Dop\dg$ is not a
commutator; configurations that a Hamiltonian interaction cannot close are
closed by a dissipative one, and the enumeration must be redone accordingly.

\section{Thermal, squeezed, and finite-temperature eigenoperators}
\label{app:thermal}

This section gives the derivations summarized in Sec.~\msec{sec:singlemode}
of the main text and the finite-temperature two-level eigenoperators omitted
from Sec.~\msec{sec:tls}.

\subsection{Vacuum loss}

For
\begin{equation}
  \Lop_0\rho = -i\omega[\ad a,\rho] + \gamma\,\Dop[a]\rho ,
  \label{eq:L0single}
\end{equation}
$c=a$, $z=\tfrac{\gamma}{2}+i\omega$, the steady state is the vacuum, and
$N=M=0$. Equation~\meqn{eq:Omn} reduces to the normal-ordered monomials
$O_{mn}=(\ad)^ma^n$ with $\Lambda_{mn}=-(zn+z^{*}m)$, Eq.~\meqn{eq:Omnvac},
which can be checked directly: the recycling term $\ad O_{mn}a$ cancels the
normal-ordered part of the anticommutator, leaving
$-\tfrac{\gamma}{2}(m+n)O_{mn}$. The right eigenoperators are the finite sums
$r_{mn}$ of Eq.~\meqn{eq:rmn}. The population--coherence operators
$\ketbra{p}{q}$ are not eigenoperators of the damped mode, because the jump
term feeds population downward; Eq.~\meqn{eq:rmn} is the resummation of that
cascade, and a worked expansion of $\ketbra{2}{2}$ is
$\ketbra{2}{2}=2r_{22}+2r_{11}+r_{00}$. Having only annihilation operators in
the jumps is not by itself sufficient for the vacuum case: a parametric term
$a^2+(\ad)^2$ in $H_0$ creates squeezing, so the full drift must be
diagonalized and the general construction of Sec.~\msec{sec:general} used.

\subsection{Thermal loss and gain}

For a mode coupled to a bath of mean occupation $\bar n$,
\begin{equation}
  \Lop_0\rho = -i\omega[\ad a,\rho] + \gamma(\bar n+1)\Dop[a]\rho
             + \gamma \bar n\,\Dop[\ad]\rho,
  \label{eq:Lthermal}
\end{equation}
the damped mode is still $c=a$ with the same $z$, and $N=\bar n$, $M=0$.
Acting on the plain monomials the generator is triangular,
\begin{equation}
  \Lop_0\dg (\ad)^ma^n = -(zn+z^{*}m)(\ad)^ma^n + \gamma \bar n\, m n\,
  (\ad)^{m-1}a^{n-1}.
\end{equation}
The ansatz $O^{\mathrm{th}}_{mn}=\sum_k c_k (\ad)^{m-k}a^{n-k}$, with
$\Lambda_{(m-k)(n-k)}-\Lambda_{mn}=k(z+z^{*})=k\gamma$, gives the recursion
$c_{k+1}=-\bar n\,c_k(m-k)(n-k)/(k+1)$, whose solution is
\begin{equation}
  O^{\mathrm{th}}_{mn} = \!\!\sum_{k=0}^{\min(m,n)}\!\!
    \frac{m!\,n!\,(-\bar n)^k}{k!\,(m-k)!\,(n-k)!}\,(\ad)^{m-k}a^{\,n-k},
  \label{eq:Othermal}
\end{equation}
the $M=0$ case of Eq.~\meqn{eq:Omn}, with unchanged eigenvalues
$\Lambda_{mn}=-(zn+z^{*}m)$: the deformation is strictly lower triangular and
cannot change eigenvalues. For $m=n=1$, $O^{\mathrm{th}}_{11}=\ad a-\bar n$ is
the photon number measured from its thermal value.

On the Schr\"odinger side the thermal state expands in the vacuum right
eigenoperators $r_{kk}$ of Eq.~\meqn{eq:rmn} as
\begin{equation}
  \rho_{\mathrm{th}} = \sum_{k=0}^{\infty}\frac{\bar n^k}{k!}\,r_{kk}
   = \frac{1}{\bar n+1}\left(\frac{\bar n}{\bar n+1}\right)^{\ad a},
  \label{eq:rhothermal}
\end{equation}
and the thermal right eigenoperators
$R^{\mathrm{th}}_{mn}=(m!\,n!)^{-1}(\mathrm{ad}_{\ad})^m(-\mathrm{ad}_a)^n\rho_{\mathrm{th}}$
of Eq.~\meqn{eq:Rmn} are, in the vacuum basis,
\begin{equation}
  R^{\mathrm{th}}_{mn} = \sum_{k=0}^{\infty} \bar n^k\,
    \frac{(m+k)!\,(n+k)!}{m!\,n!\,k!}\; r_{m+k,n+k},
  \label{eq:Rthermal}
\end{equation}
because $-\mathrm{ad}_a$ and $\mathrm{ad}_{\ad}$ raise the indices of
$r_{mn}$, $(-\mathrm{ad}_a)^k r_{mn}=\frac{(n+k)!}{n!}r_{m(n+k)}$ and
$(\mathrm{ad}_{\ad})^k r_{mn}=\frac{(m+k)!}{m!}r_{(m+k)n}$. The left thermal
sum, Eq.~\eqref{eq:Othermal}, terminates; the right sum is infinite because
$\rho_{\mathrm{th}}$ has support on every Fock state. The thermal state also
follows from detailed balance: setting the net probability current
$J_q=\gamma \bar n(q+1)p_q - \gamma(\bar n+1)(q+1)p_{q+1}$ to zero gives
$p_{q+1}/p_q = \bar n/(\bar n+1)$.

\subsection{Squeezed and squeezed-thermal baths}

A squeezed reservoir couples through the Bogoliubov mode
$b=\cosh r\,a+e^{i\phi}\sinh r\,\ad$,
\begin{equation}
  \Lop_0\rho = -i\omega[\ad a,\rho]
    + \gamma(\bar n_{\rm B}+1)\Dop[b]\rho + \gamma\bar n_{\rm B}\Dop[b\dg]\rho ,
  \label{eq:Lsqueezed}
\end{equation}
with $\bar n_{\rm B}$ the thermal occupation of the reservoir.
Equation~\meqn{eq:Ddagcompact} gives $\Dop\dg[b]a=-\tfrac12a$ and
$\Dop\dg[b\dg]a=+\tfrac12a$, so the thermal factors cancel in the drift and
the damped mode is again $c=a$ with $z=\tfrac{\gamma}{2}+i\omega$. Squeezing
and temperature enter only through the steady moments, obtained from
stationarity of $\ad a$ and $a^2$; the result is Eq.~\meqn{eq:NMsqueezed} of
the main text. The eigenoperators are Eq.~\meqn{eq:Omn} with these values,
and the right eigenoperators are the commutator ladders of Eq.~\meqn{eq:Rmn}
on the squeezed-thermal steady state. At $\omega=0$,
$\rho_{\rm ss}=S(\xi)\rho_{\rm th}(\bar n_{\rm B})S\dg(\xi)$; for
$\omega\neq0$ the rotation leaves $N$ unchanged and reduces and dephases $M$
by $\gamma/(\gamma+2i\omega)$. A squeezed vacuum has $N=\sinh^2r$.

\subsection{Two-level emitter at finite temperature}

For
$\Lop_0\rho = -i\epsilon[n_e,\rho]+\gamma(N+1)\Dop[\sigm]\rho+\gamma N\Dop[\sigp]\rho$
with $n_e=\sigp\sigm$, the rates become
\begin{equation}
  \Gamma_1 = \gamma(2N+1), \qquad \Gamma_2 = \Gamma_1/2,
\end{equation}
and the spectrum is $\{0,-\Gamma_2-i\epsilon,-\Gamma_2+i\epsilon,-\Gamma_1\}$.
The Heisenberg eigenoperators are $\{I,\sigm,\sigp,O_z\}$ with
\begin{equation}
  O_z = \frac{N+1}{2N+1}\ketbra{e}{e} - \frac{N}{2N+1}\ketbra{g}{g}
      = n_e - p_eI,
  \label{eq:tlsthermal}
\end{equation}
$p_e = N/(2N+1)$, and the Schr\"odinger eigenoperators are
$\{\rho^{(a)}_{\mathrm{th}},\ketbra{e}{g},\ketbra{g}{e},\ketbra{e}{e}-\ketbra{g}{g}\}$
with
$\rho^{(a)}_{\mathrm{th}} = \frac{N+1}{2N+1}\ketbra{g}{g}+\frac{N}{2N+1}\ketbra{e}{e}
= e^{-\beta n_e}/Z_a$. Unlike the bosonic case, the bath shifts the
eigenvalues: raising the temperature increases absorption and stimulated
emission together, so equilibrium is reached faster.

\section{The propagator: resumming versus expanding}

The choice of what to put in the line is free, and the same problem can be
organized either way. A thermal bath may be handled by diagonalizing to the
Wick-ordered eigenoperators of Sec.~\msec{sec:singlemode}, which makes the
propagator a scalar at the cost of $N$-dependent contraction vertices; or by
keeping the plain normal-ordered monomials, in which $\Lop_0\dg$ is
triangular and its resolvent is a path sum along the lowering rungs, which
keeps the elementary interaction moves bath-independent at the cost of a
decorated propagator. The
two agree, and the counting matches: the eigenoperator sum
Eq.~\eqref{eq:Othermal} has $\min(m,n)+1$ terms, and a ladder path starting
at $(m,n)$ can carry at most $\min(m,n)$ rungs before it reaches the boundary
of the lattice. Which is preferable depends on whether one values a
bath-independent move algebra or a scalar propagator. For machinery meant to
be reused, the bath-independent algebra is usually the better trade.

A term can also sit on either side of the split, as the Kerr nonlinearity of
Sec.~\msec{sec:kerr} illustrates. Writing $H_K = \tfrac{U}{2}\hat
n(\hat n-1)$ with $\hat n = \ad a$ and using
$f(\hat n)O_{mn}=O_{mn}f(\hat n+m-n)$ together with
$O_{mn}\hat n = O_{(m+1)(n+1)}+nO_{mn}$,
\begin{align}
  i[H_K,O_{mn}] = iU(m-n)\Big[&O_{(m+1)(n+1)} \nonumber\\
     &+ \tfrac{1}{2}(m+n-1)\,O_{mn}\Big].
  \label{eq:kerrband}
\end{align}
It has a diagonal part and a single $(+1,+1)$ rung: in the labels $O_{mn}$ the
Kerr term is itself a one-rung ladder, and could be resummed into the line
rather than expanded in. In
Sec.~\msec{sec:kerr} we treat the Kerr term as a vertex and expand in $U$; we
could equally have resummed it into the line and expanded in something else.
The elementary Kerr move is the same algebraic object either way; what changes
is whether it is iterated explicitly as a perturbative vertex or absorbed into
the propagator.

\section{Finite-time diagrams in the time domain}

The Feynman rules of the main text are stated in Laplace space and hold for an
arbitrary observable label and an arbitrary state label; the steady state is
the special case in which only $r_{00}$ survives. Away from it the simple poles
invert to sums of exponentials. At first order, for instance, a diagram
$mn\to pq$ contributes, with $\lambda_{mn}=zn+z^{*}m$,
\begin{equation}
  [\Lint\dg O_{mn}]_{pq}\,\Tr[O_{pq}\rho_0]\,
  \frac{e^{-\lambda_{mn}t}-e^{-\lambda_{pq}t}}{\lambda_{pq}-\lambda_{mn}},
  \label{eq:timedomain}
\end{equation}
which is the residue sum of the two propagators the diagram carries: the
incoming one of the observable label and the closing one of the state label.
The degenerate case $\lambda_{pq}=\lambda_{mn}$ gives $t\,e^{-\lambda_{mn}t}$ in
the usual way. At higher order the nested sums are best left to a computer
(Sec.~\ref{app:numerics}); an example is worked in Sec.~\ref{sec:kerrtime}.

\section{The Kerr cavity: linearized theory and the switch-on}

\subsection{Comparison with linearized theory}
\label{sec:linearized}

Equation~\meqn{eq:kerr2} can be checked independently. Expanding
$H_K$ about the exact mean field $\bar\alpha$ and keeping quadratic terms gives
\begin{equation}
  H_{\mathrm{quad}} = 2U|\bar\alpha|^2\delta\ad\delta a
    + \frac{U}{2}\left(\bar\alpha^2\delta\ad{}^2 + \bar\alpha^{*2}\delta a^2\right),
\end{equation}
a degenerate parametric interaction with squeezing amplitude
$\lambda = U\bar\alpha^2/2$. Solving the resulting second-moment equations in
steady state to leading order in $\lambda$ gives
$\avg{\delta\ad\delta a} = 2|\lambda|^2/|z|^2 = U^2|\alpha|^4/2|z|^2$.
Since $b$ is measured from the \emph{unperturbed} $\alpha$ rather than from
$\bar\alpha$, the diagrammatic result must also contain the displacement
$|\avg{b}^{(1)}|^2 = U^2|\alpha|^6/|z|^2$ from Eq.~\meqn{eq:kerr1}. The sum,
\begin{equation}
  \frac{U^2|\alpha|^6}{|z|^2} + \frac{U^2|\alpha|^4}{2|z|^2}
   = \frac{U^2|\alpha|^4}{|z|^2}\left(|\alpha|^2+\frac{1}{2}\right),
\end{equation}
agrees with Eq.~\meqn{eq:kerr2} term by term. The diagrammatic expansion thus
recovers both the classical Kerr shift and the Bogoliubov fluctuation number
from a single set of rules, and, unlike the linearized calculation, it
continues systematically to higher order.

\subsection{Switching on the drive}
\label{sec:kerrtime}

\begin{figure}[t]
\centering
\includegraphics[width=\columnwidth]{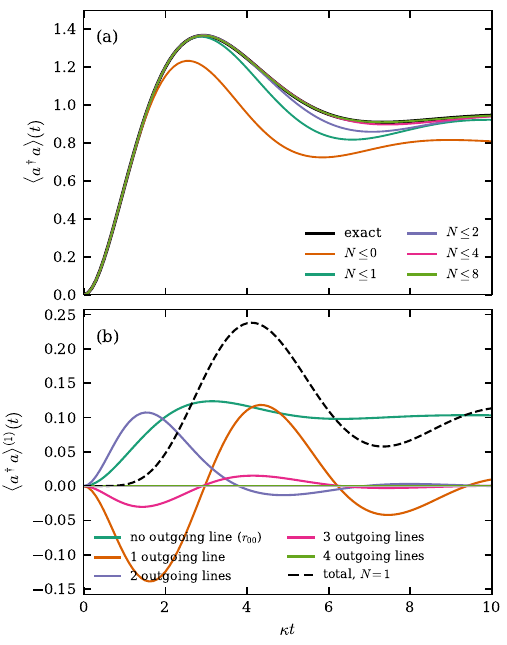}
\caption{Kerr cavity switched on from vacuum at $t=0$, with $\Delta_c=-\kappa$,
$\eta=\kappa$, and $U=0.1\kappa$. (a)~Photon number: exact evolution and partial
sums of Eq.~\meqn{eq:laplace} through order $N$. (b)~First-order correction
resolved by the number of lines leaving the diagram, i.e.\ by the degree
$p+q$ of the initial-state term $r_{qp}$ it closes on, together with their sum.}
\label{fig:kerrtime}
\end{figure}

At finite time every diagram of Fig.~\mfig{fig:kerrdiagrams} contributes. Take
the cavity empty at $t=0$ and switch on the drive, $\rho_0=\ketbra{0_a}{0_a}$.
In the displaced frame this is a coherent state of amplitude $-\alpha$, so
every initial-state term appears, with weight
$\Tr[O_{pq}\rho_0]=(-\alpha^{*})^p(-\alpha)^q$, and Eq.~\meqn{eq:laplace} sums the order-$N$ diagrams over all of them. Figure~\ref{fig:kerrtime}(a) compares
the partial sums with the exact photon number for $U=0.1\kappa$. Zeroth order
is the ring-up of the linear cavity to $|\alpha|^2$; the diagrams through $N=4$ agree with
the exact evolution to within $1.3\times10^{-2}$ over $0\leq\kappa t\leq10$,
and through $N=8$ to within $1.5\times10^{-3}$. Panel (b) resolves the
first-order correction by the number of outgoing lines. The diagram closing on
$r_{00}$ approaches the steady-state value
$2\,\mathrm{Re}[\alpha^{*}\avg{b}^{(1)}]$ at long times, while the diagrams with
outgoing lines, which carry the memory of the initial state, dominate the
transient and decay. No single initial-state term is a good approximation on
its own; the sum over them is.

\section{Chain pictures of the Kerr and bare-emitter expansions}
\label{app:chains}

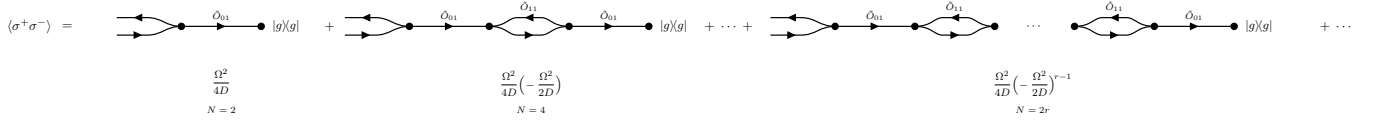
\begin{figure*}[t]
\centering
\resizebox{\textwidth}{!}{%
\begin{tikzpicture}[x=1cm,y=1cm]
\node[anchor=center,font=\normalsize] at (-4,0) {$\avg{\sigp\sigm}$};
\node[anchor=center,font=\normalsize] at (-3,0) {$=$};
\begin{scope}[shift={(0,0)}]
  \draw[ld sp,ld mark at=0.68] (0,0) .. controls (-0.42,0) and (-0.5,0.23) .. (-0.9,0.23) -- (-1.7,0.23);
  \draw[ld sm,ld mark at=0.68] (0,0) .. controls (-0.42,0) and (-0.5,-0.23) .. (-0.9,-0.23) -- (-1.7,-0.23);
  \draw[ld sm,ld mark at=0.5] (0,0) -- (2.1,0);
  \node[ld label,anchor=south,inner sep=1pt] at (1.05,0.12) {\scriptsize $\tilde O_{01}$};
  \node[ld vtx] at (0,0) {};
  \node[ld vtx] at (2.1,0) {};
  \node[ld ext,anchor=west] at (2.32,0) {$\ketbra{g}{g}$};
\end{scope}
\node[anchor=center,font=\normalsize] at (3.9,0) {$+$};
\begin{scope}[shift={(6,0)}]
  \draw[ld sp,ld mark at=0.68] (0,0) .. controls (-0.42,0) and (-0.5,0.23) .. (-0.9,0.23) -- (-1.7,0.23);
  \draw[ld sm,ld mark at=0.68] (0,0) .. controls (-0.42,0) and (-0.5,-0.23) .. (-0.9,-0.23) -- (-1.7,-0.23);
  \draw[ld sm,ld mark at=0.5] (0,0) -- (2.1,0);
  \node[ld label,anchor=south,inner sep=1pt] at (1.05,0.12) {\scriptsize $\tilde O_{01}$};
  \draw[ld sp,ld mark at=0.5] (2.1,0) .. controls (2.55,0) and (2.6,0.23) .. (2.835,0.23) -- (3.465,0.23) .. controls (3.7,0.23) and (3.75,0) .. (4.2,0);
  \draw[ld sm,ld mark at=0.5] (2.1,0) .. controls (2.55,0) and (2.6,-0.23) .. (2.835,-0.23) -- (3.465,-0.23) .. controls (3.7,-0.23) and (3.75,0) .. (4.2,0);
  \node[ld label,anchor=south,inner sep=1pt] at (3.15,0.35) {\scriptsize $\tilde O_{11}$};
  \draw[ld sm,ld mark at=0.5] (4.2,0) -- (6.3,0);
  \node[ld label,anchor=south,inner sep=1pt] at (5.25,0.12) {\scriptsize $\tilde O_{01}$};
  \node[ld vtx] at (0,0) {};
  \node[ld vtx] at (2.1,0) {};
  \node[ld vtx] at (4.2,0) {};
  \node[ld vtx] at (6.3,0) {};
  \node[ld ext,anchor=west] at (6.52,0) {$\ketbra{g}{g}$};
\end{scope}
\node[anchor=center,font=\normalsize] at (14.4,0) {$+\;\cdots\;+$};
\begin{scope}[shift={(17.2,0)}]
  \draw[ld sp,ld mark at=0.68] (0,0) .. controls (-0.42,0) and (-0.5,0.23) .. (-0.9,0.23) -- (-1.7,0.23);
  \draw[ld sm,ld mark at=0.68] (0,0) .. controls (-0.42,0) and (-0.5,-0.23) .. (-0.9,-0.23) -- (-1.7,-0.23);
  \draw[ld sm,ld mark at=0.5] (0,0) -- (2.1,0);
  \node[ld label,anchor=south,inner sep=1pt] at (1.05,0.12) {\scriptsize $\tilde O_{01}$};
  \draw[ld sp,ld mark at=0.5] (2.1,0) .. controls (2.55,0) and (2.6,0.23) .. (2.835,0.23) -- (3.465,0.23) .. controls (3.7,0.23) and (3.75,0) .. (4.2,0);
  \draw[ld sm,ld mark at=0.5] (2.1,0) .. controls (2.55,0) and (2.6,-0.23) .. (2.835,-0.23) -- (3.465,-0.23) .. controls (3.7,-0.23) and (3.75,0) .. (4.2,0);
  \node[ld label,anchor=south,inner sep=1pt] at (3.15,0.35) {\scriptsize $\tilde O_{11}$};
  \node[anchor=center] at (5.25,0) {$\cdots$};
  \draw[ld sp,ld mark at=0.5] (6.3,0) .. controls (6.75,0) and (6.8,0.23) .. (7.035,0.23) -- (7.665,0.23) .. controls (7.9,0.23) and (7.95,0) .. (8.4,0);
  \draw[ld sm,ld mark at=0.5] (6.3,0) .. controls (6.75,0) and (6.8,-0.23) .. (7.035,-0.23) -- (7.665,-0.23) .. controls (7.9,-0.23) and (7.95,0) .. (8.4,0);
  \node[ld label,anchor=south,inner sep=1pt] at (7.35,0.35) {\scriptsize $\tilde O_{11}$};
  \draw[ld sm,ld mark at=0.5] (8.4,0) -- (10.5,0);
  \node[ld label,anchor=south,inner sep=1pt] at (9.45,0.12) {\scriptsize $\tilde O_{01}$};
  \node[ld vtx] at (0,0) {};
  \node[ld vtx] at (2.1,0) {};
  \node[ld vtx] at (4.2,0) {};
  \node[ld vtx] at (6.3,0) {};
  \node[ld vtx] at (8.4,0) {};
  \node[ld vtx] at (10.5,0) {};
  \node[ld ext,anchor=west] at (10.72,0) {$\ketbra{g}{g}$};
\end{scope}
\node[anchor=center,font=\normalsize] at (30.4,0) {$+\;\cdots$};
\node[anchor=center,font=\footnotesize] at (1.05,-1.45) {$\dfrac{\Omega^2}{4D}$};
\node[anchor=center,font=\footnotesize] at (9.2,-1.5) {$\dfrac{\Omega^2}{4D}\Big(\!-\dfrac{\Omega^2}{2D}\Big)$};
\node[anchor=center,font=\footnotesize] at (22.4,-1.5) {$\dfrac{\Omega^2}{4D}\Big(\!-\dfrac{\Omega^2}{2D}\Big)^{\!r-1}$};
\node[anchor=center,font=\scriptsize] at (1.05,-2.2) {$N=2$};
\node[anchor=center,font=\scriptsize] at (9.2,-2.2) {$N=4$};
\node[anchor=center,font=\scriptsize] at (22.4,-2.2) {$N=2r$};
\end{tikzpicture}}
\caption{The bare driven emitter as a ladder. Each vertex is one action of the
drive, so only even orders survive: the chain alternates between the population
label $\tilde O_{11}$ and the coherence $\tilde O_{01}$ and closes on
$\ketbra{g}{g}$. Each pair of vertices contributes the same factor
$-\Omega^2/2D$, so the order-$2r$ chain is the $r$th term of the geometric
series of Eq.~\meqn{eq:pebare}, and summing the ladder gives
$\avg{\sigp\sigm}=(\Omega^2/4D)/(1+\Omega^2/2D)=p_e$. The dressed lines of
Sec.~\msec{sec:drivenatom} perform this resummation once and for all, which is
why the dressed expansion has no radius restriction where the bare one is
limited to $\Omega^2<2D$.}
\label{fig:laddermaster}
\end{figure*}

The single-vertex diagrams of the main text carry the order of perturbation
theory as a mark on one vertex. The same expansions can be drawn with one
vertex per order, as chains whose intermediate labels are summed. The two
pictures are related by gluing, Eq.~\meqn{eq:gluing}: the single-vertex tables
are the label-resolved content of the chains. Figure~\ref{fig:kerrmaster}
shows the chain for the photon number of the driven Kerr cavity of
Sec.~\msec{sec:kerr}, and Fig.~\ref{fig:laddermaster} the chain for the
population of the bare driven emitter of Sec.~\msec{sec:emitter}, which is
the geometric ladder of Eq.~\meqn{eq:pebare}.

\begin{figure*}[t]
\centering
\resizebox{\textwidth}{!}{%
\begin{tikzpicture}[x=1cm,y=1cm]
\node[anchor=center,font=\normalsize] at (-3.7,0) {$\avg{\bd b}$};
\node[anchor=center,font=\normalsize] at (-2.85,0) {$=$};
\begin{scope}[shift={(0,0)}]
  \draw[ld bd,ld mark at=0.68] (0,0) .. controls (-0.42,0) and (-0.5,0.23) .. (-0.9,0.23) -- (-1.7,0.23);
  \draw[ld b,ld mark at=0.68] (0,0) .. controls (-0.42,0) and (-0.5,-0.23) .. (-0.9,-0.23) -- (-1.7,-0.23);
  \node[ld vtx] at (0,0) {};
  \node[ld ext,anchor=west] at (0.22,0) {$r_{00}$};
\end{scope}
\node[anchor=center,font=\normalsize] at (1.4,0) {$+$};
\begin{scope}[shift={(3.6,0)}]
  \draw[ld bd,ld mark at=0.68] (0,0) .. controls (-0.42,0) and (-0.5,0.23) .. (-0.9,0.23) -- (-1.7,0.23);
  \draw[ld b,ld mark at=0.68] (0,0) .. controls (-0.42,0) and (-0.5,-0.23) .. (-0.9,-0.23) -- (-1.7,-0.23);
  \draw[ld bd,ld mark at=0.5] (0,0) .. controls (0.45,0) and (0.5,0.23) .. (0.805,0.23) -- (1.495,0.23) .. controls (1.8,0.23) and (1.85,0) .. (2.3,0);
  \draw[ld b,ld mark at=0.5] (0,0) .. controls (0.45,0) and (0.5,-0.23) .. (0.805,-0.23) -- (1.495,-0.23) .. controls (1.8,-0.23) and (1.85,0) .. (2.3,0);
  \node[ld label,anchor=south,inner sep=1pt] at (1.15,0.35) {\scriptsize $O_{kl}$};
  \node[ld vtx] at (0,0) {};
  \node[ld vtx] at (2.3,0) {};
  \node[ld ext,anchor=west] at (2.52,0) {$r_{00}$};
\end{scope}
\node[anchor=center,font=\normalsize] at (8.8,0) {$+$};
\begin{scope}[shift={(11,0)}]
  \draw[ld bd,ld mark at=0.68] (0,0) .. controls (-0.42,0) and (-0.5,0.23) .. (-0.9,0.23) -- (-1.7,0.23);
  \draw[ld b,ld mark at=0.68] (0,0) .. controls (-0.42,0) and (-0.5,-0.23) .. (-0.9,-0.23) -- (-1.7,-0.23);
  \draw[ld bd,ld mark at=0.5] (0,0) .. controls (0.45,0) and (0.5,0.23) .. (0.805,0.23) -- (1.495,0.23) .. controls (1.8,0.23) and (1.85,0) .. (2.3,0);
  \draw[ld b,ld mark at=0.5] (0,0) .. controls (0.45,0) and (0.5,-0.23) .. (0.805,-0.23) -- (1.495,-0.23) .. controls (1.8,-0.23) and (1.85,0) .. (2.3,0);
  \node[ld label,anchor=south,inner sep=1pt] at (1.15,0.35) {\scriptsize $O_{kl}$};
  \draw[ld bd,ld mark at=0.5] (2.3,0) .. controls (2.75,0) and (2.8,0.23) .. (3.105,0.23) -- (3.795,0.23) .. controls (4.1,0.23) and (4.15,0) .. (4.6,0);
  \draw[ld b,ld mark at=0.5] (2.3,0) .. controls (2.75,0) and (2.8,-0.23) .. (3.105,-0.23) -- (3.795,-0.23) .. controls (4.1,-0.23) and (4.15,0) .. (4.6,0);
  \node[ld label,anchor=south,inner sep=1pt] at (3.45,0.35) {\scriptsize $O_{k'l'}$};
  \node[ld vtx] at (0,0) {};
  \node[ld vtx] at (2.3,0) {};
  \node[ld vtx] at (4.6,0) {};
  \node[ld ext,anchor=west] at (4.82,0) {$r_{00}$};
\end{scope}
\node[anchor=center,font=\normalsize] at (18.4,0) {$+\;\cdots$};
\node[anchor=center,font=\footnotesize] at (0,-1.3) {$N=1$};
\node[anchor=center,font=\footnotesize] at (4.75,-1.3) {$N=2$};
\node[anchor=center,font=\footnotesize] at (13.3,-1.3) {$N=3$};
\node[anchor=center,font=\scriptsize] at (0,-1.85) {Fig.~\mfig{fig:kerrdiagrams}(a)--(f)};
\node[anchor=center,font=\scriptsize] at (9,-2.5) {one factor per vertex, intermediate label summed};
\end{tikzpicture}}
\caption{The expansion of $\avg{\bd b}$ for the driven Kerr cavity drawn with
one vertex per order. The order-$N$ amplitude has $N$ vertices, each carrying
one factor $[\Lint\dg O]_{kl}$ from the first-order table, joined by
propagators for the intermediate labels $O_{kl}$, which are summed over
subject to rule~6 of the main text; the chain closes on $r_{00}$. Reading the
$N=1$ term label by label gives Fig.~\mfig{fig:kerrdiagrams}(a)--(f) of the
main text, and gluing pairs of those by Eq.~\meqn{eq:gluing} gives the $N=2$
term, panels (g)--(n): the single-vertex tables are the label-resolved
content of the chains drawn here.}
\label{fig:kerrmaster}
\end{figure*}

\section{A cubic nonlinearity}
\label{sec:cubic}

The Kerr vertex has even degree. Our second example has odd degree, and its two
monomials contribute different vertex factors to the same diagram. Take
\begin{align}
  \Lop\rho &= -i\omega[\ad a,\rho] + \gamma\Dop[a]\rho
     + \tfrac12[f\ad - f^{*}a,\rho] \nonumber\\
     &\quad - i[H_{\mathrm{int}},\rho], \nonumber\\
  H_{\mathrm{int}} &= -\frac{ig}{2}\left[(\ad)^2a - \ad a^2\right].
  \label{eq:cubicH}
\end{align}
The drive corresponds to $\eta=f/2$ in Eq.~\meqn{eq:driveH}, so the
displacement is $\alpha=f/2z$ with $z=\gamma/2+i\omega$, and the unperturbed
problem is again the damped mode in $b=a-\alpha$.

\begin{figure*}[t]
\centering
\resizebox{0.9\textwidth}{!}{%
\begin{tikzpicture}[x=1cm,y=1cm]
\begin{scope}[shift={(0,0)}]
  \draw[ld b,ld mark at=0.5] (0,0) -- (-2,0);
  \node[ld vtx] at (0,0) {};
  \node[ld order,fill=white,fill opacity=0.85,text opacity=1,rounded corners=1pt] at (0,0.13) {$(1)$};
  \node[ld ext,anchor=east] at (-2.06,0) {$b$};
  \node[ld ext,anchor=west] at (0.22,0) {$r_{00}$};
  \node[font=\footnotesize,anchor=north] at (0,-1.05) {(a)\ $g(\alpha^2/2-|\alpha|^2)/z$};
\end{scope}
\begin{scope}[shift={(5.9,0)}]
  \draw[ld b,ld mark at=0.5] (0,0) -- (-2,0);
  \draw[ld b,ld mark at=0.5] (2,0) -- (0,0);
  \node[ld vtx] at (0,0) {};
  \node[ld order,fill=white,fill opacity=0.85,text opacity=1,rounded corners=1pt] at (0,0.13) {$(1)$};
  \node[ld ext,anchor=east] at (-2.06,0) {$b$};
  \node[ld ext,anchor=west] at (2.06,0) {$r_{10}$};
  \node[font=\footnotesize,anchor=north] at (0,-1.05) {(b)\ $g(\alpha-\alpha^{*})/z$};
\end{scope}
\begin{scope}[shift={(11.8,0)}]
  \draw[ld b,ld mark at=0.5] (0,0) -- (-2,0);
  \draw[ld bd,ld mark at=0.5] (2,0) -- (0,0);
  \node[ld vtx] at (0,0) {};
  \node[ld order,fill=white,fill opacity=0.85,text opacity=1,rounded corners=1pt] at (0,0.13) {$(1)$};
  \node[ld ext,anchor=east] at (-2.06,0) {$b$};
  \node[ld ext,anchor=west] at (2.06,0) {$r_{01}$};
  \node[font=\footnotesize,anchor=north] at (0,-1.05) {(c)\ $-g\alpha/z$};
\end{scope}
\begin{scope}[shift={(0,-2.55)}]
  \draw[ld b,ld mark at=0.5] (0,0) -- (-2,0);
  \draw[ld b,ld mark at=0.32] (2,0.26) -- (0.92,0.26) .. controls (0.5,0.26) and (0.42,0) .. (0,0);
  \draw[ld b,ld mark at=0.32] (2,-0.26) -- (0.92,-0.26) .. controls (0.5,-0.26) and (0.42,0) .. (0,0);
  \node[ld vtx] at (0,0) {};
  \node[ld order,fill=white,fill opacity=0.85,text opacity=1,rounded corners=1pt] at (0,0.13) {$(1)$};
  \node[ld ext,anchor=east] at (-2.06,0) {$b$};
  \node[ld ext,anchor=west] at (2.06,0) {$r_{20}$};
  \node[font=\footnotesize,anchor=north] at (0,-1.05) {(d)\ $g/2z$};
\end{scope}
\begin{scope}[shift={(5.9,-2.55)}]
  \draw[ld b,ld mark at=0.5] (0,0) -- (-2,0);
  \draw[ld bd,ld mark at=0.32] (2,0.26) -- (0.92,0.26) .. controls (0.5,0.26) and (0.42,0) .. (0,0);
  \draw[ld b,ld mark at=0.32] (2,-0.26) -- (0.92,-0.26) .. controls (0.5,-0.26) and (0.42,0) .. (0,0);
  \node[ld vtx] at (0,0) {};
  \node[ld order,fill=white,fill opacity=0.85,text opacity=1,rounded corners=1pt] at (0,0.13) {$(1)$};
  \node[ld ext,anchor=east] at (-2.06,0) {$b$};
  \node[ld ext,anchor=west] at (2.06,0) {$r_{11}$};
  \node[font=\footnotesize,anchor=north] at (0,-1.05) {(e)\ $-g/z$};
\end{scope}
\begin{scope}[shift={(11.8,-2.55)}]
  \draw[ld b,ld mark at=0.5] (0,0) -- (-2,0);
  \node[ld vtx] at (0,0) {};
  \node[ld order,fill=white,fill opacity=0.85,text opacity=1,rounded corners=1pt] at (0,0.13) {$(2)$};
  \node[ld ext,anchor=east] at (-2.06,0) {$b$};
  \node[ld ext,anchor=west] at (0.22,0) {$r_{00}$};
  \node[font=\footnotesize,anchor=north] at (0,-1.05) {(f)\ Eq.~\eqref{eq:cubic2}};
\end{scope}
\end{tikzpicture}}
\caption{Diagrams for $\avg{b}$ with the cubic interaction of
Eq.~\eqref{eq:cubicH}. (a)--(e)~The complete first-order table,
Eq.~\eqref{eq:cubictable}; only (a) contributes to the steady state at first
order. (f)~The second-order steady-state diagram, Eq.~\eqref{eq:cubic2},
obtained by gluing (b)--(d) to the diagrams that close them.}
\label{fig:cubicdiagrams}
\end{figure*}
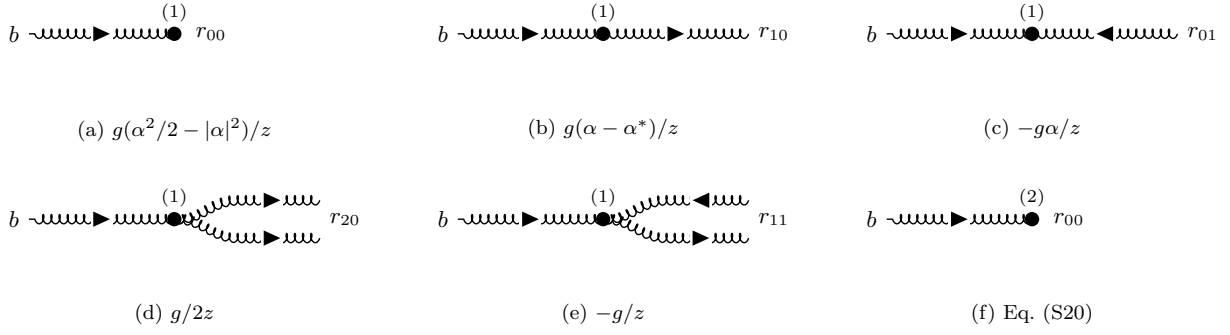

\emph{First order.} A single application of the interaction gives
\begin{align}
  \Lint\dg O_{01} &= \frac{g}{2}\left(a^2-2\ad a\right) \nonumber\\
   &= g\Big[-O_{11} - \alpha\,O_{10} + (\alpha-\alpha^{*})\,O_{01}
      + \tfrac12 O_{02} \nonumber\\
   &\qquad\ + \left(\tfrac{\alpha^2}{2}-|\alpha|^2\right)O_{00}\Big],
  \label{eq:cubictable}
\end{align}
the five first-order diagrams of Fig.~\ref{fig:cubicdiagrams}(a)--(e). The
constant term comes from both monomials: $(\ad)^2a$ absorbs the line with
either of its two creation operators and leaves $\alpha^{*}\alpha$, and
$\ad a^2$ absorbs it with its single creation operator and leaves $\alpha^2$.
In the steady state, panel (a) gives
\begin{equation}
  \avg{a} = \alpha + \frac{g}{z}\left(\frac{\alpha^2}{2}-|\alpha|^2\right) + O(g^2).
  \label{eq:cubic1}
\end{equation}

\emph{Second order.} Gluing panels (b), (c), and (d) to the first-order diagrams
that take $O_{01}$, $O_{10}$, and $O_{02}$ to $r_{00}$ gives (panel (e),
$O_{11}$, cannot be closed, by Eq.~\meqn{eq:killrule})
\begin{align}
  \avg{b}^{(2)} = {}&\frac{g^2}{z^2}\left(|\alpha|^2\alpha^{*}
      - \frac{3}{2}|\alpha|^2\alpha + \frac{\alpha^3}{2} - \frac{\alpha}{4}\right)
      \nonumber\\
    &+ \frac{g^2}{|z|^2}\left(|\alpha|^2\alpha - \frac{1}{2}|\alpha|^2\alpha^{*}\right),
  \label{eq:cubic2}
\end{align}
the value of Fig.~\ref{fig:cubicdiagrams}(f). Only the route through $O_{02}$
creates two lines and has $L=1$. By the counting of Sec.~\msec{sec:loops} with
$d=3$ it carries $|\alpha|^{N+E-2L}=|\alpha|$ against $|\alpha|^3$ for the
loop-free routes, and it is the source of the term $-\alpha/4$.

\begin{figure}[t]
\centering
\includegraphics[width=\columnwidth]{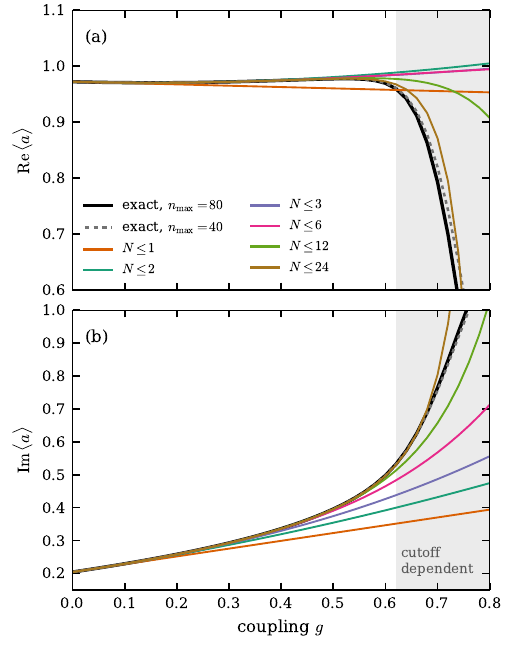}
\caption{Field amplitude for the cubic interaction with $\gamma=2.108$,
$\omega=2.153$, and $f=1.17+4.617i$ (so $\alpha=0.972+0.204i$), versus coupling
$g$. Exact steady states with Fock cutoffs $n_{\max}=80$ (black) and $40$
(dotted), and partial sums of the diagrammatic series through order $N$. In
the shaded region the two exact results differ by more than $10^{-3}$: the
truncated steady state has not converged in the cutoff, and its photon number
grows with $n_{\max}$.}
\label{fig:cubicconv}
\end{figure}

\emph{Higher orders.} Figure~\ref{fig:cubicconv} compares partial sums of
the series with exact steady states. Below $g\approx0.6$ the
agreement is systematic: at twenty-fourth order the error is
$2\times10^{-11}$ at $g=0.3$, $2\times10^{-5}$ at $g=0.5$, and $3\times10^{-3}$
at $g=0.6$, where low orders are already off by $10\%$. The exact results
require care of their own. Above $g\approx0.62$ the truncated steady state
depends on the Fock cutoff: between $n_{\max}=40$, $60$, and $80$ the mean
photon number at $g=0.8$ grows from $4.3$ to $5.5$ to $6.8$, with the
population of the highest retained level decaying only slowly, so the rapid
variation of $\avg{a}$ seen in truncated numerics there is not a converged
property of the model. This is consistent with the coefficients: the root-test
estimate $|c_N|^{-1/N}$ falls from $1.07$ at $N=10$ to $0.66$ at $N=60$ and is
still decreasing. A Hermitian cubic interaction is not bounded by the quadratic
part at large amplitude, and a plausible reading, which we have not
established, is that the state found at small $g$ is metastable and the series
asymptotic, as for the cubic anharmonic oscillator. Either way, the
diagrammatic partial sums track the converged exact results wherever those
exist.

\section{Cumulant closures for the three-site ring}
\label{app:closures}

\subsection{Construction of the closures}

The comparison of Sec.~\msec{sec:cumulant} uses the uniform three-site ring
of Sec.~\msec{sec:ring} ($\Delta=-\kappa$, $J=0.4\kappa$, $\eta=\kappa$ on
every site) with $U/\kappa$ between $0$ and $0.3$, and the on-site correlators
$g^{(2)}(0)=\avg{(a_0\dg)^2a_0^2}/\avg{a_0\dg a_0}^2$ and
$g^{(3)}(0)=\avg{(a_0\dg)^3a_0^3}/\avg{a_0\dg a_0}^3$.

The closure code treats a general bosonic lattice with Lindblad dissipation
and does not use translation symmetry. A normal-ordered monomial
$\prod_i(a_i\dg)^{m_i}\prod_i a_i^{n_i}$ on $K$ sites is stored as the
multi-index pair $(\bvec m,\bvec n)$, and operators are sparse dictionaries
of such keys. Products are brought back to normal order site by site with
$a^n(a\dg)^m=\sum_k k!\binom{n}{k}\binom{m}{k}(a\dg)^{m-k}a^{n-k}$. For every
monomial $O$ of total order $1\leq m+n\leq N_c$ the equation of motion
$d\avg{O}/dt=\avg{i[H,O]+\sum_j\kappa(a_j\dg Oa_j-\tfrac12\{a_j\dg a_j,O\})}$
is a linear combination of moments of order up to $N_c+2$. Each moment of
order $p>N_c$ is replaced by its cumulant expansion with every joint cumulant
of order larger than $N_c$ set to zero,
\begin{equation}
  \avg{X_1\cdots X_p} = -\sum_{\pi,\ |\pi|\geq2}(-1)^{|\pi|-1}(|\pi|-1)!
     \prod_{B\in\pi}\avg{X_B},
  \label{eq:cumclosure}
\end{equation}
where the sum runs over set partitions $\pi$ of the $p$ factors; blocks whose
order still exceeds $N_c$ are expanded recursively, so the closed moments are
polynomials in the retained variables. At $N_c=2$ this is the Gaussian
(Hartree--Fock--Bogoliubov) theory of the main text; at $N_c=3$ and $4$ it is
the standard higher-order truncation of
Refs.~\cite{kira2008cluster,plankensteiner2022quantumcumulants}. For $K=3$
the closures contain $27$, $83$, and $209$ complex moment variables. The
observables $\avg{(a_0\dg)^2a_0^2}$ and $\avg{(a_0\dg)^3a_0^3}$ are either
variables or, when their order exceeds $N_c$, evaluated from the same cumulant
polynomials; at $N_c=2$ this reduces $g^{(3)}$ to Wick's theorem.

The steady state is obtained by integrating the closed equations from the
product coherent state of the linear ($U=0$) problem to $t=300/\kappa$
(DOP853, relative tolerance $10^{-10}$, absolute tolerance $10^{-12}$) and
polishing the end point with a Newton-type root search (MINPACK hybrd). A
solution is accepted when the residual of the closed equations is below
$10^{-8}$; at every $U$ it was below $3\times10^{-12}$. Linear stability is
checked through the finite-difference Jacobian at the fixed point, and the
solution is translation invariant although this was never imposed.

\subsection{Exact reference}

The reference is the steady state of the full Liouvillian computed in the
displaced frame $a_j=\alpha_j+b_j$, with $\alpha_j$ the classical steady
state, and truncated at total fluctuation number $\sum_jn_j(b)\leq N_t$. The
displacement is a unitary change of frame, so only the truncation differs
from the laboratory basis. Because the state is close to coherent
($\avg{\sum_jb_j\dg b_j}=0.038$ at $U=0.3\kappa$), the truncation error falls
much faster with $N_t$ than in the undisplaced basis. With $P$ the projector
on the truncated space, every normal-ordered product obeys
$P(a\dg)^ka^lP=(Pa\dg P)^k(PaP)^l$, so the Hamiltonian, the jump terms, and
the observables are the exact projections. The density matrix is evolved to
its fixed point with a fixed-step fourth-order Runge--Kutta scheme applied
matrix-free, the step set from a power-iteration estimate of the spectral
radius; the Runge--Kutta map is a polynomial $p(h\Lop)$ with $p(0)=1$, so its
fixed point is the exact null vector of $\Lop$ for any stable step. The
iteration stops when $\max|\Lop\rho|/\max|\rho|<10^{-11}$, and $N_t$ is
raised, warm-starting from the previous cutoff, until successive $g^{(3)}$
values agree to $10^{-8}$ ($10^{-10}$ for $U\leq0.125\kappa$). The final
cutoffs are $N_t=11$ ($U=0.1\kappa$), $13$ ($0.15\kappa$), $16$
($0.2\kappa$), $19$ ($0.25\kappa$), and $21$ ($0.3\kappa$). The larger of the
last two changes in $g^{(3)}$ is at most $10^{-8}$ up to $U=0.25\kappa$ and
$8.5\times10^{-7}$ at $U=0.3\kappa$, at least three orders of magnitude below
any method error discussed here. At $U=0.02\kappa$ and $0.025\kappa$ the
result agrees with an independent calculation in the undisplaced per-site
cube $n_j<10$ to $2.4\times10^{-7}$ in $g^{(3)}$, within that calculation's
own cutoff error of $4\times10^{-6}$. The cutoff matters at the level of
the method errors compared in Fig.~\mfig{fig:compare}: a reference with a
lower cutoff gives $g^{(3)}=2.52$ at $U=0.3\kappa$, against the converged
$2.52728$.

\subsection{Validation and stability of the closures}

At $N_c=2$ the general code reproduces the momentum-space
Hartree--Fock--Bogoliubov closure to $4\times10^{-13}$ in $n$ and
$5\times10^{-12}$ in $g^{(2)}$ over all fourteen couplings. The Gaussian
three-photon moment is $\avg{(b\dg)^3b^3}=6n^3+9n|m|^2$ by Wick's theorem,
with $n=\avg{b\dg b}$ and $m=\avg{bb}$; the coefficient of $n|m|^2$ was
checked against a displaced squeezed thermal state, since the common slip
$18n|m|^2$ would make the Gaussian $g^{(3)}$ too large by $6\times10^{-5}$
at $U=0.1\kappa$ and $1.2\times10^{-2}$ at $U=0.3\kappa$, comparable to
the method errors under discussion. At $U=0$ all three orders reproduce the coherent state
($|g^{(2)}-1|$ and $|g^{(3)}-1|$ below $2\times10^{-14}$). At $U=0.02\kappa$
the relative error in $g^{(3)}$ is $3.3\times10^{-4}$, $1.1\times10^{-6}$,
and $3.6\times10^{-8}$ at $N_c=2$, $3$, and $4$, so each order gains between
one and three decades in the weakly nonlinear limit.

The known failure mode of cumulant truncations beyond second order, a loss of
positivity that produces unstable or unphysical fixed points, did not occur in
this parameter range. The $N_c=3$ and $4$ closures converge from the
coherent-state start at every $U$ up to $0.3\kappa$, the imaginary parts of
the on-site moments are below $4\times10^{-13}$, and the slowest Jacobian eigenvalue
stays at $\mathrm{Re}\,\lambda=-0.50$ to $-0.53\,\kappa$. What does appear is
a loss of convergence in the order: from $U=0.25\kappa$ on, the fourth-order
$g^{(3)}$ is worse than the third-order one, with errors of
$2.0\times10^{-2}$ and $2.7\times10^{-2}$ at $U=0.25\kappa$ and of
$1.8\times10^{-2}$ and $6.8\times10^{-2}$ at $U=0.3\kappa$. The $g^{(2)}$ of
the fourth-order closure remains the most accurate of all methods up to
$U=0.25\kappa$.

\subsection{Accuracy and cost}

Table~\ref{tab:closures} gives the relative errors against the exact steady
state at the couplings of Fig.~\mfig{fig:compare}. Against the second-order
closure, the diagrams through fourth order reduce the $g^{(3)}$ error by a
factor of $12$ at $U=0.1\kappa$ and by factors of $2.8$ to $4.8$ for
$U=0.15$ to $0.3\kappa$. Against the higher-order closures the comparison
reverses: on $g^{(3)}$ the better of the third- and fourth-order closures is
more accurate than the fourth-order diagrams at every coupling, by factors of
$5.5$, $3.7$, $2.4$, $1.6$, and $1.25$ at $U=0.1$, $0.15$, $0.2$, $0.25$, and
$0.3\kappa$. On $g^{(2)}$ the fourth-order closure is better by factors of
$20$ to $190$ up to $U=0.25\kappa$; only at $U=0.3\kappa$ are the diagrams
ahead on $g^{(2)}$, by a factor of $2.6$. Carried to tenth order, the
diagrams have a $g^{(3)}$ error smaller than that of the best closure by
factors of $103$, $11$, $2.8$, $1.8$, and $1.05$ at $U=0.1$, $0.15$, $0.2$,
$0.225$, and $0.25\kappa$, and larger by factors of $1.75$ and $2.8$ at
$U=0.275$ and $0.3\kappa$. On $g^{(2)}$ they are better than the fourth-order
closure only up to $U=0.125\kappa$ (by a factor of $2.8$ there), and on the
photon number only up to $U=0.1\kappa$; from $U=0.15\kappa$ the
fourth-order closure is more accurate on $g^{(2)}$, by a factor of $1.06$ at
$U=0.15\kappa$ rising to $22$ at $U=0.2\kappa$, and from $U=0.125\kappa$ the
better of the third- and fourth-order closures is more accurate on the photon
number, by factors of $6$ to $16$ for $U=0.15$ to $0.25\kappa$.
The tenth-order series (orders $0$ to $10$ of $n$,
$\avg{(a_0\dg)^2a_0^2}$, and $\avg{(a_0\dg)^3a_0^3}$) costs $54$~s per
coupling with the symbolic recursion and $4.2$~s for all couplings at once
in the sparse-matrix form of Sec.~\ref{app:sparse}, less than one
fourth-order closure evaluation.

\begin{table*}[t]
\caption{Relative errors of $g^{(3)}(0)$ and $g^{(2)}(0)$ on the three-site
ring against the converged exact steady state, for the cumulant closures at
order $N_c$, and the diagrams through order $N$. The notation $1.2(-2)$ means $1.2\times10^{-2}$. Cost on one
core: $0.13$, $0.6$, and $5$~s per coupling for the closures; $0.21$ and
$1.1$~s per coupling for the diagrams through second and fourth order with
the pruning of Sec.~\msec{sec:ring} ($54$~s for orders $0$ to $10$); and
$0.03$, $0.11$, and $4.2$~s through second, fourth, and tenth order for all
couplings at once in the sparse-matrix form of Sec.~\ref{app:sparse}.}
\label{tab:closures}
\begin{ruledtabular}
\begin{tabular}{llccccccc}
$U/\kappa$ & & exact & $N_c=2$ & $N_c=3$ & $N_c=4$ & $N\leq2$ & $N\leq4$ & $N\leq10$ \\ \hline
$0.10$ & $g^{(3)}$ & $1.312582$ & $1.2(-2)$ & $8.5(-4)$ & $1.8(-4)$ & $1.7(-2)$ & $1.0(-3)$ & $1.8(-6)$ \\
       & $g^{(2)}$ & $1.093641$ & $1.0(-3)$ & $9.6(-5)$ & $1.3(-6)$ & $2.9(-3)$ & $1.1(-4)$ & $1.5(-7)$ \\
       & $n$       & $0.296458$ & $4.1(-5)$ & $4.6(-7)$ & $2.8(-8)$ & $8.6(-4)$ & $3.2(-5)$ & $1.7(-8)$ \\[2pt]
$0.15$ & $g^{(3)}$ & $1.538157$ & $3.1(-2)$ & $4.7(-3)$ & $1.8(-3)$ & $5.3(-2)$ & $6.5(-3)$ & $1.6(-4)$ \\
       & $g^{(2)}$ & $1.150869$ & $3.9(-3)$ & $6.2(-4)$ & $1.4(-5)$ & $9.8(-3)$ & $7.5(-4)$ & $1.4(-5)$ \\
       & $n$       & $0.302914$ & $2.5(-4)$ & $4.7(-6)$ & $1.3(-7)$ & $3.2(-3)$ & $2.6(-4)$ & $2.0(-6)$ \\[2pt]
$0.20$ & $g^{(3)}$ & $1.831380$ & $6.1(-2)$ & $1.3(-2)$ & $8.3(-3)$ & $1.1(-1)$ & $2.0(-2)$ & $3.0(-3)$ \\
       & $g^{(2)}$ & $1.216912$ & $9.6(-3)$ & $2.2(-3)$ & $1.3(-5)$ & $2.2(-2)$ & $2.4(-3)$ & $2.8(-4)$ \\
       & $n$       & $0.310913$ & $9.4(-4)$ & $7.9(-6)$ & $1.0(-5)$ & $8.5(-3)$ & $1.2(-3)$ & $5.5(-5)$ \\[2pt]
$0.25$ & $g^{(3)}$ & $2.181138$ & $8.8(-2)$ & $2.0(-2)$ & $2.7(-2)$ & $1.8(-1)$ & $3.2(-2)$ & $1.9(-2)$ \\
       & $g^{(2)}$ & $1.290142$ & $1.7(-2)$ & $5.3(-3)$ & $1.7(-4)$ & $3.9(-2)$ & $3.7(-3)$ & $1.7(-3)$ \\
       & $n$       & $0.321042$ & $2.3(-3)$ & $3.1(-4)$ & $7.2(-5)$ & $1.8(-2)$ & $3.3(-3)$ & $5.8(-4)$ \\[2pt]
$0.30$ & $g^{(3)}$ & $2.527280$ & $8.9(-2)$ & $1.8(-2)$ & $6.8(-2)$ & $2.2(-1)$ & $2.3(-2)$ & $5.1(-2)$ \\
       & $g^{(2)}$ & $1.363850$ & $2.3(-2)$ & $1.2(-2)$ & $2.4(-3)$ & $5.4(-2)$ & $9.1(-4)$ & $2.6(-3)$ \\
       & $n$       & $0.333841$ & $3.7(-3)$ & $1.6(-3)$ & $1.0(-7)$ & $3.3(-2)$ & $7.0(-3)$ & $3.0(-3)$ \\
\end{tabular}
\end{ruledtabular}
\end{table*}

On one core the time per closure evaluation is $0.13$, $0.6$, and $5$~s at
$N_c=2$, $3$, and $4$, including generation of the equations and the
steady-state solve. For the same three-site ring, the pruned symbolic diagram
recursion takes $0.21$~s per coupling through second order and $1.1$~s per
coupling through fourth order, counting the $n$, $g^{(2)}$, and $g^{(3)}$
series. In the sparse-matrix representation, constructing and evaluating the
same series for the full coupling sweep takes $0.03$~s through second order,
$0.11$~s through fourth order, and $4.2$~s through tenth order. Thus the
third-order closure costs about $0.6$ times the pruned fourth-order diagram
calculation per coupling, whereas the fourth-order closure costs about $4.7$
times as much; the sparse representation amortizes the vertex construction
over all couplings. Shared-node timings varied with load, so these values are
reported from the fixed-load benchmark passes stored with the code. The exact
steady state takes $90$ to $600$~s on one core in the displaced basis and
grows exponentially with $K$.

A dimer tuned so that its uniform mode sits at the mean-field bistability
edge ($\Delta=-0.6\kappa$, $J=0.4\kappa$, $\eta=\kappa$) degrades the
Gaussian closure and the diagrams together: at $U=0.2\kappa$ the exact
per-site photon number is $1.374$, the closure gives $1.325$ ($4\%$) and the
diagrams through seventh order, beyond which their terms grow, give $1.272$
($7\%$). That is where the perturbative regime ends.

\section{Exact references, label count, and runtimes}
\label{app:truncation}

Every diagrammatic result in the main text is the partial sum through a
stated order: $N\leq16$ for the single Kerr cavity, $N\leq10$ for the
three-site ring, $N\leq8$ and $N\leq6$ for the counting cumulants of
Table~\mtab{tab:fcs}, and $N\leq6$ for the eight-site chain. The partial sums
at every order and, where an exact result exists, the error at every order
are stored with the code. The exact references themselves must be converged
before such errors mean anything. For the three-site counting problem the
errors of $c_3/c_1$ are $0.087$ at $N\leq3$ and $0.040$ at $N\leq4$; against
a laboratory-basis reference with the cutoffs of Sec.~\ref{app:exactref}, or
with the recursion truncated as described in Sec.~\ref{app:fcs}, the order
of these two errors would appear reversed.

\subsection{Exact references in the displaced frame}
\label{app:exactref}

In the laboratory Fock basis the truncation error of a steady state falls off
only as the amplitude of the highest retained level, not as its population.
At $U=0$ the ring steady state is exactly coherent, yet with seven levels per
site the laboratory-basis solver gives $g^{(2)}-1=-4.5\times10^{-5}$ and
$g^{(3)}-1=-5.6\times10^{-4}$, and at $U=0.15\kappa$ the same truncation
error amounts to $7.4\times10^{-4}$ in $g^{(2)}$ and $1.0\times10^{-2}$ in
$g^{(3)}$, larger than the fourth-order diagram error in
Figs.~\mfig{fig:ring} and \mfig{fig:compare}. All exact references are
therefore computed in the displaced frame $a_j=\alpha_j+b_j$, with $\alpha_j$ either the linear or
the self-consistent mean-field displacement, and only the $b$ modes are
truncated; this is exact at $U=0$ for any cutoff. With eight $b$ levels per
site, the change from seven levels is at most $10^{-6}$ in $g^{(2)}$ for
$U\leq0.15\kappa$ and $3\times10^{-4}$ at $U=0.3\kappa$; for $g^{(3)}$ it is
$8\times10^{-6}$ at $0.15\kappa$ and $5\times10^{-3}$ at $0.3\kappa$. For the
three-site counting problem of Table~\mtab{tab:fcs}, laboratory cutoffs of
$(9,6,6)$ and $(11,7,7)$ levels give $c_3/c_1=1.6037$ and $1.6358$, whereas
the displaced frame with $b$ cutoffs from $(8,5,5)$ to $(9,6,6)$ converges to
$c_3/c_1=1.63925$, stable to $6\times10^{-6}$. A laboratory-basis
calculation with cutoffs $(9,6,6)$ and five-point finite differences at
$h=0.05$ gives $1.6116$, low by $0.028$, mostly because of the cutoff.

\subsection{Label count}

A label is a normal-ordered monomial in the $2K$ operators $b_q$ and
$b_q\dg$. Each application of the quartic Kerr vertex followed by the
propagator changes the degree of a label by $+2$, $+1$, $0$, $-1$, or $-2$,
and only labels of degree $0$ contribute to the steady state. For an
observable of degree $d_0$ carried to order $N$, a label after $s$ vertices
can still reach degree $0$ only if its degree is at most $2(N-s)$, and it
cannot exceed $d_0+2s$. The largest degree ever needed is therefore
\begin{equation}
  D_{\max}=\max_s\min\big(d_0+2s,\ 2(N-s)\big)=2\lfloor N/2+d_0/4\rfloor .
  \label{eq:Dmax}
\end{equation}
The number of labels is bounded by the number of monomials of degree at most
$D_{\max}$ in $2K$ operators, $\binom{2K+D_{\max}}{D_{\max}}$, which grows as
$(2K)^{D_{\max}}/D_{\max}!$. For the on-site $g^{(2)}$ numerator ($d_0=4$)
the label count grows as $K^4$ at $N=2$ and $3$, as $K^6$ at $N=4$ and $5$,
and in general as $K^{2\lfloor N/2\rfloor+2}$; for $\avg{n}$ ($d_0=2$) it
grows as $K^2$ at $N=2$ and as $K^4$ at $N=3$ and $4$. On the
translation-invariant ring the vertex conserves lattice momentum and only
labels of zero net momentum can reach the steady state, which removes one
power of $K$. The momentum engine was instrumented for $K=3$ to $14$ and the
tensor engine for $K=3$ to $24$. Apart from the roundoff labels described
below, the largest label degree reached equals $D_{\max}$ in every case. For
$\avg{n}$ at $N=2$ the visited label count equals the bound exactly ($435$
at $K=14$); for the $g^{(2)}$ numerator at $N=4$ it is about $40\%$ of the
bound ($29\,717$ against $74\,613$ at $K=8$). The exponents fitted over the
largest three $K$ lie below the asymptotic values because the binomial bound
is still pre-asymptotic there: for the $g^{(2)}$ numerator at $N=4$ the
fitted exponent is $4.5$ for all labels and $3.6$ for zero-momentum labels,
against $4.8$ and $3.8$ for the bound over the same range.

Both engines also retain labels that carry only roundoff. The uncontracted
terms of the commutator $[H_{\mathrm{int}},X]$ cancel identically for
normal-ordered products, but the engines form both orderings and subtract
them, which leaves labels of degree $d+4$ with relative weight of order
$10^{-17}$ that the degree pruning keeps. Restricting the commutator to terms
with at least one contraction changes no result by more than $10^{-17}$
relative. In the tensor engine at fourth order it is five times faster at
$K=14$ ($6.1$~s instead of $30$~s), and for the three-site ring at tenth
order it keeps the memory of the momentum engine bounded, where the original
code grew to $0.6$~GB per worker without finishing.

\subsection{Runtimes}

All runtimes were measured on one physical core of the same shared node
(one CPU per task, one BLAS thread). Wall time equals CPU time for every
point, so the spread between passes reflects contention for cache and memory
bandwidth from other jobs; load averages were recorded for each point. The
values in Fig.~\mfig{fig:ring}(b) come from the pass at load average $23$; a
pass at load $80$ to $99$ was slower by factors of up to $5$. The observable
is the on-site $g^{(2)}$ of a uniform ring with $\Delta=-\kappa$,
$J=0.4\kappa$, $\eta=\kappa$, and $U=0.15\kappa$.

At second order the tensor engine takes $0.0031$~s at $K=4$ and $2.1$~s at
$K=32$; above $K=16$ the time grows as $K^{5.9}$, where construction of the
vertex tensor dominates, and the series evaluation alone grows as $K^{4.0}$.
At fourth order it takes $30$~s at $K=14$ and grows as $K^{6.0}$; with the
commutator restricted to contracted terms this becomes $6.1$~s, growing as
$K^{5.9}$. The momentum engine at second order grows as $K^{5.7}$ over $K=3$
to $6$ ($2.4$~s at $K=6$). The exact steady state by sparse time evolution
with five levels per site grows by a factor of $45$ per added site ($0.16$,
$5.4$, and $328$~s at $K=2$, $3$, and $4$), and with four levels per site by
a factor of $25$ ($1426$~s at $K=5$). Timings on the shared node vary by
up to a factor of two to five with its load; the values quoted in the main
text are from a single pass at recorded load, and the load averages are
stored with the raw timings in the public repository.

\subsection{Sparse-matrix evaluation of the recursion}
\label{app:sparse}

The recursion $X_N=\mathcal{V}\mathcal{G}_0X_{N-1}$ of
Eq.~\meqn{eq:recursion} is linear in $X$, so the map
$\mathcal{V}\mathcal{G}_0$ is a matrix on the space of labels. The momentum
engine rebuilds it symbolically at every order, for every observable, and for
every coupling. Building it once removes all three redundancies. The
reachable labels are enumerated layer by layer from the observable, keeping
after vertex $N$ only labels of degree at most $2(N_{\max}-N)$ and of zero net
momentum, since the vertex conserves momentum and only zero-momentum labels
can reach the label $O_{00}$; the column of $\mathcal{V}\mathcal{G}_0$ for
each label is computed once with the same algebra and stored; and the series
is $N_{\max}$ sparse matrix-vector products, the layer masks reproducing the
pruning exactly. Because the vertex is proportional to $U$, the matrix is
built at $U=1$ and the $N$th-order term is $c_NU^N$, so one run gives the
partial sums at every coupling. For the $K=3$ ring the label sets are small
($136$, $408$, $691$, $312$, $72$, $10$, and $1$ labels after vertices $0$
to $6$ for $\avg{(a_0\dg)^3a_0^3}$ at $N_{\max}=6$; $3702$ at most at
$N_{\max}=10$), and the three moments through order $2$, $4$, and $10$ take
$0.03$, $0.11$, and $4.2$~s on one core for all couplings, against $0.21$
and $1.1$~s per coupling for the pruned symbolic recursion through order $2$
and $4$ and $54$~s per coupling through order $10$. The results agree with
the symbolic recursion to $10^{-15}$ (Table~\ref{tab:checks2}). Table~\ref{tab:pruning}
shows what the two selection rules remove: without them the label set grows
geometrically with the order, with the return-to-$O_{00}$ bound it peaks at
mid order and collapses to the single label $O_{00}$ at the last vertex, and
the momentum rule removes a further two thirds. The cost is
dominated by the construction of the columns, which grows as $K^{4.9}$ at
second order and $K^{6.4}$ at fourth order over $K=6$ to $12$ ($7.5$~s and
$240$~s at $K=12$), steeper than the tensor implementation, which is
therefore kept for large arrays; the sparse form is the tool for high order
on small rings. The engine is \texttt{engines/sparse\_ring.py} in the
public repository.

\begin{table}[t]
\caption{Labels kept after each vertex for $\avg{(a_0\dg)^3a_0^3}$ on the
$K=3$ ring through order $8$, with no selection, with the
return-to-steady-state bound (degree at most $2(N_{\max}-N)$ after vertex
$N$), and with the zero-momentum rule added; the three give the same
coefficients to all digits. The time is for building the sparse matrix and
evaluating the series on one core.}
\label{tab:pruning}
\begin{ruledtabular}
\begin{tabular}{lrrr}
 & none & degree bound & + momentum \\ \hline
after vertex 0 & 400 & 400 & 136 \\
1 & 1222 & 1222 & 408 \\
2 & 3133 & 3133 & 1045 \\
3 & 7053 & 5096 & 1704 \\
4 & 14\,397 & 2913 & 975 \\
5 & 27\,222 & 924 & 312 \\
6 & 48\,397 & 210 & 72 \\
7 & 81\,793 & 28 & 10 \\
8 & 1 & 1 & 1 \\ \hline
total & 183\,618 & 13\,927 & 4663 \\
time (s) & 86 & 4.4 & 1.3 \\
\end{tabular}
\end{ruledtabular}
\end{table}

\section{Counting statistics: reference methods, validation, and the large-$\chi$ transition}
\label{app:fcs}

\subsection{Validation against the exact tilted eigenvalue}

The tilted expansion of Sec.~\msec{sec:fcs} was checked against the exact
leading eigenvalue of the tilted Liouvillian, obtained by direct
diagonalization. Three points deserve emphasis, because each is a
plausible simplification that changes the counting statistics by more than
the differences discussed in the main text.

First, the renormalization terms
$-\sum_{k=1}^{N-1}\theta^{(k)}\psi_{N-k}$ of Eq.~\meqn{eq:thetaN} cannot be
dropped. With them, the order-$8$ value of $\theta(\chi)$ for the single
driven Kerr cavity agrees with exact diagonalization to $7\times10^{-8}$;
without them it agrees only to $3\times10^{-6}$. They first matter at
order $4$ for $K=1$ and at order $3$ for $K>1$; they leave $c_1$ unchanged
and shift $c_2$ and $c_3$. For the eight-site chain of Fig.~\mfig{fig:fcs}
the values of $c_3/c_1$ through sixth order are $1.0820$, $1.1962$, $1.3632$,
$1.6174$, and $2.0134$ at $U=0.02$ to $0.10\kappa$, against $1.0822$,
$1.1991$, $1.3754$, $1.6542$, and $2.1043$ from the recursion without the
renormalization terms; at every $U$ the shift exceeds the sixth-order
increment $|S_6-S_5|$, by a factor of $74$ at $U=0.02\kappa$ falling to $1.7$
at $U=0.10\kappa$. At $N=3$ the
shift in $c_3/c_1$ is $2.5\times10^{-2}$ at $U=0.1\kappa$. The spectral projection in
Eq.~\meqn{eq:thetaN} must remove the $Y_{00}$ direction rather than the
$O_{00}$ label; removing the label instead leaves a systematic error of a few
percent that converges smoothly to the wrong answer.

Second, the cumulants are obtained from $\theta(\chi)$ by a discrete Cauchy
integral on a circle $|\chi|=r$ with $M=16$ or $32$ points, using
$\theta(\chi^{*})=\theta(\chi)^{*}$, and not by finite differences on the
real axis: five-point differences with $h=0.05$ bias $c_3/c_1$ of the exact
single cavity by $0.001$ at $U=0.05\kappa$ and by $0.07$ at $U=0.1\kappa$,
because the higher cumulants are large (the ratio $c_5/c_1$ at
$U=0.1\kappa$ is about $98$). For the exact single
cavity at $U=0.10\kappa$ the leading eigenvalue of the tilted Liouvillian
nearly crosses the next one at $\chi\approx+0.2$: the gap falls to $0.13\kappa$,
which limits the radius of convergence, and the circle radius must stay below
it. With $r=0.03$ to $0.08$ the cumulant ratios agree to $10^{-8}$; at
$r=0.15$ the value of $c_3/c_1$ is already off by $3.6\times10^{-4}$, and at
$r=0.2$ the contour result is meaningless. The diagram partial sums have no
such singularity, and their values at $r=0.1$ and $0.2$ agree to all printed
digits. For the Gaussian row, $\alpha$, $n$, and $m$ are frozen at their
untilted Hartree--Fock--Bogoliubov values and the resulting linear tilted
Liouvillian is diagonalized; its $\theta(\chi)$ is analytic, and the contour
values agree with real-axis finite differences extrapolated in $h$ ($h=0.05$
to $0.00625$) to $10^{-6}$.

Third, the exact references are computed in the displaced frame and
converged in the fluctuation cutoff, Sec.~\ref{app:exactref}. The multimode
implementation reproduces the single-mode one at $K=1$ to four digits, and
for two coupled cavities with independent counting fields the Gaussian
prediction $\theta_0=\sum_k\kappa_k|\alpha_k|^2(e^{\chi_k}-1)$ matches the
numerically computed leading eigenvalue to six digits.

The expansion in the counting field has a limited range. At $U=0.1\kappa$
the exact $\theta(\chi)$ of the single cavity rises from $0.101$ at
$\chi=0.1$ to $0.344$ at $\chi=0.2$ and $0.998$ at $\chi=0.3$, while the
perturbative branch continues smoothly; the exact values are converged in the
cutoff, so this is physical. The tilted dynamics at large $\chi$ favours
histories with many emissions, and the dominant eigenvector switches to the
bright mean-field branch: a dynamical transition in the counting field, and
the same nonperturbative structure that limits the expansion in $U$. The
cumulants quoted in the main text are derivatives at $\chi=0$, obtained on
contours of radius $r\leq0.1$ inside the smooth region.

\subsection{Method hierarchy}

Four treatments are compared in the main text. \emph{Mean field} keeps the
coherent amplitude only ($n=m=0$) and emits Poissonian light,
$\theta_{\mathrm{MF}}(\chi)=\kappa|\alpha_c|^2(e^\chi-1)$; it is the
Poissonian baseline and should not be identified with the second-order
cumulant closure. For the single cavity the self-consistent classical
solution is used; for the three-site ring of Table~\mtab{tab:fcs} the $U=0$
displacement is used, as marked there. The \emph{Gaussian} treatment solves
the untilted Hartree--Fock--Bogoliubov equations for $\alpha$, $n$, $m$,
freezes them into the quadratic fluctuation Hamiltonian
\begin{align}
  H_{\mathrm G}\supset\sum_j\Big[&2U\big(|\alpha_j|^2+n_{jj}\big)\delta a_j\dg\delta a_j
    \nonumber\\
    &+ \tfrac{U}{2}\big((\alpha_j^2+m_{jj})\delta a_j^{\dagger2}+\mathrm{H.c.}\big)\Big],
\end{align}
and computes the counting statistics of that quadratic model. The anomalous
term makes the Gaussian approximation squeezed, and therefore not
Poissonian. In the single-cavity implementation the drive of the frozen model
is set to $\eta_{\mathrm{eff}}=(\kappa/2+i\Delta_c)\bar\alpha$, which makes its
steady mean and covariance coincide with the HFB solution to five digits, and
its counting statistics are then obtained by diagonalizing the tilted
generator. For arrays the same quadratic model is solved instead through the
tilted covariance (Riccati) equations, Eq.~\meqn{eq:riccati} of the main
text, whose cost is $O(K^2)$ variables; the two routes agree to five digits at
$K=1$ and, at $K=2$, to the Fock-cutoff accuracy of the diagonalization. The
\emph{diagrams} are the expansion of the main text with the full recursion
Eq.~\meqn{eq:thetaN}, and the \emph{exact} results are leading eigenvalues of
the tilted Liouvillian in the displaced frame, converged in the fluctuation
cutoff.

\subsection{Convergence and checks}

For the three-site ring the exact tilted eigenvalue in the laboratory basis
converges slowly with the cutoff, as described in Sec.~\ref{app:exactref};
the values in Table~\mtab{tab:fcs} are from the displaced frame with $b$
cutoffs $(9,6,6)$, stable to $6\times10^{-6}$ in $c_3/c_1$. The acceptance
test applied to every tilted calculation is $c_1=\kappa\langle a_c\dg
a_c\rangle$, which the Schr\"odinger-picture generator satisfies only with
the correct sign convention $\Lop\rho=-i[H,\rho]+\Dop\rho$. The test is
not redundant: an implementation with the opposite sign still reproduces
$\theta(\chi)$ for the single Kerr cavity, because there the sign flip maps
the problem onto its complex conjugate, and is exposed only once the
Gaussian model introduces a complex anomalous term.

\subsection{Disorder realization of the eight-site chain}
\label{app:disorder}

The chain of Fig.~\mfig{fig:fcs} has $K=8$ sites with nearest-neighbour
hopping $J=0.4\kappa$, uniform loss $\kappa$, and detunings
$\Delta_j=-\kappa+0.3\kappa\,\xi_j$ drawn once from a fixed realization of the
disorder. The values used, in units of $\kappa$ and in site order
$j=0,\ldots,7$, are
\begin{align}
  \Delta_j/\kappa = {}& -1.039031,\ -0.715488,\ -0.761393,\ -0.793461,
  \nonumber\\
  & -1.064557,\ -1.004186,\ -0.893986,\ -1.263518 .
  \label{eq:disorder}
\end{align}
The coherent drive $\eta=\kappa$ acts on site $0$ only, and the counting
field is attached to the loss channel of the same site. All results in
Fig.~\mfig{fig:fcs} refer to this single realization. The diagrammatic and
Gaussian cumulants of the chain are evaluated as described above: the
diagrams on a contour $|\chi|=0.1$ (checked at $0.05$) with the full
recursion Eq.~\meqn{eq:thetaN}, the Gaussian theory by Richardson
extrapolation of real-axis differences at $h=0.05$, $0.025$, and $0.0125$,
because the tilted Riccati equations contain complex conjugates and cannot be
continued off the real axis.

\subsection{Quantum-jump benchmark of the eight-site chain}
\label{app:traj}

\emph{Unravelling.} The Lindblad equation of the chain is simulated with the
Monte Carlo wave-function method in the physical unravelling, so the recorded
jumps of $a_0$ are the photon counts whose cumulants Fig.~\mfig{fig:fcs}
shows. The no-jump evolution $\dot\psi=-iH_{\mathrm{eff}}\psi$ with
$H_{\mathrm{eff}}=H-\tfrac{i}{2}\sum_ja_j\dg a_j$ is integrated with
fixed-step fourth-order Runge--Kutta ($\delta t=0.05/\kappa$) in a truncated
Fock basis, with $512$ trajectories held as the columns of one state matrix.
Jumps follow the waiting-time rule: a jump occurs when $\|\psi\|^2$ falls
below a uniform random threshold, and the crossing time inside the step is
found by log-linear interpolation of the norm with two secant refinements, so
jump times are not tied to the step grid. The jump site is drawn with
probability proportional to $\avg{a_j\dg a_j}$. Trajectories start in the
vacuum, and the interval $t<40/\kappa$ is discarded; the linearized dynamics
relaxes at $\kappa/2$, so this is twenty relaxation times.

\emph{Estimator.} For a stationary Markovian counting process the cumulants
of the counts in a window of length $T$ obey $\kappa_n(T)=c_nT+b_n+O(e^{-gT})$,
with $g$ the Liouvillian gap, so the rate $\kappa_n(T)/T$ is biased by
$b_n/T$; for the single mode at $U=0.1\kappa$ the bias is $-0.09$ on
$c_3/c_1$ at $T=10/\kappa$. The two-window difference
$c_n=[\kappa_n(2T_0)-\kappa_n(T_0)]/T_0$ with $T_0=10/\kappa$ cancels $b_n$.
The exact finite-window cumulants of the single mode, computed from
$\Tr\,e^{\Lop_\chi T}\rho_{\mathrm{ss}}$, show that the residual bias at this
$T_0$ is below $10^{-3}$ on $c_3/c_1$; for the chain, estimates at
$T_0=7.5$, $10$, $15$, and $20/\kappa$ agree within their errors. Window
cumulants are pooled over overlapping windows with start spacing
$0.5/\kappa$, which lowers the variance by about a quarter compared with
disjoint windows. Standard errors are from a nonparametric bootstrap over
trajectories ($1000$ resamples); the scatter of twelve disjoint sub-ensembles
gives the same errors within its own uncertainty. As a check on the jump
record, $c_1$ from the counts is compared with $\kappa\avg{a_0\dg a_0}$
averaged along the same trajectories; the difference
$N_T-\int_0^T\kappa\avg{a_0\dg a_0}_t\,dt$ is a martingale with variance
$c_1T$, which fixes the standard error of that comparison.

\emph{Validation.} For the single mode ($\Delta_c=-\kappa$, $\eta=\kappa$,
cutoff $14$; $2304$ trajectories of length $5000/\kappa$ per run) and for the
three-site ring of Table~\mtab{tab:fcs} ($U=0.05\kappa$, cutoffs $(10,7,7)$;
$6400$ trajectories of length $5000/\kappa$) the estimator reproduces the
exact long-time cumulants at the same cutoff. Over the five runs (single mode
at $U=0.05$ and $0.1\kappa$, each with $\delta t=0.05$ and $0.1/\kappa$, and
the ring) the fifteen deviations in $c_1$, $c_2/c_1$, and $c_3/c_1$ all lie
within $1.6$ standard errors, with $\sum z^2=13.2$. For the ring the
trajectories give $c_1=1.1456(2)\kappa$, $c_2/c_1=1.1626(15)$, and
$c_3/c_1=1.605(15)$ against the exact values $1.14561$, $1.16356$, and
$1.62754$ at the same cutoff (the laboratory-basis cutoff $(10,7,7)$ lowers
the Fano factor by $0.0017$ and $c_3/c_1$ by $0.012$ relative to the
converged values of Table~\mtab{tab:fcs}). The deviations of $c_3/c_1$ are
negative in four of the five runs (mean $z=-0.9$), so a residual bias of
order half a standard error cannot be excluded; at the precision of the
eight-site runs it is below $0.02$. For the single mode the plain rates
$\kappa_n(T)/T$ also agree with the exact finite-window values at $T=5$,
$10$, and $20/\kappa$, all within $1.9$ standard errors, which tests the
simulation independently of the estimator.

\emph{Fock cutoff.} The cutoff must be set by the cumulants, not by the
population of the top level. For the single mode at $U=0.1\kappa$, which has
the occupation of site $0$ of the chain, a cutoff of $8$ leaves a top-level
population of $1.6\times10^{-4}$ but shifts $c_2/c_1$ by $-0.024$ and
$c_3/c_1$ by $-0.14$; cutoffs $12$ and $14$ give $-0.018$ and $-0.004$ on
$c_3/c_1$. At $U=0.08\kappa$ cutoff $12$ is converged to $4\times10^{-4}$.
The slow convergence comes from the Kerr-shifted single-photon resonance at
$n\approx|\Delta|/U$, which carries little weight but contributes to the
higher cumulants. A two-site model made of the first two chain sites fixes
the cutoff of site $1$: raising it from $5$ to $6$ changes $c_3/c_1$ by
$-0.034$ at $U=0.1\kappa$ and by $-0.0096$ at $U=0.08\kappa$, and from $6$ to
$7$ by $-0.0016$ at $U=0.08\kappa$, so $7$ is used. The production cutoffs
are $(10,6,5,4,3,3,2,2)$ at $U=0.04\kappa$, $(12,7,5,4,3,3,2,2)$ at
$U=0.08\kappa$, and $(14,7,5,4,3,3,2,2)$ at $U=0.1\kappa$ (dimensions
$43\,200$, $60\,480$, and $70\,560$); the cutoffs of sites $2$ to $7$, whose
occupations are below $0.05$, were not varied. At the production cutoffs the
population of the top level of site $0$ is below $3\times10^{-8}$.

\emph{Result.} The three production runs give, at $U=0.04\kappa$ ($1216$
trajectories, total counting time $1.22\times10^{6}/\kappa$),
$c_1=0.8507(9)\kappa$, $c_2/c_1=1.0635(71)$, and $c_3/c_1=1.286(58)$; at
$U=0.08\kappa$ ($3072$ trajectories, $3.07\times10^{6}/\kappa$),
$0.9009(6)\kappa$, $1.1541(49)$, and $1.585(45)$; and at $U=0.1\kappa$
($3552$ trajectories, $3.55\times10^{6}/\kappa$), $0.9309(6)\kappa$,
$1.2363(49)$, and $2.054(49)$. The counted $c_1$ agrees with
$\kappa\avg{a_0\dg a_0}$ from the same trajectories ($0.8497$, $0.9003$, and
$0.9306\kappa$) to within $1.1$, $1.0$, and $0.3$ martingale standard
errors ($0.0008$, $0.0005$, and $0.0005$). At $U=0.04\kappa$ the two-window estimates of $c_3/c_1$ drift
upward with the window, from $1.182(30)$ at $T_0=5/\kappa$ to $1.368(116)$
at $T_0=20/\kappa$, and a fit over $T_0=5$ to $20/\kappa$ gives $1.252(46)$;
the primary value at $T_0=10/\kappa$ is quoted. The endpoint and fitted
estimates are statistically compatible with it when their uncertainties are
combined (the largest discrepancy is $1.6$ combined standard errors), even
though the full numerical spread is wider than the primary one-standard-error
bar. At $U=0.1\kappa$ the same estimates agree within $0.04$. A
run at $U=0.1\kappa$ with the cutoff of site $0$ at $8$ instead of $14$
($1024$ trajectories) gives $0.9283(10)\kappa$, $1.2098(84)$, and
$1.742(85)$, low by $2.6$, $3.2$, and $3.7$ of its own standard errors,
which is the cutoff bias quantified above. The comparison with the diagrams and the Gaussian
theory is given in Sec.~\msec{sec:fcs} of the main text.

\section{A two-level atom in a weakly driven cavity}

\begin{figure}[t]
\centering
\resizebox{\columnwidth}{!}{%
\begin{tikzpicture}[x=1cm,y=1cm]
\begin{scope}[shift={(0,0)}]
  \draw[ld sp,ld mark at=0.5] (0,0) -- (-2,0);
  \draw[ld sp,ld mark at=0.32] (2,0.26) -- (0.92,0.26) .. controls (0.5,0.26) and (0.42,0) .. (0,0);
  \draw[ld sp,ld mark at=0.32] (2,-0.26) -- (0.92,-0.26) .. controls (0.5,-0.26) and (0.42,0) .. (0,0);
  \node[ld vtx] at (0,0) {};
  \node[ld order,fill=white,fill opacity=0.85,text opacity=1,rounded corners=1pt] at (0,0.13) {$(1)$};
  \node[ld ext,anchor=east] at (-2.06,0) {$\sigp$};
  \node[ld ext,anchor=west] at (2.06,0) {$r_{02}$};
  \node[font=\footnotesize,anchor=north] at (0,-1.05) {(a)\ atom, $\sigp$ from the left: $0$};
\end{scope}
\begin{scope}[shift={(5.9,0)}]
  \draw[ld ad,ld mark at=0.5] (0,0) -- (-2,0);
  \draw[ld ad,ld mark at=0.32] (2,0.26) -- (0.92,0.26) .. controls (0.5,0.26) and (0.42,0) .. (0,0);
  \draw[ld ad,ld mark at=0.32] (2,-0.26) -- (0.92,-0.26) .. controls (0.5,-0.26) and (0.42,0) .. (0,0);
  \node[ld vtx] at (0,0) {};
  \node[ld order,fill=white,fill opacity=0.85,text opacity=1,rounded corners=1pt] at (0,0.13) {$(1)$};
  \node[ld ext,anchor=east] at (-2.06,0) {$c\dg$};
  \node[ld ext,anchor=west] at (2.06,0) {$r_{02}$};
  \node[font=\footnotesize,anchor=north] at (0,-1.05) {(b)\ dipole, $c\dg$ from the left: $1$};
\end{scope}
\begin{scope}[shift={(0,-2.35)}]
  \draw[ld sp,ld mark at=0.5] (0,0) -- (-2,0);
  \draw[ld sp,ld mark at=0.32] (2,0.26) -- (0.92,0.26) .. controls (0.5,0.26) and (0.42,0) .. (0,0);
  \draw[ld sm,ld mark at=0.32] (2,-0.26) -- (0.92,-0.26) .. controls (0.5,-0.26) and (0.42,0) .. (0,0);
  \node[ld vtx] at (0,0) {};
  \node[ld order,fill=white,fill opacity=0.85,text opacity=1,rounded corners=1pt] at (0,0.13) {$(1)$};
  \node[ld ext,anchor=east] at (-2.06,0) {$\sigp$};
  \node[ld ext,anchor=west] at (2.06,0) {$r_{11}$};
  \node[font=\footnotesize,anchor=north] at (0,-1.05) {(c)\ atom, $\sigm$ from the left: $-1$};
\end{scope}
\begin{scope}[shift={(5.9,-2.35)}]
  \draw[ld ad,ld mark at=0.5] (0,0) -- (-2,0);
  \draw[ld ad,ld mark at=0.32] (2,0.26) -- (0.92,0.26) .. controls (0.5,0.26) and (0.42,0) .. (0,0);
  \draw[ld a,ld mark at=0.32] (2,-0.26) -- (0.92,-0.26) .. controls (0.5,-0.26) and (0.42,0) .. (0,0);
  \node[ld vtx] at (0,0) {};
  \node[ld order,fill=white,fill opacity=0.85,text opacity=1,rounded corners=1pt] at (0,0.13) {$(1)$};
  \node[ld ext,anchor=east] at (-2.06,0) {$c\dg$};
  \node[ld ext,anchor=west] at (2.06,0) {$r_{11}$};
  \node[font=\footnotesize,anchor=north] at (0,-1.05) {(d)\ dipole, $c$ from the left: $+1$};
\end{scope}
\begin{scope}[shift={(0,-4.7)}]
  \draw[ld sp,ld mark at=0.5] (0,0) -- (-2,0);
  \node[ld vtx] at (0,0) {};
  \node[ld order,fill=white,fill opacity=0.85,text opacity=1,rounded corners=1pt] at (0,0.13) {$(1)$};
  \node[ld ext,anchor=east] at (-2.06,0) {$\sigp$};
  \node[ld ext,anchor=west] at (0.22,0) {$r_{00}$};
  \node[font=\footnotesize,anchor=north] at (0,-1.05) {(e)\ atom, $\sigm$ from the left: $1$};
\end{scope}
\begin{scope}[shift={(5.9,-4.7)}]
  \draw[ld ad,ld mark at=0.5] (0,0) -- (-2,0);
  \node[ld vtx] at (0,0) {};
  \node[ld order,fill=white,fill opacity=0.85,text opacity=1,rounded corners=1pt] at (0,0.13) {$(1)$};
  \node[ld ext,anchor=east] at (-2.06,0) {$c\dg$};
  \node[ld ext,anchor=west] at (0.22,0) {$r_{00}$};
  \node[font=\footnotesize,anchor=north] at (0,-1.05) {(f)\ dipole, $c$ from the left: $1$};
\end{scope}
\end{tikzpicture}}
\caption{Where an atom and a classical dipole part ways. First-order moves on
a single left-arrow line for the atom (left column, solid lines,
Eq.~\meqn{eq:emittermoves} with $p_e=0$) and for a harmonic dipole $c$ with the
same rates (right column, wavy lines,
Eqs.~\meqn{eq:move1}--\meqn{eq:move4}).
(a),(b)~Creating a second line of the same orientation is forbidden for the
atom. (c),(d)~The creation term that accompanies an absorption has opposite
signs. (e),(f)~Absorption itself is identical. Every other element of the
diagrams is the same for both.}
\label{fig:atomdipole}
\end{figure}
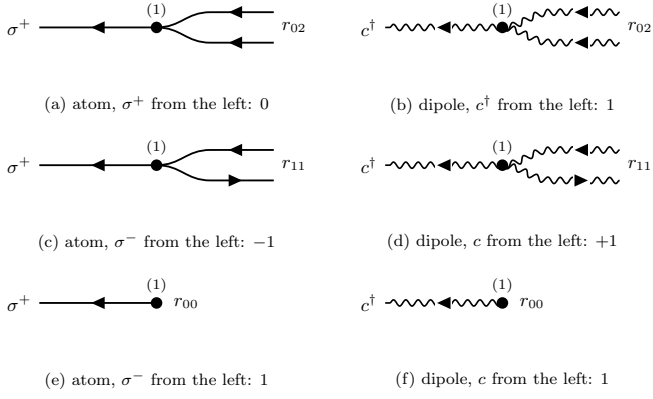

The emitter moves of Eq.~\meqn{eq:emittermoves} differ from the bosonic moves of Eqs.~\meqn{eq:move1}--\meqn{eq:move4} in exactly two ways: exclusion
forbids a second line of the same orientation, and $\sigm\sigp=1-\sigp\sigm$
gives the creation term the opposite sign. Everything that distinguishes a
two-level atom from a classical, harmonic dipole must therefore appear in the
diagrams through those two changes, and nowhere else. The comparison is
concrete: replace $\sigm$ by a bosonic mode $c$ with the same damping $\gamma$
and detuning, which is the Lorentz oscillator model and the weak-excitation
approximation $\sigma_z\approx-I$ of
Sec.~\msec{sec:drive}. The unperturbed
generator, its rates, and the shape of every diagram are then identical, and
only the vertex weights change (Fig.~\ref{fig:atomdipole}). We compare the two
in the setting where the difference is subtlest (an undriven atom seen through
a weakly driven cavity) and in the setting where it is largest (an atom driven
directly and strongly).

\subsection{Antibunching from exclusion}
\label{sec:atomcavity}

A resonant atom sits in a cavity that is driven by a laser at detuning $\Delta$
from both,
\begin{align}
  H &= \Delta\left(\ad a + \sigp\sigm\right) + g\left(\ad\sigm + a\sigp\right)
       + i(\eta\ad-\eta^{*}a), \nonumber\\
  \Lop\rho &= -i[H,\rho] + \kappa\Dop[a]\rho + \gamma\Dop[\sigm]\rho .
  \label{eq:atomcavityH}
\end{align}
We take $\Lop_0$ to be the displaced cavity of
Sec.~\msec{sec:drive} together with
the undriven atom, so that $b=a-\alpha$ with $\alpha=\eta/z$ and
$z=\kappa/2+i\Delta$ carries coil lines, the atom carries $\sigma^\mp$ lines
with $z_a=\gamma/2+i\Delta$, and the whole coupling is the perturbation,
$\Lint\dg O = ig[\bd\sigm+b\sigp+\alpha^{*}\sigm+\alpha\sigp,\,O]$. The last two
terms drive the atom through the cavity's mean field.

The first diagrams are common to the atom and the dipole. At first order the
$\alpha\sigp$ vertex closes a $\sigm$ line, giving
$\avg{\sigm}^{(1)}=-ig\alpha/z_a$, and at second order the photon line is
converted into an atomic line and closed the same way, giving
$\avg{b}^{(2)}=-g^2\alpha/(zz_a)$. For the dipole the diagrams continue as a
geometric series in $-g^2/(zz_a)$ and sum to the coherent-state result
$\avg{a}=\alpha\,zz_a/(zz_a+g^2)$, with $g^{(2)}(0)=1$ exactly: the boson moves
never generate a non-Gaussian correction. For the atom the same routes are
present, plus corrections from the two modified moves. In the photon number
these carry an extra factor of the atom's excitation, $|g\alpha/z_a|^2$, and are
negligible at weak drive. In the two-photon correlation
$\avg{(\ad)^2a^2}$ they are not: the most direct routes transfer both photons
of $b^2$ onto the atom, $b^2\to b\,\sigm\to\sigm\sigm$, and the last step is
forbidden by exclusion [Fig.~\ref{fig:atomdipole}(a)]. The atom cannot absorb
two photons at once, so
\begin{equation}
  g^{(2)}(0) = \frac{\avg{(\ad)^2a^2}}{\avg{\ad a}^2}
  = 1 + O(g^4)
  \label{eq:g2atom}
\end{equation}
for the atom, while it equals one at every order for the dipole. Both
quantities are assembled from the same diagram sets by
Sec.~\msec{sec:recombine}: with $N=M=0$ in the displaced frame, only
Eq.~\meqn{eq:undisplace} acts, and $\avg{\ad a}$ and $\avg{(\ad)^2a^2}$
differ by the weights it assigns to $O_{01}$, $O_{10}$, $O_{11}$, and the
second-order labels.

\begin{figure}[t]
\centering
\includegraphics[width=\columnwidth]{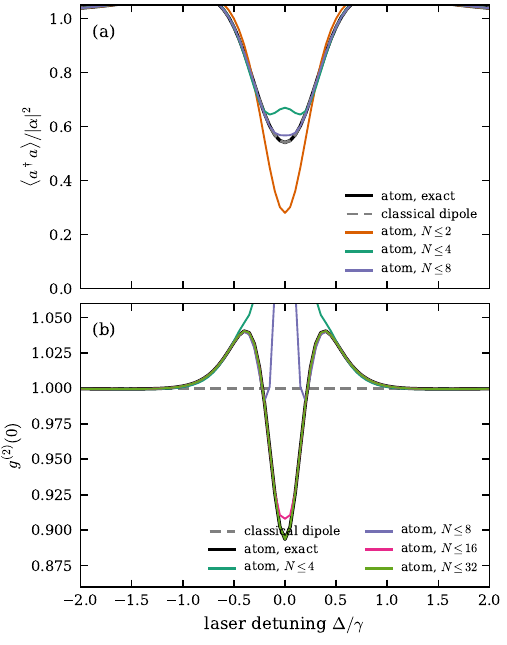}
\caption{An atom in a weakly driven cavity, Eq.~\eqref{eq:atomcavityH}, with
$\kappa=\gamma$, $g=0.3\gamma$, and $\eta=0.02\gamma$, versus the common laser
detuning. (a)~Transmitted intensity normalized by the empty cavity: exact atom,
classical dipole (indistinguishable from it on this scale), and diagrammatic
partial sums for the atom. (b)~Second-order correlation of the transmitted
light: the classical dipole gives $g^{(2)}(0)=1$ identically; the atom
antibunches on resonance and bunches on either side, and the partial sums
converge to the exact result.}
\label{fig:atomcavity}
\end{figure}

Figure~\ref{fig:atomcavity} shows both quantities for $\kappa=\gamma$,
$g=0.3\gamma$, and weak drive. In the intensity the atom and the dipole agree
to $4\times10^{-4}$ across the whole scan; transmission alone cannot tell them
apart. In $g^{(2)}(0)$ the dipole stays at one while the atom gives
$g^{(2)}(0)=0.894$ on resonance and bunching up to $1.04$ at
$\Delta\approx\pm0.4\gamma$, the single-atom photon statistics of
cavity-enhanced absorption (cf.\ Ref.~\cite{Rice1988}). The partial sums converge to the
exact result: at thirty-second order the error is below $1.1\times10^{-5}$
everywhere, and away from resonance, $|\Delta|>0.5\gamma$, sixteenth order is
already good to $4\times10^{-8}$. Near resonance convergence is slower because
the expansion parameter there is $g^2/|zz_a|=4g^2/\kappa\gamma=0.36$, the ratio
that controls how strongly atom and cavity hybridize.

\section{Dyson series without \texorpdfstring{$e^{-\Lop_0\Delta t}$}{exp(-L0 dt)}}
\label{app:dyson}

The derivation of
Eq.~\meqn{eq:dyson} in Sec.~\msec{sec:diagrams} multiplies by
$e^{-\Lop_0\Delta t}$, which is unbounded for the Gaussian generators used
here. The same result follows without it. Taylor expanding
$e^{(\Lop_0+\Lint)\Delta t}$ about $\Lint=0$ and using
\begin{equation}
  \frac{d}{dx}e^{A(x)} = \int_0^1 e^{(1-s)A(x)}\,A'(x)\,e^{sA(x)}\,ds
\end{equation}
repeatedly generates, at $N$th order, an $N$-fold integral over the ordered
simplex $0\leq s_N\leq\cdots\leq s_1\leq1$ of
$e^{\Lop_0(1-s_1)\Delta t}\Lint e^{\Lop_0(s_1-s_2)\Delta t}\cdots\Lint
e^{\Lop_0 s_N\Delta t}$, times $(\Delta t)^N$. Substituting
$\tau_i = s_i\Delta t$ reproduces
Eq.~\meqn{eq:dyson} exactly, with every
exponential carrying a nonnegative argument.

\section{Automated evaluation and numerical checks}
\label{app:numerics}

\subsection{Automated evaluation}

Equation~\meqn{eq:recursion} was evaluated with operators stored as maps from
labels $(m,n)$ to coefficients; for the emitter examples the coefficients are
$2\times2$ matrices acting on the emitter. Products of normal-ordered
monomials use
$(\bd)^mb^n(\bd)^pb^q=\sum_k k!\binom{n}{k}\binom{p}{k}(\bd)^{m+p-k}b^{n+q-k}$,
so each vertex is a commutator computed exactly on the label lattice. The
propagator map acts on the $(m,n)$ block as $(zn+z^{*}m-\Lop_a\dg)^{-1}$, which
reduces to $(zn+z^{*}m)^{-1}$ without an emitter. On the $(0,0)$ block it is
the group inverse $-(\Lop_a\dg-\mathcal{P})^{-1}(1-\mathcal{P})$ with
$\mathcal{P}X=\Tr[X\rho_{\mathrm{ss}}]\,I$, which implements rule~6; without an
emitter it is zero. A vertex of degree $d$ lowers $m+n$ by at most $d$, so
after the $N$th of $N_{\max}$ vertices every label with $m+n>d(N_{\max}-N)$
can be discarded without error. Finite-time contributions were obtained by
exponentiating the block-bidiagonal generator with $\Lop_0\dg$ on the diagonal
and $\Lint\dg$ on the superdiagonal, restricted to labels with
$m+n\leq d_0+2N_{\max}$; its $(0,N)$ block is the order-$N$ term of the Dyson
series, which is then paired with $\Tr[O_{pq}\rho_0]$ for every label,
Eq.~\meqn{eq:laplace}.

Exact steady states were obtained as the normalized null vector of the
Liouvillian in a truncated Fock basis, or, for the Kerr ring of
Sec.~\msec{sec:ring}, by evolving the master equation to stationarity in the
displaced frame (Sec.~\ref{app:exactref}). For
the atom in a driven cavity a cutoff of $10$ photons was used, and the
classical dipole was obtained both in closed form and with a truncated
oscillator of $12$ levels in place of the atom in the same code. For the cubic
interaction the cutoff dependence is the subject of Sec.~\ref{sec:cubic}.

The multimode diagram code stores labels as sparse multi-indices over the
momentum modes. The disordered chain uses a second implementation in which
the labels of a given degree sector $(p,q)$ are stored as a symmetric
$K^{p+q}$-component tensor over the mode index, the normal-ordered product of
two such tensors contracts $k$ annihilation indices of the first with $k$
creation indices of the second with the weight $k!\binom{q_1}{k}\binom{p_2}{k}$,
and sectors whose degree exceeds twice the number of remaining vertices are
never formed, since a Kerr vertex lowers the degree by at most two. The site
operators are $a_j=\alpha_j+\sum_qV_{jq}\beta_q$ with $V$ the orthonormal
eigenvectors of the hopping-plus-detuning matrix and $\alpha_j$ the linear
steady state. The site-basis cumulant closure used for comparison propagates
the mean field and the full $K\times K$ normal and anomalous covariances with
the on-site Hartree--Fock--Bogoliubov self-energies; the momentum-space
closure used for the uniform ring was integrated in time to stationarity,
and its Gaussian predictions for $g^{(2)}$ and $g^{(3)}$ follow from the
converged $\alpha$, $n$, and $m$ by Wick's theorem. The Gaussian counting
statistics of an array are obtained from the tilted moment equations quoted
in the main text, integrated to stationarity. Runtimes and label counts are
reported in Sec.~\ref{app:truncation}.

\subsection{Numerical checks}
\label{app:checks}

Table~\ref{tab:checks} collects the checks performed on the constructions of
Secs.~\msec{sec:gaussian} and \msec{sec:diagrams} of the main text and on the
diagram sums of Secs.~\msec{sec:examples} and \msec{sec:ring};
Table~\ref{tab:checks2} collects those on the exact references, the cumulant
closures, and the counting statistics. Each row names the quantity checked,
the independent reference it was compared with, and the level of agreement.
The validation of the tilted (counting-field) expansion against the exact
tilted eigenvalue is given in Sec.~\ref{app:fcs}.

\begin{table*}[t]
\caption{Numerical and symbolic checks on the constructions of Secs.~\msec{sec:gaussian}
and \msec{sec:diagrams} and on the diagram sums of Secs.~\msec{sec:examples} and
\msec{sec:ring}. Equation, table, and figure numbers without an S prefix refer to
the main text. ``Symbolic'' means exact agreement of algebraic expressions;
numerical agreement is quoted as the observed residual or relative difference.}
\label{tab:checks}
\begin{ruledtabular}
\begin{tabular}{p{0.36\textwidth}p{0.33\textwidth}p{0.25\textwidth}}
\raggedright Check & \raggedright Reference & \raggedright Agreement \tabularnewline \hline
\raggedright Single-mode Gaussian eigenoperators and bath parameters, Eqs.~\meqn{eq:Omn}--\meqn{eq:symplectic}, squeezed-thermal bath & \raggedright Numerical diagonalization of $\Lop_0\dg$ in a truncated Fock basis & \raggedright Machine precision \tabularnewline
\raggedright Multimode right eigenoperators, Eq.~\meqn{eq:Rmultimode}, two modes with overlapping decay channels and non-normal $\Gamma$ & \raggedright Numerical eigenvalue equation and biorthogonality & \raggedright Verified numerically \tabularnewline
\raggedright Dressed-emitter eigenoperators and inversion, Eqs.~\meqn{eq:blochmatrix}--\meqn{eq:dressedexpand}, on and off resonance & \raggedright Numerical diagonalization of $\Lop_a\dg$ & \raggedright Eigenvalue residuals at machine precision; $\sigma^{\mp}$ reconstructed to $6\times10^{-17}$ \tabularnewline
\raggedright General single-mode moves, Eqs.~\meqn{eq:gmove1}--\meqn{eq:gmove4}, arbitrary contractions & \raggedright Symbolic Wick multiplication & \raggedright Symbolic \tabularnewline
\raggedright Vertex composition for a squeezed-thermal steady state, Eq.~\meqn{eq:selfcontract} & \raggedright Direct multiplication of the operators & \raggedright Agrees \tabularnewline
\raggedright Vertex tables, Tables~\mtab{tab:interactions} and \mtab{tab:vertices} & \raggedright Symbolic evaluation of the commutator & \raggedright Symbolic \tabularnewline
\raggedright Inverse Wick map, Eq.~\meqn{eq:unwick}, squeezed-thermal steady state & \raggedright Numerical evaluation of both sides & \raggedright Verified numerically \tabularnewline
\raggedright Emitter inversions, Eqs.~\meqn{eq:emitterinvert} and \meqn{eq:dressedinvert} & \raggedright Dual basis of a resonantly and off-resonantly driven emitter & \raggedright Verified numerically \tabularnewline
\raggedright Loop-free part of $\avg{b}^{(N)}$, Kerr cavity, $N\leq5$ & \raggedright $U^N$ term of the classical Kerr equation $z\bar\alpha+iU|\bar\alpha|^2\bar\alpha=\eta$ & \raggedright Coincides (direct enumeration) \tabularnewline
\raggedright Order-$N$ recursion, Eq.~\meqn{eq:recursion}, Kerr cavity, $N\leq14$ & \raggedright Schr\"odinger-picture expansion $\rho^{(N)}=-\Lop_0^{+}\Lint\rho^{(N-1)}$ of the truncated Liouvillian & \raggedright Agrees \tabularnewline
\raggedright Closed forms Eqs.~\meqn{eq:kerr1}, \meqn{eq:kerr2}, \meqn{eq:kerrphoton}, \meqn{eq:emitter2}, \eqref{eq:cubic1}, \eqref{eq:cubic2}, and the diagram enumeration behind Figs.~\mfig{fig:kerrdiagrams} and \ref{fig:cubicdiagrams} & \raggedright Automated evaluation of Eq.~\meqn{eq:recursion} & \raggedright Agrees \tabularnewline
\raggedright Classical dipole, $\avg{c\dg c}$ at orders $4$--$10$ & \raggedright Closed form $\Omega^2/4D$ & \raggedright All higher-order terms zero \tabularnewline
\raggedright Kerr cavity switched on from vacuum, $U=0.1\kappa$, $0\leq\kappa t\leq10$ (Fig.~\ref{fig:kerrtime}) & \raggedright Direct integration of the master equation, cutoff $30$ & \raggedright $1.3\times10^{-2}$ ($N\leq4$), $1.5\times10^{-3}$ ($N\leq8$) \tabularnewline
\end{tabular}
\end{ruledtabular}
\end{table*}

\begin{table*}[t]
\caption{Numerical checks on the exact references, the cumulant closures, and
the counting statistics. Notation as in Table~\ref{tab:checks}.}
\label{tab:checks2}
\begin{ruledtabular}
\begin{tabular}{p{0.36\textwidth}p{0.33\textwidth}p{0.25\textwidth}}
\raggedright Check & \raggedright Reference & \raggedright Agreement \tabularnewline \hline
\raggedright Exact Kerr steady state, $\avg{\ad a}$ at $U=0.3\kappa$ & \raggedright Fock cutoffs $32$ vs.\ $45$ & \raggedright $<10^{-14}$ \tabularnewline
\raggedright Exact emitter steady state & \raggedright Fock cutoffs $8$ vs.\ $14$ & \raggedright $<10^{-10}$ \tabularnewline
\raggedright Exact $K=3$ Kerr-ring steady state, displaced frame & \raggedright $7$ vs.\ $8$ fluctuation levels per site & \raggedright $\leq10^{-6}$ in $g^{(2)}$ for $U\leq0.15\kappa$, $3\times10^{-4}$ at $U=0.3\kappa$ \tabularnewline
\raggedright Displaced-frame exact $g^{(3)}$ at $U=0.02$, $0.025\kappa$ & \raggedright Undisplaced per-site cube $n_j<10$ & \raggedright $2.4\times10^{-7}$ (within the cube's cutoff error $4\times10^{-6}$) \tabularnewline
\raggedright Displaced-frame exact steady state at $U=0$ & \raggedright Coherent state & \raggedright $|g^{(2)}-1|$, $|g^{(3)}-1|<10^{-15}$ (laboratory basis, 7 levels: $4.5\times10^{-5}$, $5.6\times10^{-4}$) \tabularnewline
\raggedright Sparse-matrix recursion (Sec.~\ref{app:sparse}), $K=3$, orders $0$--$10$, thirteen couplings & \raggedright Symbolic recursion with the same pruning & \raggedright $9\times10^{-16}$ in the partial sums of $n$, $g^{(2)}$, $g^{(3)}$ \tabularnewline
\raggedright General cumulant-closure code at $N_c=2$ & \raggedright Momentum-space HFB closure & \raggedright $4\times10^{-13}$ in $n$, $5\times10^{-12}$ in $g^{(2)}$ \tabularnewline
\raggedright Cumulant closures $N_c=2,3,4$ at $U=0$ & \raggedright Coherent state & \raggedright $|g^{(2)}-1|$, $|g^{(3)}-1|<2\times10^{-14}$ \tabularnewline
\raggedright Gaussian three-photon moment $\avg{(b\dg)^3b^3}=6n^3+9n|m|^2$ & \raggedright Displaced squeezed thermal state (QuTiP) & \raggedright Exact coefficient \tabularnewline
\raggedright Tilted recursion with renormalization terms, $K=1$, order $8$ & \raggedright Exact tilted eigenvalue & \raggedright $7\times10^{-8}$ ($3\times10^{-6}$ without the terms) \tabularnewline
\raggedright Contour cumulants, exact $K=1$ & \raggedright Radius scan $r=0.03$ to $0.08$ & \raggedright Ratios stable to $10^{-8}$ \tabularnewline
\raggedright Contour cumulants, Gaussian row & \raggedright Richardson-extrapolated finite differences & \raggedright $10^{-6}$ \tabularnewline
\raggedright Exact $K=3$ tilted eigenvalue, displaced frame & \raggedright $b$ cutoffs $(8,5,5)$ to $(9,6,6)$ & \raggedright $c_3/c_1$ stable to $6\times10^{-6}$ \tabularnewline
\raggedright Multimode diagram code at $K=1$, through fourth order & \raggedright Single-mode diagram code & \raggedright Machine precision \tabularnewline
\raggedright Tensor implementation on the uniform ring and at $K=1$ & \raggedright Momentum-space and single-mode codes & \raggedright All digits \tabularnewline
\raggedright Momentum-space cumulant closure at $K=1$ & \raggedright Single-mode closure & \raggedright Twelve digits \tabularnewline
\raggedright Site-basis cumulant closure & \raggedright Uniform-ring and single-mode closures & \raggedright Exact reduction \tabularnewline
\raggedright Gaussian counting statistics from Eq.~\meqn{eq:riccati}, $K=1$ & \raggedright Fock-space diagonalization of the tilted quadratic model, same derivative rule & \raggedright Five digits at $U=0.05\kappa$ and $0.1\kappa$ \tabularnewline
\raggedright Trajectory counting estimator, $K=1$ ($U=0.05$, $0.1\kappa$; $\delta t=0.05$, $0.1/\kappa$) and $K=3$ ring & \raggedright Exact long-time cumulants at the same cutoff & \raggedright 15 deviations within $1.6$ standard errors, $\sum z^2=13.2$ \tabularnewline
\raggedright Trajectory plain window rates $\kappa_n(T)/T$, $K=1$ & \raggedright Exact finite-window cumulants at $T=5$, $10$, $20/\kappa$ & \raggedright Within $1.9$ standard errors \tabularnewline
\raggedright Counted $c_1$ of the eight-site chain, $U=0.04$, $0.08$, $0.1\kappa$ & \raggedright $\kappa\avg{a_0\dg a_0}$ along the same trajectories & \raggedright $0.8506(8)$, $0.9008(5)$, $0.9308(5)$ vs.\ $0.8497$, $0.9003$, $0.9306$ (martingale errors) \tabularnewline
\raggedright Gaussian counting statistics from Eq.~\meqn{eq:riccati}, $K=2$ & \raggedright Fock-space diagonalization at cutoff $8$ & \raggedright $(1.0293,1.1533,1.4985)$ vs.\ $(1.0274,1.1446,1.4940)$; residual is the reference's truncation error, whose $c_1=1.0274$ is $0.2\%$ below $\kappa\avg{a_c\dg a_c}=1.0287$ \tabularnewline
\end{tabular}
\end{ruledtabular}
\end{table*}

The $K=2$ comparison in Table~\ref{tab:checks2} is the one at which the
multimode index structure of the tilt terms first matters: a transposition
of one of the two multimode tilt terms in the tilted moment equations is
invisible at $K=1$ and is caught only by this comparison. The
scripts that generate every numerical figure and table are provided with the
Supplemental Material.

\end{document}